\documentclass[usenatbib, onecolumn]{mn2e}

\usepackage{graphicx}
\usepackage{color}
\usepackage{amsmath,amssymb}
\usepackage{dcolumn}
\usepackage{bm}
\usepackage{comment}

\newcommand{\beq}{\begin{equation}}
\newcommand{\beqa}{\begin{eqnarray}}
\newcommand{\eeq}{\end{equation}}
\newcommand{\eeqa}{\end{eqnarray}}

\newcommand{\simgt}{\lower.5ex\hbox{$\; \buildrel > \over \sim \;$}}
\newcommand{\simlt}{\lower.5ex\hbox{$\; \buildrel < \over \sim \;$}}

\newcommand{\bd}[1]{\mbox{\boldmath $#1$}}

\title[Probing cosmology with weak lensing selected clusters I]
{
Probing cosmology with weak lensing selected clusters I:
Halo approach and all-sky simulations
}

\author[M. Shirasaki et al.]
{Masato Shirasaki$^{1}$\thanks{E-mail: masato.shirasaki@nao.ac.jp},
Takashi Hamana$^{1}$
and
Naoki Yoshida$^{2,3,4}$
\\
$^{1}$National Astronomical Observatory of Japan, 
Mitaka, Tokyo 181-8588, Japan \\
$^{2}$Department of Physics, School of Science, The University of Tokyo, 
7-3-1 Hongo, Bunkyo, Tokyo 113-0033, Japan \\
$^{3}$Kavli Institute for the Physics and Mathematics of the Universe (WPI), 
Todai Institutes for Advanced Study, \\
The University of Tokyo, Kashiwa, Chiba 277-8583, Japan\\
$^{4}$CREST, Japan Science and Technology Agency, 4-1-8 Honcho, Kawaguchi, Saitama, 332-0012, Japan\\
}
\begin{document}

\date{}

\volume{453} \pagerange{3043--3067} \pubyear{2015}

\maketitle

\label{3043}

\begin{abstract}
We explore a variety of statistics of clusters selected with cosmic shear measurement 
by utilizing both analytic models and large numerical simulations. 
We first develop a halo model to predict the abundance and 
the clustering of weak lensing selected clusters. 
Observational effects such as galaxy shape noise are included in our model. 
We then generate realistic mock weak lensing catalogs to test the accuracy of our analytic model. 
To this end, we perform full-sky ray-tracing simulations that allow us 
to have multiple realizations of a large continuous area. 
We model the masked regions on the sky using the actual positions of bright stars, 
and generate 200 mock weak lensing catalogs with sky coverage of $\sim$1000 squared degrees. 
We show that our theoretical model agrees well with the ensemble average of statistics 
and their covariances calculated directly from the mock catalogues. 
With a typical selection threshold, 
ignoring shape noise correction causes overestimation of the clustering 
of weak lensing selected clusters with a level of about 10\%, 
and shape noise correction boosts the cluster abundance by a factor of a few. 
We calculate the cross-covariances using the halo model with accounting 
for the effective reduction of the survey area due to masks. 
The covariance of the cosmic shear auto power spectrum is 
affected by the mode-coupling effect that originates from sky masking. 
Our model and the results can be readily used for cosmological analysis 
with ongoing and future weak lensing surveys.
\end{abstract}

\begin{keywords} 
gravitational lensing: weak 
--- cosmological parameters cosmology: theory 
--- large-scale structure in the universe
\end{keywords}

\section{INTRODUCTION}

The accelerating expansion of the universe is now established by
an array of astronomical observations such as
Type Ia supernovae 
\citep[e.g.,][]{2014A&A...568A..22B}, 
measurement of baryon acoustic oscillations in galaxy surveys
\citep[e.g.,][]{2011MNRAS.416.3017B, 2011MNRAS.418.1707B, 2014MNRAS.441...24A},
anisotropies of the cosmic microwave background (CMB)
\citep[e.g.,][]{Hinshaw2013, 2014A&A...571A..16P},
large-scale galaxy distribution
\citep[e.g.,][]{2012MNRAS.425..415S, 2014MNRAS.443.1065B},
and weak gravitational lensing \citep[e.g.,][]{Kilbinger2013}. 
In order to realize the cosmic acceleration within the
theory of general relativity, 
an exotic form of energy 
needs to be postulated to dominate in the present-day universe. 
There is another possibility to explain the cosmic acceleration without
dark energy, e.g., modified gravity theory.
Modified gravity models
do not assume an unknown energy but change 
essentially the basic equation of gravitational action.
Observationally, measurement of the growth of matter density fluctuations
will help us to distinguish the models including the Einstein gravity
with dark energy,
because the modification of gravity induce characteristic 
clustering patterns 
in matter density distribution.

Gravitational lensing is a powerful
probe of the matter distribution in the universe.
Small image distortions of distant galaxies 
are caused by intervening mass distribution.
Small distortion caused by the large-scale structure
of the universe is called cosmic shear. 
It contains, in principle, rich information 
on the matter distribution at small and large scales
and the evolution over time.
Image distortion induced by gravitational lensing is,
however, very small in general. 
Therefore, we need statistical analyses of the 
cosmic shear signal by sampling a large number of distant 
galaxies in order to extract cosmological information 
from gravitational lensing. 
The conventional statistic of cosmic shear is two-point 
correlation function or its Fourier-counterpart, power spectrum.
If the cosmic shear field obeys a Gaussian distribution, 
the two-point statistics suffice to describe all the information 
of cosmic shear. 
However, this is not the case in reality 
because cosmic shear has non-Gaussian information 
caused by non-linear gravitational growth \citep{Sato2009}.
In order to extract the full information content, 
it is desirable to use other statistical quantities
that probes nonlinear structure of
length scale of $\sim$10 Mpc or less.
Clusters of galaxies are one the most reliable objects for this purpose.

The number count of clusters is expected to be highly sensitive to 
growth of matter density perturbations \citep{1992ApJ...386L..33L},
whereas the spatial correlation of the position of clusters and cosmic shear
provides the information on the matter density profile 
as well as clustering of clusters 
\citep[e.g.,][]{2012MNRAS.420.3213O,2013ApJ...769L..35O, 2014ApJ...784L..25C}.
Fortunately, cosmic shear itself provides an efficient
way of locating galaxies of clusters.
Cluster finding methods with cosmic shear are based on 
reconstruction of matter density distribution over an area of sky
\citep{Hamana2004, 2005ApJ...624...59H, 2005A&A...442..851M, 2012MNRAS.423.1711M}.
A reconstructed mass density map can be used
to identify high density regions as ``peaks''
that are mostly caused by massive collapsed objects 
such as clusters of galaxies
\citep{2007ApJ...669..714M, 2007A&A...462..875S, 2012ApJ...748...56S}.
The unique advantage of weak lensing among various techniques
is that it does not rely on uncertain physical 
state of the baryonic component in clusters.
There have been a number of studies that investigate 
cosmological information in number counts of weak lensing selected clusters
\citep{2010A&A...519A..23M, Kratochvil2010, 2010MNRAS.402.1049D, Yang2011, 2012MNRAS.426.2870H}.
Recently, \citet{2013MNRAS.432.1338M} combined other statistics 
beyond the abundance of weak lensing selected clusters.
The authors conclude that correlation analysis of weak lensing selected 
clusters allow one to derive tight constraints on 
cosmological parameters.

In the present paper, 
we study in detail the properties of a class of statistics 
of weak lensing selected clusters.
Our study is aimed at being applied to real observations
such as Subaru Hyper-Suprime-Cam Survey.
It is well-known that 
the intrinsic ellipticities of source galaxies 
induce noise to lensing shear maps. The so-called shape noise 
causes typically false detection of clusters with cosmic shear measurement.
We develop theoretical framework 
to model the correspondence of 
underlying dark matter halos and weak lensing selected 
clusters in presence of shape noise.
Sky masking causes another important observational effect
on statistical analyses of weak lensing selected 
clusters \citep{2014ApJ...784...31L}.
Since reconstructed mass density is usually 
defined by local cosmic shear signals,
boundaries of masked regions would make the reconstruction inaccurate.
In order to realize the realistic situation in galaxy imaging surveys,
we perform gravitational lensing simulations on curved full-sky.
We then utilize these simulations to create two hundreds of mock weak lensing catalogs 
with the proposed sky coverage in ongoing Hyper Suprime-Cam (HSC) survey\footnote{
{\rm{http://www.naoj.org/Projects/HSC/index.html}} 
}.
For a large set of mock HSC surveys, we generate reconstructed mass map
and identify the local maxima on each map as an indicator of weak lensing selected clusters.
These simulations enable us to 
study the statistical property of weak lensing selected clusters
in presence of shape noise and masked region.
We are also able to examine our theoretical model through 
the large set of realistic mock weak lensing observations.

The rest of the paper is organized as follows.
In Section~\ref{sec:WL}, we describe the methodology to search clusters with
cosmic shear measurement.
There, we present the statistical property of weak lensing selected clusters and 
the theoretical model of statistics of interest.
In Section~\ref{sec:sim}, we use a large set of $N$-body simulations
to perform full-sky lensing simulations and create 
mock weak lensing maps incorporated with the information of HSC surveys.
In Section~\ref{sec:res}, we provide 
the result of our measurement of statistical quantities 
over a set of full-sky and masked sky simulations.
We also compare the simulation results and our theoretical models in detail.
Conclusions and discussions are summarized in Section~\ref{sec:con}.
 
\section{Weak lensing}\label{sec:WL}
We summarize the basics of weak gravitational lensing effect in this section.
We also describe the finder algorithm of galaxy clusters with weak lensing 
measurement.

\subsection{Basics}

When considering the observed position of a source object 
as $\mbox{\boldmath $\theta$}$ 
and the true position as $\mbox{\boldmath $\beta$}$,
one can characterize the distortion of image of a source 
object by the following 2D matrix:
\beqa
A_{ij} = \frac{\partial \beta^{i}}{\partial \theta^{j}}
           \equiv \left(
\begin{array}{cc}
1-\kappa -\gamma_{1} & -\gamma_{2}  \\
-\gamma_{2} & 1-\kappa+\gamma_{1} \\
\end{array}
\right), \label{distortion_tensor}
\eeqa
where $\kappa$ is convergence and $\gamma$ is shear.

One can relate each component of $A_{ij}$ to
the second derivative of the gravitational potential as follows
\citep{Bartelmann2001, Munshi2008};
\beqa
A_{ij} &=& \delta_{ij} - \phi_{ij}, \label{eq:Aij} \\
\phi_{ij}  &=&\frac{2}{c^2}\int _{0}^{\chi}{\rm d}\chi^{\prime} g(\chi,\chi^{\prime}) \partial_{i}\partial_{j}\Phi(\chi^{\prime}), \label{eq:shear_ten}\\	
g(\chi,\chi^{\prime}) &=& \frac{r(\chi-\chi^{\prime})r(\chi^{\prime})}{r(\chi)},
\eeqa
where $\chi$ is the comoving distance and $r(\chi)$ represents the comoving angular diameter distance.
Gravitational potential $\Phi$ can be related to matter density perturbation $\delta$ 
according to Poisson equation.
Therefore, convergence can be expressed as the weighted integral of $\delta$ along the line of sight;
\beqa
\kappa = \frac{3}{2}\left(\frac{H_{0}}{c}\right)^2 \Omega_{\rm m0} \int _{0}^{\chi}{\rm d}\chi^{\prime} g(\chi,\chi^{\prime}) \frac{\delta}{a}. \label{eq:kappa_delta}
\eeqa

\subsection{Cluster finding}
\label{subsec:WL_select_cluster}
Weak lensing provides a physical method to reconstruct the projected 
matter density field.
The conventional technique for reconstruction is based 
on the smoothed map of cosmic shear.
Let us first define the smoothed convergence field as
\beqa
{\cal K} (\bd{\theta}) = \int {\rm d}^2 \theta^{\prime} \ \kappa(\bd{\theta}-\bd{\theta}^{\prime}) U(\bd{\theta^{\prime}}), \label{eq:ksm_u}
\eeqa
where $U$ is the filter function to be specified below.
Although we can calculate the same quantity by smoothing the shear field $\gamma$ 
\citep[e.g.,][]{Shirasaki2014},
we use Eq.~(\ref{eq:ksm_u}) for simplicity in the following.

Various functional form of $U$ are proposed 
in literature \citep[e.g.,][]{Hamana2004, 2005ApJ...624...59H, 
2010A&A...519A..23M, 2012MNRAS.423.1711M}.
We consider the Gaussian filter\footnote{
In practice, 
the filter function should be compensated 
in order to remove an undetermined constant convergence
\citep{1996MNRAS.283..837S}.
In Appendix \ref{appendix:compensated}, we examine 
compensated Gaussian filters when searching for weak-lensing 
clusters. There, we show that our model 
can be suitably modified for
the case of compensated Gaussian filters.
}
\beqa
U(\theta) = \frac{1}{\pi \theta_{G}^{2}} \exp \left( -\frac{\theta^2}{\theta_{G}^2} \right).
\eeqa
With this filter, we can easily model 
the statistical properties of the contaminant 
of a smoothed ${\cal K}$ map, called shape noise.
The noise in a ${\cal K}$ map would follow the Gaussian distribution 
when one 
can use a sufficient large number of source galaxies
and when source galaxies are oriented randomly.
The Gaussian properties of the noise 
makes it easy to model the lensing peak statistics,
as will be shown in the following.

We denote the shape noise contribution to a smoothed lensing map
by ${\cal N}$.
For a given smoothing scale $\theta_{G}$,
correlation function of the shape noise after Gaussian 
smoothing is given by \citep{VanWaerbeke2000}
\begin{eqnarray}
\langle{\cal N}(\bd{\theta}){\cal N}(\bd{\theta}^{\prime})\rangle 
= \frac{\sigma_{\gamma}^2}{4\pi n_{\rm gal}\theta_{G}^2}
\exp\left[-\frac{|\bd{\theta}-\bd{\theta}^{\prime}|^2}{2\theta_{G}^2}\right],\label{eq:shape_noise_sm}
\end{eqnarray}
where $\sigma_{\gamma}$ is the rms of the intrinsic ellipticity of sources 
and $n_{\rm gal}$ represents
the number density of source galaxies.
One can derive the power spectrum of noise convergence field ${\cal N}$ by
Fourier transforming of Eq.~(\ref{eq:shape_noise_sm});
\beqa
P_{\cal N}(\ell) = \frac{\sigma_{\gamma}^2}{2n_{\rm gal}}\exp\left[-\frac{1}{2} \theta_{G}^2 \ell^2\right].
\label{eq:noise_power}
\eeqa
Using Eq.~(\ref{eq:noise_power}), we define the moment of ${\cal N}$ as
\beqa
\sigma_{{\rm noise}, i} = 
\left(\int \frac{{\rm d}^2 \bd{\ell}}{(2\pi)^2}\, \ell^{2i} P_{\cal N}(\ell)\right)^{1/2}.
\label{eq:noise_moment}
\eeqa

In a smoothed lensing map, 
peaks with high signal-to-noise ratio $\nu = {\cal K}/\sigma_{{\rm noise}, 0}$ 
are likely associated with cluster of galaxies
\citep[e.g.,][]{Hamana2004}.
We first locate high peaks on a ${\cal K}$ map 
and then relate each peak with an isolated 
massive halo along the line of sight.
We assume the following universal density profile of
dark matter halos \citep{Navarro1997}:
\beqa
\rho_{h}(r) = \frac{\rho_s}{\left(r/r_s\right)\left(1+r/r_s\right)^2}, \label{eq:rho_nfw}
\eeqa
where $r_{s}$ and $\rho_s$ represent the scale radius and the scale density, respectively.
The parameters $r_s$ and $\rho_s$  
can be essentially convolved into one parameter, 
the concentration $c_{\rm vir}(M,z)$, by the use of two halo 
mass relations; 
namely, $M=4\pi r^3_{\rm vir} \Delta_{\rm vir}(z) \rho_{\rm crit}(z)/3$, 
where $r_{\rm vir}$ is the virial radius corresponding to the overdensity 
criterion $\Delta_{\rm vir}(z)$
as shown in, e.g., \citet{Navarro1997},  
and $M= \int dV \, \rho_h (r_s, \rho_s)$ with the integral performed out to $r_{\rm vir}$.
In this paper, we adopt the functional form of the concentration parameter 
in \citet{2008MNRAS.390L..64D},
\beqa
c_{\rm vir}(M, z) = 5.72 \left( \frac{M}{10^{14} h^{-1}M_{\odot}}\right)^{-0.081}(1+z)^{-0.71}.
\eeqa

The corresponding convergence can be calculated as in \citet{Hamana2004},
\beqa
\kappa_{h}(R) = \frac{2\rho_{s}r_{s}f(R/r_{s})}{\Sigma_{\rm crit}},
\label{eq:kappa_nfw}
\eeqa
where $R$ represents  the perpendicular proper distance from the center of halo 
and $f(x)$ is
\beqa
f(x)&=&
\left\{
\begin{array}{ll}
-\frac{\sqrt{c_{\rm vir}^2-x^2}}{(1-x^2)(1+c_{\rm vir})}+\frac{1}{(1-x^2)^{3/2}}{\rm arccosh}\left[\frac{x^2+c_{\rm vir}}{x(1+c_{\rm vir})}\right] & (x<1), \\
\frac{\sqrt{c_{\rm vir}^2-1}}{3(1+c_{\rm vir})}\left(1+\frac{1}{1+c_{\rm vir}}\right) & (x=1), \\
-\frac{\sqrt{c_{\rm vir}^2-x^2}}{(1-x^2)(1+c_{\rm vir})}-
\frac{1}{(1-x^2)^{3/2}}{\rm arccos}\left[\frac{x^2+c_{\rm vir}}{x(1+c_{\rm vir})}\right] & (1<x\le c_{\rm vir}), \\
0 & (x > c_{\rm vir}).
\end{array}
\right. 
\eeqa
In Eq.~(\ref{eq:kappa_nfw}), $\Sigma_{\rm crit}$ is defined by the following relation
\beqa
\Sigma_{\rm crit} = \frac{c^2}{4\pi G}\frac{D_{\rm s}}{D_{\rm l}D_{\rm ls}},
\eeqa
where $D_{\rm s}$, $D_{\rm l}$, and $D_{\rm ls}$ are the angular diameter distance
to the source, to the lens, and between the source and the lens, respectively.

In order to predict the peak height in ${\cal K}$ map, we need to take 
the following effects into account:
(i) the offset between the position of a peak and the center of the
corresponding halo 
and
(ii) the modulation of peak height due to the shape noise.
\citet{2010ApJ...719.1408F} have studied these two effects using numerical 
simulations and analytic approach.
Let us first work on the simple assumption that the peak position 
is set to be the halo center.
The peak height in absence of shape noise
is given by
\beqa
{\cal K}_{{\rm peak}, h} = \int {\rm d}^2\theta \, U(\theta; \theta_{G})\kappa_{h}(\theta). \label{eq:kpeak_nfw}
\eeqa
The actual peak height on a noisy ${\cal K}$ map 
is not given by Eq.~(\ref{eq:kpeak_nfw}), 
but it obeys a probability distribution
\citep{2010ApJ...719.1408F}.
The probability distribution function 
for a given ${\cal K}_{{\rm peak}, h}$
is calculated by 
\beqa
{\rm Prob}({\cal K}_{\rm peak, obs}|{\cal K}_{{\rm peak}, h}) 
= \frac{n_{\rm peak, N}({\cal K}_{\rm peak, obs}|{\cal K}_{{\rm peak}, h})}{\int n_{\rm peak, N}({\cal K}^{\prime}_{\rm peak, obs}|{\cal K}_{{\rm peak}, h}) {\rm d}{\cal K}^{\prime}_{\rm peak, obs}},
\label{eq:prob_peak_obs_h}
\eeqa
where ${\cal K}_{\rm peak, obs}$ is the measured peak height
and $n_{{\rm peak},{\rm N}}$ is defined 
as the expected number density of peaks with the measured 
peak height of ${\cal K}_{{\rm peak},{\rm obs}}$ 
when the halo contribution $\mathcal{K}_{{\rm peak},h}$ is known in advance.
For derivation of $n_{\rm peak, N}$,
we decompose the observed peak height ${\cal K}_{{\rm peak}, {\rm obs}}$ 
into three components:
\beqa
{\cal K}_{{\rm peak}, {\rm obs}} = {\cal N}+{\cal K}_{\rm LSS}+{\cal K}_{{\rm peak}, h},
\eeqa
where ${\cal N}$ is the noise convergence field caused by shape noise,
${\cal K}_{\rm LSS}$ and ${\cal K}_{{\rm peak}, h}$ represent 
the convergence field due to large-scale structure and foreground halos, 
respectively.
Note that ${\cal K}_{{\rm peak}, h}$ is a known quantity to derive 
$n_{{\rm peak},{\rm N}}$.
We aim at determining the relationship 
between ${\cal K}_{{\rm peak}, {\rm obs}}$
and a given ${\cal K}_{{\rm peak}, h}(z, M)$.

Following \citet{2010ApJ...719.1408F},
we assume that the noise field ${\cal N}$ is given by a Gaussian distribution 
with the power spectrum of Eq.~(\ref{eq:noise_power}).
If ${\cal K}_{\rm LSS}$ is a Gaussian random field,
at the position of peaks, the {\it total} noise field (i.e. ${\cal N}+{\cal K}_{\rm LSS}$) 
obeys the probability distribution function of Gaussian peaks.
Also, we can calculate the contribution from the (known) corresponding halo 
once the difference between the peak position and the halo center is specified.
We assume that the peak position is at the center of the corresponding halo.
although the equality does not hold in general.
We have checked that the assumption is indeed reasonable 
for peaks with high signal-to-noise ratio
in the case of $\theta_{G}\sim 2\, {\rm arcmin}$, 
$\sigma_{\gamma}=0.4$, and $n_{\rm gal} \simgt 10\, {\rm arcmin}^{-2}$.

Therefore, we set $n_{\rm peak, N}$ to be the number density of peaks 
for Gaussian field ${\cal N}+{\cal K}_{\rm LSS}$.
Using the relation of ${\cal N}+{\cal K}_{\rm LSS}={\cal K}_{{\rm peak},{\rm obs}}
-{\cal K}_{{\rm peak},h}$, we can obtain the number density $n_{\rm peak, N}$ as (see, \citet{2010ApJ...719.1408F} for details)
\beqa
n_{\rm peak, N}({\cal K}_{{\rm peak},{\rm obs}}|{\cal K}_{{\rm peak},h}) 
&=& 
\frac{1}{2\pi \theta_{*}^2}\frac{1}{\sqrt{2\pi}}
\exp\left[-\frac{1}{2}\left(\frac{{\cal K}_{{\rm peak},{\rm obs}}-{\cal K}_{{\rm peak},h}}{\sigma_{0}}\right)^2\right]
\int \frac{{\rm d}x_{N}}{\left[2\pi\left(1-\gamma_{N}^2\right)\right]^{1/2}} 
\, F(x_{N}|{\cal K}_{{\rm peak},h}) \nonumber \\
&&\times \exp \Biggl\{ -\frac{1}{2(1-\gamma_{N}^2)}
\left[x_{N}+\left({\cal K}_{{\rm peak},h}^{11}+{\cal K}_{{\rm peak},h}^{22}\right)/\sigma_{2}
-\gamma_{N}\left({\cal K}_{{\rm peak},{\rm obs}}-{\cal K}_{{\rm peak},h}\right)/\sigma_{0}
\right]^2 \Biggr\},
\label{eq:npeak_N_def}
\eeqa
where $\sigma_{i}$ is the ith moment of ${\cal N}+{\cal K}_{\rm LSS}$,
$\theta_{*}^2 = 2(\sigma_{1}/\sigma_{2})^2$,
$\gamma_{N} = \sigma_{1}^2/(\sigma_{0}\, \sigma_{2})$ and
${\cal K}_{{\rm peak},h}^{ii}$ denotes the second derivative of ${\cal K}_{{\rm peak},h}$ 
with respect to $\theta_{i}$ at the halo centre.
Here, $F(x_{N}|{\cal K}_{{\rm peak},h})$ is calculated as follows:
\beqa
F(x_{N}|{\cal K}_{{\rm peak},h}) 
&=&
\exp\left[-\left(\frac{{\cal K}_{{\rm peak},h}^{11}-{\cal K}_{{\rm peak},h}^{22}}{\sigma_{2}}\right)^2\right]
\int_{0}^{1/2}{\rm d}e_{N}\, 8(x_{N}^2 e_{N})x_{N}^2(1-4e_{N}^2)\exp\left(-4x_{N}^2e_{N}^2\right) \nonumber \\
&&
\, \, \, \, \, \, \, \, \, \, \, \, \, \,  \, \, \, \, \, \, \,
\, \, \, \, \, \, \, \, \, \, \, \, \, \,  \, \, \, \, \, \, \,
\, \, \, \, \, \, \, \, \, \, \, \, \, \,  \, \, \, \, \, \, \, 
\, \, \, \, \, \, \, \, \, \, \, \, \, \,  \, \, \, \, \, \, \, 
\, \, \, \, \, \, \, \, \, \, \, \, \, \,  \, \, \, \, \, \, \, \times
\int_{0}^{\pi}\frac{{\rm d}\theta_{N}}{\pi}\, \exp\left[-4x_{N}e_{N}\cos 2\theta_{N}\frac{{\cal K}_{{\rm peak},h}^{11}-{\cal K}_{{\rm peak},h}^{22}}{\sigma_{2}}\right].
\label{eq:npeak_N_def2}
\eeqa
In Eqs.~(\ref{eq:npeak_N_def}) and (\ref{eq:npeak_N_def2}),
$x_{N}$ and $e_{N}$ are given by
\beqa
x_{N} = \frac{\lambda_{N1}+\lambda_{N2}}{\sigma_{2}}, \, \, \, 
e_{N} = \frac{\lambda_{N1}+\lambda_{N2}}{2\sigma_{2}x_{N}},
\eeqa
where $\lambda_{Ni}$ represents the ith diagonal component 
of the second derivative tensor of ${\cal N}+{\cal K}_{\rm LSS}$.

\begin{figure*}
\centering \includegraphics[clip, width=0.4\columnwidth, viewport = 20 20 500 500]{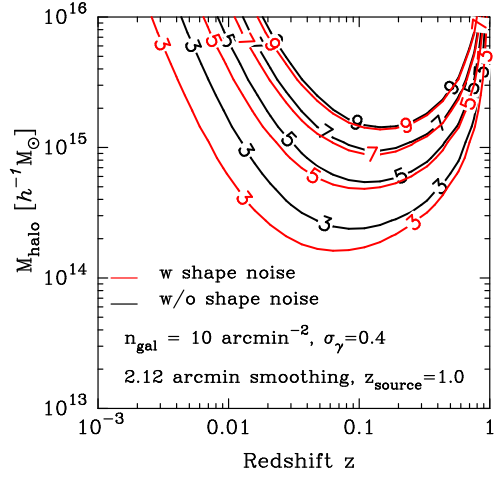}
\caption{
	The effective selection function of galaxy clusters.
	Black line shows the expected convergence signal caused 
	by an isolated dark matter halo. 
        Red line shows the modulated peak height 
	on a noisy smoothed convergence map (see the text for the definition).
	We plot each line in units of the variance of 
        shape noise $\sigma_{\rm noise, 0}=0.017$.
	\label{fig:kappa_z_m}
	}
\end{figure*}

We adopt $\sigma_{\gamma}=0.4$, $n_{\rm gal}=10\, {\rm arcmin}^{-2}$
and the source redshift is set to be $z_{\rm source}=1$.
These are the typical values for ground-based galaxy imaging surveys
\citep[e.g.,][]{2012MNRAS.427..146H}. 
Also, we employ a Gaussian smoothing with the full width 
at half maximum of 5 arcmin.
This corresponds to $\theta_{G}=5/\sqrt{8\ln 2}=2.12$ arcmin
and thus to $\sigma_{\rm noise, 0}\simeq 0.017$.
Although the smoothing scale adopted here 
is slightly larger than that in previous works 
\citep[see, e.g.,][]{Hamana2004} by a factor of about two,
the noise level on the smoothed convergence map is similar.
Gaussian smoothing with $\theta_{G}\sim 2$ arcmin with the actual data 
set is already examined in \citet{2012ApJ...748...56S}.

Let us examine the effect of shape noise on weak lensing peaks.
For this purpose, we define the mean modulation of peak 
height in a noisy ${\cal K}$ map as
follows:
\beqa
{\bar {\cal K}}_{\rm peak, obs}(z, M) =  \int {\rm d}{\cal K}\, {\cal K}\, 
{\rm Prob}({\cal K}|{\cal K}_{{\rm peak},h}(z, M)). \label{eq:mean_peak_height}
\eeqa
Figure \ref{fig:kappa_z_m} shows the comparison 
with ${\bar {\cal K}}_{\rm peak, obs}(z, M)$
and ${\cal K}_{{\rm peak},h}(z, M)$ for a given dark 
matter halo with mass of $M$ at redshift $z$.
In this figure, red line shows the contour of ${\bar {\cal K}}_{\rm peak, obs}(z, M)$ in units of
$\sigma_{\rm noise, 0}$, whereas black line indicates the contour of 
${\cal K}_{{\rm peak},h}(z, M)$.
In a noisy $\cal K$ map, 
the shape noise modulates the height of peaks and 
the number of peaks slightly increases.
We have tested the validity of our model 
against numerical simulations.
The result is shown in Appendix~\ref{appendixH}.

As shown in Figure \ref{fig:kappa_z_m}, 
the Gaussian smoothing of $\sim2$ arcmin are effective 
to search for clusters with 
mass of $\sim10^{14}h^{-1}M_{\odot}$ at $z\sim0.1-0.2$.
The selection of mass and redshift is basically determined 
by the typical angular size of dark matter halos of interest
\citep[e.g.,][]{Hamana2004}.
Naively, it is expected that lower redshift clusters are
detected with larger smoothing scales.
In order to verify this expectation, 
we have studied the statistical properties of lensing peaks 
when adopting a Gaussian smoothing with the full width 
at half maximum of 15 arcmin (corresponding to $\theta_{G}\sim 6.4$ arcmin).
In this case, we do not find one-to-one correspondence 
between selected peaks and halos, and thus our analytic model does 
not work.
This is likely caused by the so-called projection effect; 
the effective redshift of lensing in the case of $z_{\rm source}=1$ 
is $0.1-0.5$ whereas we attempt to search for clusters 
at lower redshift $z<0.1$.
We argue that our model is valid 
when the smoothing scale is set to be $1-2$ arcmin, 
corresponding to the typical angular size of
dark matter halos with mass of $\sim10^{14}h^{-1}M_{\odot}$ at $z\sim0.1-0.5$.

\subsection{Statistics}
\label{subsec:WL_stat}
We consider a set of statistics derived from weak lensing measurement.
In order to extract cosmological information from 
the number and the distribution
of massive dark matter halos, 
we utilize peaks on a smoothed lensing map as described 
in Section~\ref{subsec:WL_select_cluster}.

\subsubsection*{Convergence power spectrum}

First, we consider the power spectrum of convergence.
Under the flat sky approximation, 
the Fourier transform of convergence field is defined by
\beqa
\kappa(\bd{\theta})= \int \frac{{\rm d}^2 \ell}{(2\pi)^2}e^{i\bd{\ell}\cdot\bd{\theta}}\tilde{\kappa}(\bd{\ell}).
\eeqa
The power spectrum of convergence field $P_{\kappa\kappa}$ is defined by 
\beqa
\langle \tilde{\kappa}(\bd{\ell}_{1}) \tilde{\kappa}(\bd{\ell}_2)\rangle
=(2\pi)^2 \delta_{D}^{(2)}(\bd{\ell}_{1}-\bd{\ell}_{2}) P_{\kappa\kappa}(\ell_1),
\eeqa
where $\delta_{D}^{(2)}(\bd{\ell})$ is the Dirac delta function.
By using Limber approximation\footnote{
The validity of Limber approximation have been discussed in e.g., \citet{2009PhRvD..80l3527J}.
The typical accuracy of Limber approximaion is of a level of $\simlt$1\% for $\ell > 10$.
} \citep{Limber:1954zz,Kaiser:1991qi}
and Eq.~(\ref{eq:kappa_delta}), 
one can calculate the convergence power spectrum as follows:
\beqa
P_{\kappa\kappa}(\ell) &=& \int_{0}^{\chi_s} {\rm d}\chi \frac{W_{\kappa}(\chi)^2}{r(\chi)^2} 
P_{\delta}\left(k=\frac{\ell}{r(\chi)},z(\chi)\right)
\label{eq:kappa_power},
\eeqa
where $P_{\delta}(k)$ is the three dimensional matter power spectrum, 
$\chi_s$ is comoving distance to source galaxies and 
$W_{\kappa}(\chi)$ is the lensing weight function defined as
\beqa
W_{\kappa}(\chi) = \frac{3}{2}\left(\frac{H_{0}}{c}\right)^2 \Omega_{\rm m0}
\frac{r(\chi_s-\chi)r(\chi)}{r(\chi_s)}(1+z(\chi))
\label{eq:kappa_weight}.
\eeqa

The non-linear gravitational growth of 
density fluctuations significantly affects 
the amplitude of convergence power spectrum at the angular scales less than 1 degree
\citep{2000ApJ...530..547J, Hilbert2009, Sato2009}.
Typical weak lensing surveys
aim at measuring the angular scales larger than a few arcmin corresponding to a few Mpc.
This leads weak lensing can be one of the most powerful probes 
for constraints on dark matter distribution at Mpc scale.
Therefore, accurate theoretical prediction of non-linear matter power spectrum 
is essential for deriving 
cosmological constraints from weak lensing power spectrum.
In order to predict the non-linear evolution of $P_{\delta}(k)$ for standard $\Lambda$CDM universe,
a numerical approach based on $N$-body simulations has 
given steady results over the past few decades
\citep{1996MNRAS.280L..19P, 2003MNRAS.341.1311S, 2010ApJ...715..104H, Takahashi2012}.
We adopt the most recent model of non-linear $P_{\delta}(k)$ 
of \citet{Takahashi2012}.

\subsubsection*{Convergence peak count}

The number count of massive clusters is 
sensitive to various cosmological 
parameters such as the equation of state of dark energy
\citep[e.g.,][]{2011ARA&A..49..409A}.
In this section, we use peak counts as a cosmological probe. 
We locate the local maxima in a smoothed lensing map and 
associate each identified peak with a massive dark matter halo
along the same line of sight.

In practice, one can select a lensing peak by its peak height.
We define the signal-to-noise ratio of a peak 
by $\nu = {\cal K}_{\rm peak, obs}/\sigma_{\rm noise,0}$.
For a given threshold $\nu_{\rm thre}$, 
one can predict the surface number density of peaks with $\nu > \nu_{\rm thre}$ as follows
\citep[e.g.,][]{Hamana2004}:
\beqa
N_{\rm peak}(\nu_{\rm thre}) = 
\int {\rm d}z\, {\rm d}M\, 
\frac{{\rm d}^2V}{{\rm d}z{\rm d}\Omega} \frac{{\rm d}n}{{\rm d} M}(z, M)
\int_{\nu_{\rm thre}\sigma_{\rm noise,0}}^{\infty} 
{\rm d}{\cal K}_{\rm peak, obs}\, \, {\rm Prob}({\cal K}_{\rm peak, obs}|\, {\cal K}_{{\rm peak},h}(z, M)),  \label{eq:npeak}
\eeqa
where ${\rm d}n/{\rm d}M$ represents the mass function of dark matter halo 
and the volume element is expressed as ${\rm d}^2 V/{\rm d}z{\rm d}\Omega = \chi^2/H(z)$ for a spatially flat universe.
Here, ${\rm Prob}({\cal K}_{\rm peak, obs}|\, {\cal K}_{{\rm peak},h})$ is given by Eq.~(\ref{eq:prob_peak_obs_h}).
In the following, we adopt the model of halo mass function in \citet{2011ApJ...732..122B}.

\subsubsection*{Convergence peak auto spectrum and cross spectrum}

We next consider the auto-correlation function of peaks, and the peak-convergence 
cross correlation.
\citet{2013MNRAS.432.1338M} study cosmological information 
obtained from the statistics using a large set of numerical simulations.
They conclude that using the auto- and cross-correlation functions 
can improve the constraints on cosmological parameters 
when combined with the number of peaks.
We develop an analytic halo model
in order to predict the correlation function of peaks and 
cross correlation between peaks and convergence.

In the halo model, the number density field of weak lensing 
selected clusters is given by 
\beqa
n_{\rm cl}(\bd{x}) &=& \sum_{i}\delta_{D}^{(3)}(\bd{x}-\bd{x}_{i}) S(z_{i}, M_{i}) \nonumber \\
&=& \sum_{i} \int {\rm d}M \, S(z, M) \, \delta_{D}(M-M_{i}) \int {\rm d}^3 \bd{x}^{\prime}\, 
\delta_{D}^{(3)}(\bd{x}^{\prime}-\bd{x}_{i}) \delta_{D}^{(3)}(\bd{x}-\bd{x}^{\prime}),
\label{eq:ndens_field_cl}
\eeqa
where $S(z, M)$ represents the selection function
\citep[e.g.,][]{2007NJPh....9..446T}.
For clusters identified as lensing peaks, $S(z, M)$ is expressed as
\beqa
S(z, M|\nu_{\rm thre}) = \int_{\nu_{\rm thre}\sigma_{\rm noise,0}}^{\infty} 
{\rm d}{\cal K}_{\rm peak, obs}\, \, {\rm Prob}({\cal K}_{\rm peak, obs}|\, {\cal K}_{{\rm peak},h}(z, M)).
\eeqa
Also, underlying matter density field can be approximated as
\beqa
\rho_{m}(\bd{x}) &=& \sum_{i} \rho_{h}(\bd{x}-\bd{x}_{i}|z, M) \nonumber \\
&=& \sum_{i} \int {\rm d}M \, M\, \delta_{D}(M-M_{i}) \int {\rm d}^3 \bd{x}^{\prime}\, 
\delta_{D}^{(3)}(\bd{x}^{\prime}-\bd{x}_{i}) u_{m}(\bd{x}-\bd{x}^{\prime}|z, M),
\label{eq:dens_field_m}
\eeqa
where $\rho_{h}$ is the density profile of a massive halo 
given by Eq.~(\ref{eq:rho_nfw}), 
and $\rho_{h}(r|z,M) = M u_{m}(r|z,M)$.

In order to derive the auto power spectrum of weak lensing selected clusters 
for a given threshold $\nu_{\rm thre}$,
we first consider the auto power spectrum of $n_{\rm cl}$.
The two point correlation function of $n_{\rm cl}$ is given by
\beqa
\bar{n}_{\rm cl}^2 \xi_{cc}(\bd{x}_{1}-\bd{x}_{2}) 
&\equiv& \langle n_{\rm cl}(\bd{x}_{1})n_{\rm cl}(\bd{x}_{1})\rangle - \bar{n}_{\rm cl}^2
\nonumber \\
&=& 
\langle 
\sum_{i} S^2(M_{i}) 
\delta_{D}^{(3)}(\bd{x}_{1}-\bd{x}_{i})\delta_{D}^{(3)}(\bd{x}_{2}-\bd{x}_{i})
\rangle
+
\langle
\sum_{i,j|\, i\neq j} 
S(M_{i}) S(M_{j})
\delta_{D}^{(3)}(\bd{x}_{1}-\bd{x}_{i})\delta_{D}^{(3)}(\bd{x}_{2}-\bd{x}_{j})
\rangle
\nonumber \\ 
&=&
\langle 
\sum_{i} \int{\rm d}M \, \int{\rm d}^3 y\,
S^2(M) \delta_{D}(M-M_i) 
\delta_{D}^{(3)}(\bd{x}_{1}-\bd{y})\delta_{D}^{(3)}(\bd{x}_{2}-\bd{y})
\delta_{D}^{(3)}(\bd{y}-\bd{x}_{i})
\rangle
\nonumber \\ 
&+&
\langle 
\sum_{i,j|\, i\neq j} \int{\rm d}M \, \int{\rm d}^3 y\,
S(M) \delta_{D}(M-M_i) 
\delta_{D}^{(3)}(\bd{x}_{1}-\bd{y})\delta_{D}^{(3)}(\bd{y}-\bd{x}_{i})
\nonumber \\ 
&&\,\,\,\,\, \,\,\,\,\, \,\,\,\,\, \,\,\,\,\, \,\,\,\,\, \,\,\,\,\, \times
\int{\rm d}M^{\prime} \, \int{\rm d}^3 y^{\prime}\,
S(M) \delta_{D}(M^{\prime}-M_j) 
\delta_{D}^{(3)}(\bd{x}_{2}-\bd{y}^{\prime})\delta_{D}^{(3)}(\bd{y}^{\prime}-\bd{x}_{j})
\rangle
\nonumber \\ 
&=&
\int {\rm d}M \, \frac{{\rm d}n}{{\rm d}M} S^2(M) 
\int {\rm d}^3 y \, \delta_{D}^{(3)}(\bd{x}_{1}-\bd{y})\delta_{D}^{(3)}(\bd{x}_{2}-\bd{y})
\nonumber \\ 
&+&
\int {\rm d}M \, \frac{{\rm d}n}{{\rm d}M} S(M) 
\int {\rm d}^3 y \, \delta_{D}^{(3)}(\bd{x}_{1}-\bd{y})
\int {\rm d}M^{\prime} \, \frac{{\rm d}n}{{\rm d}M^{\prime}}S(M^{\prime})
\int {\rm d}^3 y^{\prime} \, \delta_{D}^{(3)}(\bd{x}_{2}-\bd{y}^{\prime})
\xi_{hh}(\bd{y}-\bd{y}^{\prime}; M, M^{\prime})
\nonumber \\ 
&=&
\int {\rm d}M \, \frac{{\rm d}n}{{\rm d}M}S^2(M)
\int {\rm d}^3 y \, \delta_{D}^{(3)}(\bd{x}_{1}-\bd{x}_{2})
+
\int {\rm d}M \, \frac{{\rm d}n}{{\rm d}M}S(M) 
\int {\rm d}M^{\prime} \, \frac{{\rm d}n}{{\rm d}M^{\prime}}S(M^{\prime}) 
\xi_{hh}(\bd{x}_{1}-\bd{x}_{2}; M, M^{\prime}),
\label{eq:2pcf_ndens_field_cl_derivation}
\eeqa
where $\xi_{hh}(\bd{y}-\bd{y}^{\prime}; M, M^{\prime})$ is the two point correlation function
of dark matter haloes with mass of $M$ and $M^{\prime}$.
In the above calculation, we use the following relations as
\beqa
\langle \sum_{i} \delta_{D}(M-M_i) \delta_{D}^{(3)}(\bd{x}-\bd{x}_{i})\rangle
&\equiv& \frac{{\rm d}n}{{\rm d}M}, 
\\
\langle 
\sum_{i,j|\, i\neq j} \delta_{D}(M-M_i) \delta_{D}(M^{\prime}-M_j)
\delta_{D}^{(3)}(\bd{x}_{1}-\bd{x}_{i})\delta_{D}^{(3)}(\bd{x}_{2}-\bd{x}_{j})
\rangle
&\equiv&
\frac{{\rm d}n}{{\rm d}M} \frac{{\rm d}n}{{\rm d}M^{\prime}} 
\xi_{hh}(\bd{x}_{1}-\bd{x}_{2}^{\prime}; M, M^{\prime}).
\eeqa
We further approximate $\xi_{hh}$ as $b_{h}(M)b_{h}(M^{\prime})\xi_{\delta \delta}^{L}$,
where $b_{h}$ is the linear halo bias and $\xi_{\delta \delta}^{L}$ is 
the two point correlation function of linear matter density field.
Then, we finally obtain the following equation:
\beqa
\bar{n}_{\rm cl}^2 \xi_{cc}(\bd{x}_{1}-\bd{x}_{2})
&=&
\int {\rm d}M \, \frac{{\rm d}n}{{\rm d}M}\, S^2(M) 
\delta_{D}^{(3)}(\bd{x}_{1}-\bd{x}_{2})
+
\left[\int \, {\rm d}M \, \frac{{\rm d}n}{{\rm d}M}\, S(M) b_{h}(M)\right]^2
\xi_{\delta \delta}^{L}(\bd{x}_{1}-\bd{x}_{2}).
\label{eq:2pcf_ndens_field_cl}
\eeqa
The Fourier transform of Eq.~(\ref{eq:2pcf_ndens_field_cl}) is 
the three-dimensional power spectrum of $n_{cl}$, which is expressed as
\beqa
P_{cc}(k) = \int {\rm d}^3 r \, e^{-i\bd{k}\cdot\bd{r}} \xi_{cc}(r).
\label{eq:pk_ndens_field_cl}
\eeqa
The observed surface number density of weak lensing selected clusters 
(for a given threshold $\nu_{\rm thre}$) is thus given by
\beqa
p(\bd{\theta}) = \frac{1}{N_{\rm peak}(\nu_{\rm thre})}
\int {\rm d}\chi\, \frac{{\rm d}^2 V}{{\rm d}\chi {\rm d}\Omega}
n_{\rm cl}(r(\chi)\bd{\theta}, z(\chi)),
\label{eq:def_peak_field}
\eeqa
where $N_{\rm peak}$ is defined by Eq.~(\ref{eq:npeak}).
Using the Limber approximation, we can derive the angular power spectrum of $p$ as 
\beqa
P_{\rm pp}(\ell) 
&=& \int {\rm d}\chi\, \frac{1}{r(\chi)^2}
\left(\frac{1}{N_{\rm peak}} \frac{{\rm d}^2 V}{{\rm d}\chi {\rm d}\Omega}\right)^2
\bar{n}_{\rm cl}^2 P_{cc}\left(k=\frac{\ell}{r(\chi)}, z(\chi)\right).
\label{eq:pk_pp_full}
\eeqa
Except for the shot noise term in Eq.~(\ref{eq:pk_ndens_field_cl}), 
we obtain
\beqa
P_{\rm pp}(\ell) = \int {\rm d}\chi\, \frac{1}{r(\chi)^2}  
\left(\frac{1}{N_{\rm peak}} \frac{{\rm d}^2 V}{{\rm d}\chi {\rm d}\Omega}\right)^2
\left[\int {\rm d}M\, 
\frac{{\rm d}n}{{\rm d} M}(z, M) S(z, M|\nu_{\rm thre}) b_{h}(z, M)
\right]^2\, P^{L}_{m}\left(k=\frac{\ell}{r(\chi)}, z(\chi)\right),
\eeqa
where $P^{L}_{m}$ is the linear matter power spectrum.
Throughput this paper, we adopt the functional form of $b_{h}$ 
proposed in \citet{2011ApJ...732..122B}.

The similar derivation can be applied to the cross power spectrum of weak lensing selected clusters
and lensing convergence.
Let us first consider the three-dimensional cross correlation function of $n_{\rm cl}$ 
and $\rho_{m}$:
\beqa
\bar{\rho}_{m} \xi_{c\delta}(\bd{x}_{1}-\bd{x}_{2}) 
&=& \langle n_{\rm cl}(\bd{x}_{1})\rho_{m}(\bd{x}_{1})\rangle - \bar{n}_{\rm cl}\bar{\rho}_{m}
\nonumber \\
&=&
\int {\rm d}M \, \frac{{\rm d}n}{{\rm d}M}\, S(M) M 
u_{m}(\bd{x}_{1}-\bd{x}_{2}|M)
+
\left[\int \, {\rm d}M \, \frac{{\rm d}n}{{\rm d}M}\, S(M)b_{h}(M)\right] \bar{\rho}_{m}
\xi_{\delta \delta}^{L}(\bd{x}_{1}-\bd{x}_{2}).
\label{eq:2pcf_rho_ndens_field_cl}
\eeqa
Through the derivation of Eq.~(\ref{eq:2pcf_rho_ndens_field_cl}),
we use the following fact that
\beqa
\bar{\rho}_{m} =
\int {\rm d}M\, \frac{{\rm d}n}{{\rm d}M}\, b_{h}(M) M
\int {\rm d}^3 x^{\prime} \, u_{m}(\bd{x}-\bd{x}^{\prime}|M).
\eeqa
Then, we can obtain the angular cross power spectrum of $p$ and lensing convergence $\kappa$
by the similar calculation as Eq.~(\ref{eq:pk_pp_full}).
The cross power spectrum $P_{{\rm p}\kappa}$ is given by
\beqa
P_{{\rm p}\kappa}(\ell) 
=\int {\rm d}\chi\, \frac{W_{\kappa}(\chi)}{r(\chi)^2}
\left(\frac{1}{N_{\rm peak}} \frac{{\rm d}^2 V}{{\rm d}\chi {\rm d}\Omega}\right)
P_{c\delta}\left(k=\frac{\ell}{r(\chi)}, z(\chi)\right),
\label{eq:pk_pk_full}
\eeqa
where $P_{c\delta}(k)$ represents the three-dimensional 
cross power spectrum of $n_{\rm cl}$ and matter overdensity field $\delta$, i.e., 
\beqa
P_{c\delta}(k) = \int {\rm d}^3 r \, e^{-i\bd{k}\cdot\bd{r}} \xi_{c\delta}(r).
\label{eq:pk_ndens_field_cl}
\eeqa


Finally, the cross power spectrum between peaks and convergence 
is given by 
(also, see \citet{2011PhRvD..83b3008O})
\beqa
P_{{\rm p}\kappa}(\ell) &=& P^{1h}_{{\rm p}\kappa}(\ell)+P^{2h}_{{\rm p}\kappa}(\ell), \\
P^{1h}_{{\rm p}\kappa}(\ell) &=& 
\int {\rm d}\chi\, 
\frac{W_{\kappa}(\chi)}{r(\chi)^2}
\left(\frac{1}{N_{\rm peak}} \frac{{\rm d}^2 V}{{\rm d}\chi {\rm d}\Omega}\right)
\int {\rm d}M\, \frac{{\rm d}n}{{\rm d}M} S(z, M|\nu_{\rm thre}) \, 
\left(\frac{M}{\bar{\rho}_{m}(z)}\right)
\tilde{u}_{m}\left(k=\frac{\ell}{r(\chi)}\Bigg|z(\chi), M\right), \\
P^{2h}_{{\rm p}\kappa}(\ell) &=&
\int {\rm d}\chi\, \frac{W_{\kappa}(\chi)}{r(\chi)^2}
\left(\frac{1}{N_{\rm peak}} \frac{{\rm d}^2 V}{{\rm d}\chi {\rm d}\Omega}\right)
\left[\int {\rm d}M\, 
\frac{{\rm d}n}{{\rm d} M}(z, M) S(z, M|\nu_{\rm thre}) b_{h}(z, M)
\right]\,  
P^{L}_{m}\left(k=\frac{\ell}{r(\chi)}, z(\chi)\right),
\eeqa
where $\tilde{u}_{m}$ is the Fourier transform of $u_{m}(r|z,M)$.

\subsection{Covariances between statistics}
\label{subsec:cov}

We summarize covariance matrices between statistics of interest in 
the following.
As a first approximation, we can use 
the Gaussian covariance between three binned spectra 
$P_{\kappa\kappa}$, $P_{{\rm p}\kappa}$ and $P_{\rm pp}$ as,
\beqa
{\rm Cov}[P_{XY}(\ell), P_{AB}(\ell^{\prime})]
= \frac{1}{(2\ell+1)\Delta \ell f_{\rm sky}}
\left[
P^{\rm obs}_{XA}(\ell)P^{\rm obs}_{YB}(\ell)+
P^{\rm obs}_{XB}(\ell)P^{\rm obs}_{YA}(\ell)\right]\delta_{\ell \ell^{\prime}},
\eeqa
where $\Delta \ell$ is the width of binning in multipole 
and $f_{\rm sky}$ represents the observed sky fraction.
The observed spectra $P^{\rm obs}_{XY}(\ell)$ are then defined by
\beqa
P^{\rm obs}_{\kappa \kappa} = P_{\kappa\kappa}+
\frac{\sigma_{\gamma}^2}{2n_{\rm gal}}, \,
P^{\rm obs}_{\rm pp} = P_{\rm pp} + \frac{1}{N_{\rm peak}}, \,
P^{\rm obs}_{{\rm p}\kappa} = P_{{\rm p}\kappa}.
\eeqa
The non-linear gravitational growth causes mode-coupling of
the density fluctuations with different wavelengths.
The mode-coupling then induces the correlation of weak lensing statistics 
between different multipoles, i.e., 
we can not use the Gaussian approximation to covariances 
\citep[e.g.,][]{Sato2009}.
Modeling of the non-Gaussian covariances is still being developed
\citep[e.g.,][]{2001ApJ...554...56C, Takada2004, 2007NJPh....9..446T}.
We here present a theoretical model of non-Gaussian covariance 
matrices between
$P_{\kappa\kappa}$, $P_{{\rm p}\kappa}$ and $N_{\rm peak}$.

Let us consider the following set of four-point correlation functions 
in Fourier space:
\beqa
\langle \tilde{\delta}_m(\bd{k}_1)\tilde{\delta}_m(\bd{k}_2)
\tilde{\delta}_m(\bd{k}_3)\tilde{\delta}_m(\bd{k}_4)\rangle
&=&(2\pi)^3 \delta^{(3)}_{D}(\bd{k}_{1234})
T_{\delta \delta \delta \delta}(\bd{k}_1, \bd{k}_2, \bd{k}_3, \bd{k}_4) \\
\langle \tilde{\delta}_{cl}(\bd{k}_1)\tilde{\delta}_m(\bd{k}_2)
\tilde{\delta}_{cl}(\bd{k}_3)\tilde{\delta}_m(\bd{k}_4)\rangle
&=&(2\pi)^3 \delta^{(3)}_{D}(\bd{k}_{1234})
T_{c \delta c \delta}(\bd{k}_1, \bd{k}_2, \bd{k}_3, \bd{k}_4) \\
\langle \tilde{\delta}_{cl}(\bd{k}_1)\tilde{\delta}_m(\bd{k}_2)
\tilde{\delta}_m(\bd{k}_3)\tilde{\delta}_m(\bd{k}_4)\rangle
&=&(2\pi)^3 \delta^{(3)}_{D}(\bd{k}_{1234})
T_{c \delta \delta \delta}(\bd{k}_1, \bd{k}_2, \bd{k}_3, \bd{k}_4),
\eeqa
where $\bd{k}_{ij\cdots n} = \bd{k}_{i}+\bd{k}_{j}+\cdots+\bd{k}_{n}$,
$\delta_{m}$ and $\delta_{cl}$ represent over density of matter 
and weak lensing selected clusters, respectively.
In the flat sky approximation, we can relate three tri-spectra 
$(T_{\delta \delta \delta \delta}, T_{c \delta c \delta}, T_{c \delta \delta \delta})$
with the non-Gaussian part of 
the covariance matrix of weak lensing statistics as follows:
\beqa
{\rm Cov}\left[P_{\kappa\kappa}(\ell), P_{\kappa\kappa}(\ell^{\prime})\right]_{\rm NG}
&=& \frac{1}{4\pi} \int \frac{{\rm d} \phi}{2\pi} 
T_{\kappa \kappa \kappa \kappa}(\bd{\ell},-\bd{\ell}, 
\bd{\ell}^{\prime},-\bd{\ell}^{\prime}; \phi), \\
{\rm Cov}\left[P_{{\rm p}\kappa}(\ell), P_{{\rm p}\kappa}(\ell^{\prime})\right]_{\rm NG}
&=& \frac{1}{4\pi} \int \frac{{\rm d} \phi}{2\pi} 
T_{{\rm p} \kappa {\rm p} \kappa}(\bd{\ell},-\bd{\ell}, 
\bd{\ell}^{\prime},-\bd{\ell}^{\prime}; \phi), \\
{\rm Cov}\left[P_{{\rm p}\kappa}(\ell), P_{\kappa\kappa}(\ell^{\prime})\right]_{\rm NG}
&=& \frac{1}{4\pi} \int \frac{{\rm d} \phi}{2\pi} 
T_{{\rm p} \kappa \kappa \kappa}(\bd{\ell},-\bd{\ell}, 
\bd{\ell}^{\prime},-\bd{\ell}^{\prime}; \phi), 
\eeqa
where $\phi$ is the angle between two vectors $\bd{\ell}$ and $\bd{\ell}^{\prime}$.
In practice, the integral over $\phi$ is often simplified as, e.g.,
\beqa
\frac{1}{4\pi} \int \frac{{\rm d} \phi}{2\pi} 
T_{\kappa \kappa \kappa \kappa}(\bd{\ell},-\bd{\ell}, 
\bd{\ell}^{\prime},-\bd{\ell}^{\prime}; \phi)
\simeq
T_{\kappa \kappa \kappa \kappa}(\ell,\ell, \ell^{\prime}, {\ell}^{\prime}),
\eeqa
and so on.
With Limber approximation, 
$T_{\kappa \kappa \kappa \kappa}$, 
$T_{{\rm p} \kappa {\rm p} \kappa}$, and 
$T_{{\rm p} \kappa \kappa \kappa}$ are given by 
\beqa
T_{\kappa \kappa \kappa \kappa}(\bd{\ell}_{1}, \bd{\ell}_{2}, \bd{\ell}_{3}, \bd{\ell}_{4})
&=& \int_{0}^{\chi_{s}}{\rm d}\chi\, \frac{W_{\kappa}^4}{r(\chi)^6} 
T_{\delta \delta \delta \delta}(\bd{k}_1, \bd{k}_2, \bd{k}_3, \bd{k}_4; z(\chi)), \\
T_{{\rm p}  \kappa {\rm p} \kappa}(\bd{\ell}_{1}, \bd{\ell}_{2}, \bd{\ell}_{3}, \bd{\ell}_{4})
&=& \int_{0}^{\chi_{s}}{\rm d}\chi\, \frac{W_{\kappa}^2}{r(\chi)^6} 
\left(\frac{1}{N_{\rm peak}} \frac{{\rm d}^2 V}{{\rm d}\chi {\rm d}\Omega}\right)^2
T_{c \delta c \delta}(\bd{k}_1, \bd{k}_2, \bd{k}_3, \bd{k}_4; z(\chi)), \\
T_{{\rm p} \kappa \kappa \kappa}(\bd{\ell}_{1}, \bd{\ell}_{2}, \bd{\ell}_{3}, \bd{\ell}_{4})
&=& \int_{0}^{\chi_{s}}{\rm d}\chi\, \frac{W_{\kappa}^3}{r(\chi)^6} 
\left(\frac{1}{N_{\rm peak}} \frac{{\rm d}^2 V}{{\rm d}\chi {\rm d}\Omega}\right)
T_{c \delta \delta \delta}(\bd{k}_1, \bd{k}_2, \bd{k}_3, \bd{k}_4; z(\chi)),
\eeqa
where $k_{i} = l_{i}/\chi$,
$\chi_{s}$ is the comoving distance to sources,
the window function $W_{\kappa}$ is given by Eq.~(\ref{eq:kappa_weight})
and $N_{\rm peak}$ is defined by Eq.~(\ref{eq:npeak}).

Previous works show that 
the dominant contribution of the non-Gaussian covariance is 
the so-called one-halo term of the relevant tri-spectrum at $\ell \simgt 100$
\citep[e.g.,][]{Sato2009}.
One halo term arises from the four point correlation 
between different modes $\bd{k}_{i} \, (i=1,2,3,4)$ in a 
single dark matter halo.
Thus, one halo term of the underlying tri-spectra between $\delta_{m}$ and $\delta_{c}$ 
can be calculated as
\beqa
T_{\delta \delta \delta \delta}^{1h}(\bd{k}_1, \bd{k}_2, \bd{k}_3, \bd{k}_4; z)
&=&
\int {\rm d}M\, \frac{{\rm d}n}{{\rm d}M}(z, M)\, 
\left(\frac{M}{\bar{\rho}_{m}(z)}\right)^4 
\tilde{u}_{m}(\bd{k}_{1}|z, M)
\tilde{u}_{m}(\bd{k}_{2}|z, M)
\tilde{u}_{m}(\bd{k}_{3}|z, M)
\tilde{u}_{m}(\bd{k}_{4}|z, M), \\
T_{c \delta c \delta}^{1h}(\bd{k}_1, \bd{k}_2, \bd{k}_3, \bd{k}_4; z)
&=&
\int {\rm d}M\, \frac{{\rm d}n}{{\rm d}M}(z, M)\, 
\left(\frac{M}{\bar{\rho}_{m}(z)}\right)^2 
S(z, M)
\tilde{u}_{m}(\bd{k}_{2}|z, M)
\tilde{u}_{m}(\bd{k}_{4}|z, M), \\
T_{c \delta \delta \delta}^{1h}(\bd{k}_1, \bd{k}_2, \bd{k}_3, \bd{k}_4; z)
&=&
\int {\rm d}M\, \frac{{\rm d}n}{{\rm d}M}(z, M)\, 
\left(\frac{M}{\bar{\rho}_{m}(z)}\right)^3 
S(z, M)
\tilde{u}_{m}(\bd{k}_{2}|z, M)
\tilde{u}_{m}(\bd{k}_{3}|z, M)
\tilde{u}_{m}(\bd{k}_{4}|z, M).
\eeqa
Another important contribution of covariance at degree scales or less 
is so-called halo sampling variance (HSV) \citep{Sato2009, 2013MNRAS.429..344K}.
This term describes the mode-coupling between the measured 
Fourier modes and larger modes of length-scales comparable 
to survey volume.
It is expected to be important
when the number of massive haloes found in a finite region 
is correlated with the overall mass density fluctuation in the region 
\citep{2003ApJ...584..702H}.
Following \citet{2013MNRAS.429..344K}, 
we model the HSV of weak lensing statistics as
\beqa
{\rm Cov}\left[P_{\kappa\kappa}(\ell), P_{\kappa\kappa}(\ell^{\prime})\right]_{\rm HSV}
&=&
\int_{0}^{\chi_{s}}
{\rm d}\chi \, \left(\frac{{\rm d}^2 V}{{\rm d}\chi {\rm d}\Omega}\right)^2 \,
\left[
\int {\rm d}M\, \frac{{\rm d}n}{{\rm d} M}b_{h}(M) |\tilde{\kappa}_{h}(\ell|M)|^2
\right]
\nonumber \\
&\times&
\left[
\int {\rm d}M^{\prime}\, \frac{{\rm d}n}{{\rm d} M^{\prime}}b_{h}(M^{\prime}) |\tilde{\kappa}_{h}(\ell^{\prime}|M^{\prime})|^2
\right]
\left[
\int_{0}^{\infty}
\frac{k {\rm d}k}{2\pi}
P^{L}_{m}(k)|\tilde{W}(k\chi \Theta_{\rm survey})|^2
\right], \\
{\rm Cov}\left[P_{{\rm p} \kappa}(\ell), P_{{\rm p} \kappa}(\ell^{\prime})\right]_{\rm HSV}
&=&
\int_{0}^{\chi_{s}}
{\rm d}\chi \, \left(\frac{{\rm d}^2 V}{{\rm d}\chi {\rm d}\Omega}\right)^2 \,
\left[
\int {\rm d}M\, \frac{{\rm d}n}{{\rm d} M}b_{h}(M) 
S(z, M) \tilde{\kappa}_{h}(\ell|M)
\right]
\nonumber \\
&\times&
\left[
\int {\rm d}M^{\prime}\, \frac{{\rm d}n}{{\rm d} M^{\prime}}b_{h}(M^{\prime})
S(z, M^{\prime}) \tilde{\kappa}_{h}(\ell^{\prime}|M^{\prime})
\right]
\left[
\int_{0}^{\infty}
\frac{k {\rm d}k}{2\pi}
P^{L}_{m}(k)|\tilde{W}(k\chi \Theta_{\rm survey})|^2
\right], \\
{\rm Cov}\left[P_{{\rm p} \kappa}(\ell), P_{\kappa \kappa}(\ell^{\prime})\right]_{\rm HSV}
&=&
\int_{0}^{\chi_{s}}
{\rm d}\chi \, \left(\frac{{\rm d}^2 V}{{\rm d}\chi {\rm d}\Omega}\right)^2 \,
\left[
\int {\rm d}M\, \frac{{\rm d}n}{{\rm d} M}b_{h}(M) 
S(z, M) \tilde{\kappa}_{h}(\ell|M)
\right]
\nonumber \\
&\times&
\left[
\int {\rm d}M^{\prime}\, \frac{{\rm d}n}{{\rm d} M^{\prime}}b_{h}(M^{\prime})
|\tilde{\kappa}_{h}(\ell^{\prime}|M^{\prime})|^2
\right]
\left[
\int_{0}^{\infty}
\frac{k {\rm d}k}{2\pi}
P^{L}_{m}(k)|\tilde{W}(k\chi \Theta_{\rm survey})|^2
\right],
\eeqa
where 
$\kappa_{h}$ is the Fourier transforming of Eq.~(\ref{eq:kappa_nfw}).
Here, $\tilde{W}(k\chi \Theta_{\rm survey})$ represents the window function 
of the survey region in Fourier space with $\Theta_{\rm survey}$ 
denoting the squared root of the survey area.
We use the circular function of $\tilde{W}(x) = J_{1}(x)/x$.

Cross covariance between the number count of weak lensing selected clusters
and the lensing spectra can be naturally incorporated in the halo model 
\citep{2007NJPh....9..446T, 2014MNRAS.441.2456T}.

The covariance of $N_{\rm peak}$ with two different thresholds 
is given by
\beqa
{\rm Cov}[N_{\rm peak}(\nu_{{\rm thre},1}), N_{\rm peak}(\nu_{{\rm thre},2})]
&=&
\int {\rm d}^2\theta \, W({\bd \theta}) 
\int {\rm d}\theta^{\prime} \, W({\bd \theta}^{\prime})
\int {\rm d}\chi \, \frac{{\rm d}^2 V}{{\rm d}\chi{\rm d}\Omega}
\int {\rm d}\chi^{\prime} \, \frac{{\rm d}^2 V}{{\rm d}\chi^{\prime}{\rm d}\Omega^{\prime}} \nonumber \\
&&
\times
\left[
n_{\rm cl}(\chi{\bd \theta}, z(\chi)|\nu_{{\rm thre}, 1})
n_{\rm cl}(\chi^{\prime}{\bd \theta}^{\prime}, z(\chi^{\prime})|\nu_{{\rm thre}, 2})
-\bar{n}_{\rm cl}(z(\chi)|\nu_{{\rm thre}, 1})\bar{n}_{\rm cl}(z(\chi^{\prime})|\nu_{{\rm thre}, 2})
\right],
\eeqa
where $W({\bd \theta}) $ represents the window function in real space.
Performing the similar calculation as in Eq.~(\ref{eq:2pcf_ndens_field_cl_derivation}),
one can find that (see also the appendix in \citet{2007NJPh....9..446T})
\beqa
{\rm Cov}[N_{\rm peak}(\nu_{{\rm thre},1}), N_{\rm peak}(\nu_{{\rm thre},2})]
&=& \delta_{12}
\Biggl\{
\frac{N_{\rm peak}(\nu_{{\rm thre},1})}{4\pi f_{\rm sky}}
+
\int {\rm d}\chi \, \left(\frac{{\rm d}^2 V}{{\rm d}\chi {\rm d}\Omega}\right)^2\, 
r(\chi)^{-2} \, 
\left[\int {\rm d}M\, \frac{{\rm d}n}{{\rm d} M}b_{h}(M) 
S(z, M | \nu_{{\rm thre}, 1})
\right]^2
\nonumber \\
&&
\,\,\,\,\,\,\,\,\,\,\,\,\,\,
\,\,\,\,\,\,\,\,\,\,\,\,\,\,
\,\,\,\,\,\,\,\,\,\,\,\,\,\,
\,\,\,\,\,\,\,\,\,\,\,\,\,\,
\,\,\,\,\,\,\,\,\,\,\,\,\,\,
\,\,\,\,\,\,\,\,\,\,\,\,\,\,
\,\,\,\,\,\,\,\,\,\,\,\,\,\,
\times
\left[
\int_{0}^{\infty}
\frac{k {\rm d}k}{2\pi}
P^{L}_{m}(k)|\tilde{W}(k\chi \Theta_{\rm survey})|^2
\right]
\Biggr\},
\eeqa
where $f_{\rm sky}$ is the sky fraction for survey of interest.
In order to derive the cross covariance between $N_{\rm peak}$ 
and $P_{\kappa\kappa}$ or $P_{{\rm p} \kappa}$, 
we first define the estimator of power spectrum as
\beqa
P_{\kappa\kappa}^{\rm est}(\ell) 
=
\frac{1}{4\pi f_{\rm sky}N_{p}(\ell)} 
\sum_{{\bd \ell}}{\tilde \kappa}({\bd \ell}){\tilde \kappa}(-{\bd \ell}), \, \, \, 
P_{{\rm p}\kappa}^{\rm est}(\ell|\nu_{\rm thre}) 
= 
\frac{1}{4\pi f_{\rm sky}N_{p}(\ell)} 
\sum_{{\bd \ell}}{\tilde p}({\bd \ell}|\nu_{\rm thre}){\tilde \kappa}(-{\bd \ell}),
\label{eq:est_pk}
\eeqa
where the summation is taken over all the Fourier modes in the range of 
$[\ell-\Delta\ell/2, \ell+\Delta\ell/2]$ and $\Delta\ell$ is the width of multipoles.
Also, $N_{p}(\ell)$ represents the number of modes 
to estimate of power spectrum with the multipole of $\ell$.
With Eq.~(\ref{eq:est_pk}), 
we can express the cross covariance of $N_{\rm peak}$ and $P_{\kappa\kappa}$ as
\beqa
{\rm Cov}[N_{\rm peak}(\nu_{\rm thre}), P_{\kappa \kappa}(\ell)]
&=& 
\frac{1}{4\pi f_{\rm sky} N_{p}(\ell)}
\sum_{\bd \ell}
\int {\rm d}^2\theta \, W({\bd \theta})
\langle
{\tilde \kappa}({\bd \ell}){\tilde \kappa}(-{\bd \ell})
\left[
n_{\rm cl}(\chi{\bd \theta}, z(\chi)|\nu_{{\rm thre}})
-\bar{n}_{\rm cl}(z(\chi)|\nu_{\rm thre})
\right]
\rangle, \nonumber \\
&=&
\frac{1}{4\pi f_{\rm sky} N_{p}(\ell)}
\int {\rm d}^2\theta \, W({\bd \theta})
N_{\rm peak}(\nu_{\rm thre})
\sum_{\bd \ell}
\sum_{{\bd \ell}^{\prime}}
\langle
{\tilde \kappa}({\bd \ell}){\tilde \kappa}(-{\bd \ell})
{\tilde p}({\bd \ell}^{\prime}|\nu_{\rm thre})
\rangle
e^{-i{\bd \ell}^{\prime}\cdot{\bd \theta}},
\eeqa
where we use Eq.~(\ref{eq:def_peak_field}) through the derivation.
Similarly, the cross covariances of $N_{\rm peak}$ and $P_{{\rm p}\kappa}$ is given by
\beqa
{\rm Cov}[N_{\rm peak}(\nu_{{\rm thre},1}), P_{{\rm p} \kappa}(\ell|\nu_{{\rm thre},2})]
&=&
\frac{1}{4\pi f_{\rm sky} N_{p}(\ell)}
\int {\rm d}^2\theta \, W({\bd \theta})
N_{\rm peak}(\nu_{{\rm thre}, 2})
\nonumber \\
&&
\,\,\,\,\,\,\,\,\,\,\,\,\,\,
\,\,\,\,\,\,\,\,\,\,\,\,\,\,
\,\,\,\,\,\,\,\,\,\,\,\,\,\,
\,\,\,\,\,\,\,\,\,\,\,\,\,\,
\,\,\,\,\,\,\,\,\,\,\,\,\,\,
\times
\sum_{\bd \ell}
\sum_{{\bd \ell}^{\prime}}
\langle
{\tilde p}({\bd \ell}|\nu_{{\rm thre},1}){\tilde \kappa}(-{\bd \ell})
{\tilde p}({\bd \ell}^{\prime}|\nu_{{\rm thre}, 2})
\rangle
e^{-i{\bd \ell}^{\prime}\cdot{\bd \theta}}.
\eeqa
Therefore, the cross covariances of $N_{\rm peak}$ and lensing spectra 
include the three point correlation of the relevant field $p$ or $\kappa$.
We can thus summarize the relevant covariance are 
\beqa
{\rm Cov}[N_{\rm peak}(\nu_{\rm thre}), P_{\kappa \kappa}(\ell)]
&=&
\frac{1}{4\pi f_{\rm sky}}
\int {\rm d}\chi \, \frac{{\rm d}^2 V}{{\rm d}\chi {\rm d}\Omega} \,
\frac{W_{\kappa}^2}{r(\chi)^4} \, B_{c\delta \delta} (0, \ell/\chi, \ell/\chi; z(\chi)), 
\\
{\rm Cov}[N_{\rm peak}(\nu_{{\rm thre}, 1}), P_{{\rm p} \kappa}(\ell |\nu_{{\rm thre},2})]
&=&
\frac{1}{4\pi f_{\rm sky}}
\int {\rm d}\chi \, \left(\frac{{\rm d}^2 V}{{\rm d}\chi {\rm d}\Omega}\right)^2 \,
\frac{W_{\kappa}}{r(\chi)^4} \, B_{c(1)c(2)\delta} (0, \ell/\chi, \ell/\chi; z(\chi)),
\eeqa
where $B_{c\delta \delta}$ and $B_{c(1)c(2) \delta}$ are defined by
\beqa
\langle \tilde{\delta}_{cl}(\bd{k}_1)\tilde{\delta}_m(\bd{k}_2)
\tilde{\delta}_{m}(\bd{k}_3)\rangle
&=&(2\pi)^3 \delta^{(3)}_{D}(\bd{k}_{123})
B_{c \delta \delta}({\bd k}_{1}, {\bd k}_{2}, {\bd k}_3), \\
\langle \tilde{\delta}_{cl}(\bd{k}_1|\nu_{{\rm thre},1})
\tilde{\delta}_{cl}(\bd{k}_2|\nu_{{\rm thre},2})
\tilde{\delta}_{m}(\bd{k}_3)\rangle
&=&(2\pi)^3 \delta^{(3)}_{D}(\bd{k}_{123})
B_{c(1)c(2)\delta}({\bd k}_{1}, {\bd k}_{2}, {\bd k}_3).
\eeqa
At $\ell \simgt 500$, the main contributor to the covariances is 
the one-halo term of the above bi-spectra.
Similarly to the case of tri-spectra, 
the corresponding terms are expressed as
\beqa
B_{c \delta \delta}^{1h}(\bd{k}_1, \bd{k}_2, \bd{k}_3; z)
&=&
\int {\rm d}M\, \frac{{\rm d}n}{{\rm d}M}(z, M)\, 
\left(\frac{M}{\bar{\rho}_{m}(z)}\right)^2 
S(z, M|\nu_{\rm thre})
\tilde{u}_{m}(\bd{k}_{2}|z, M)
\tilde{u}_{m}(\bd{k}_{3}|z, M), \\
B_{c(1)c(2) \delta}^{1h}(\bd{k}_1, \bd{k}_2, \bd{k}_3; z)
&=&
\int {\rm d}M\, \frac{{\rm d}n}{{\rm d}M}(z, M)\, 
\left(\frac{M}{\bar{\rho}_{m}(z)}\right) 
S(z, M|\nu_{{\rm thre}, 1})
S(z, M|\nu_{{\rm thre}, 2}) 
\tilde{u}_{m}(\bd{k}_{3}|z, M).
\eeqa

\section{NUMERICAL SIMULATION}\label{sec:sim}

In order to study in detail the  weak lensing statistics considered 
in this paper, we use mock weak lensing catalogs
generated from a set of full-sky weak gravitational simulations.
The $N$-body simulations reproduce the gravity-driven 
non-Gaussianities in the underlying matter density field,
and realistic mask regions are pasted on our mock catalogues.
Thus we can compare our model directly with the simulated catalogues.

\subsection{$N$-body simulation}
\label{subsec:nbody}
We first run a number of cosmological $N$-body simulations 
to generate three-dimensional matter density fields. 
We use the parallel Tree-Particle Mesh code {\tt Gadget2}
\citep{Springel2005}. 
We arrange the simulation boxes
to cover the past light-cone of a hypothetical observer 
with an angular extent of $4\pi/8$ steradian.
A set of simulations consist of six different boxes
and cover one-eighth of the sky.
In order to cover the full-sky,
we use the same boxes 
by adopting periodic boundary condition (see Figure 2).

The simulations are run with $1024^3$ dark matter particles
in six different volumes:
the box side length ranges from $450\, h^{-1}$Mpc to $2700\, h^{-1}$Mpc
with increments of $450\, h^{-1}$Mpc.
The largest volume simulations with $2700\, h^{-1}$Mpc on a side
enable us to simulate the gravitational lensing effect 
with source redshift of $\sim 1$.
We generate the initial conditions using a parallel code 
developed by \citet{2009PASJ...61..321N} and
\citet{2011A&A...527A..87V}, which employs the 
second-order Lagrangian perturbation theory 
\cite[e.g.][]{2006MNRAS.373..369C}.
We set slightly different initial redshift $z_{\rm init}$
as the box size increases.
In order to generate the initial conditions, 
we calculate the linear matter transfer function 
using {\tt CAMB} \citep{Lewis2000}.
Our fiducial cosmological model is characterized by the following parameters:
matter density $\Omega_{\rm m0}=0.279$, 
dark energy density $\Omega_{\Lambda 0}=0.721$, 
the density fluctuation amplitude
$\sigma_{8}=0.823$,
the parameter of the equation of state of dark energy $w_{0} = -1$,
Hubble parameter $h=0.700$ and 
the scalar spectral index $n_s=0.972$.
These parameters are consistent with 
the WMAP nine-year results \citep{Hinshaw2013}.
The parameter of our $N$-body simulations are summarized in Table~\ref{tab:n-body}.

\begin{table*}
\begin{tabular}{@{}lcccc|}
\hline
\hline
$L_{\rm box}\, [h^{-1}{\rm Mpc}]$ & $z_{\rm init}$ & No. of sim. & output redshift \\ \hline
450 & 72 & 10 & 0.025, 0.076, 0.129\\
900 & 36 & 10 & 0.182, 0.237, 0.294\\
1350 & 24 & 10 & 0.352, 0.412, 0.475\\
1800 & 18 & 10 & 0.540, 0.607, 0.677\\
2250 & 15 & 10 & 0.751, 0.827, 0.901\\
2700 & 12 & 10 & 0.990, 1.077, 1.169\\
\hline
\end{tabular}
\caption{
	Parameters used for $N$-body simulations. 
	Each simulation was run with $1024^3$ dark matter particles.
	The output redshift of each simulation corresponds 
	to the comoving distance to the center of lens-shells.
	We adopted the standard $\Lambda$CDM model, which 
	is consistent with WMAP nine-year results \citep{Hinshaw2013}.
	}
\label{tab:n-body}
\end{table*}

\subsection{Ray-tracing simulation}\label{subsec:RT}

\begin{figure*}
\centering \includegraphics[clip, width=0.4\columnwidth, viewport = 20 20 530 530]{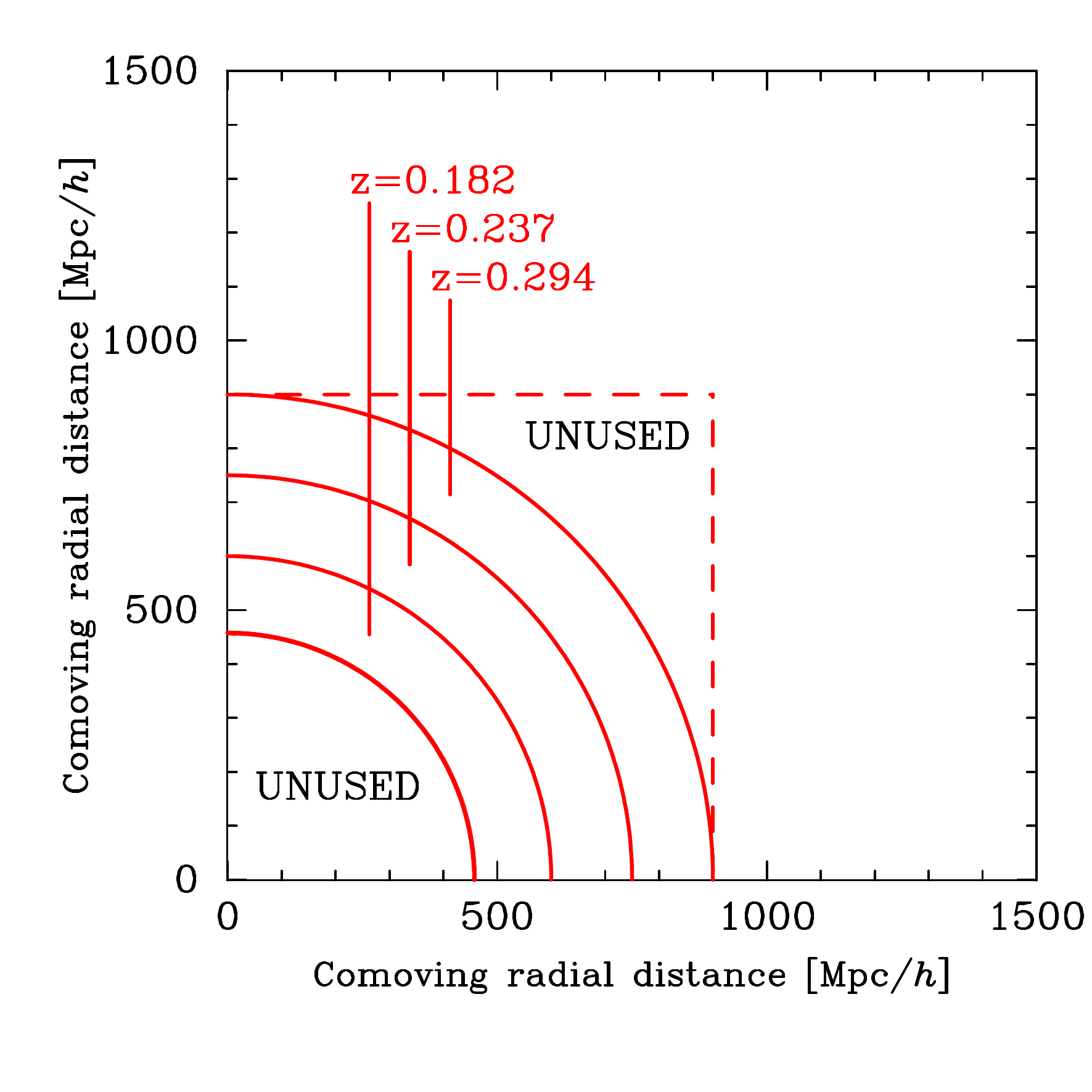}
\centering \includegraphics[clip, width=0.38\columnwidth, viewport = 20 20 530 530]{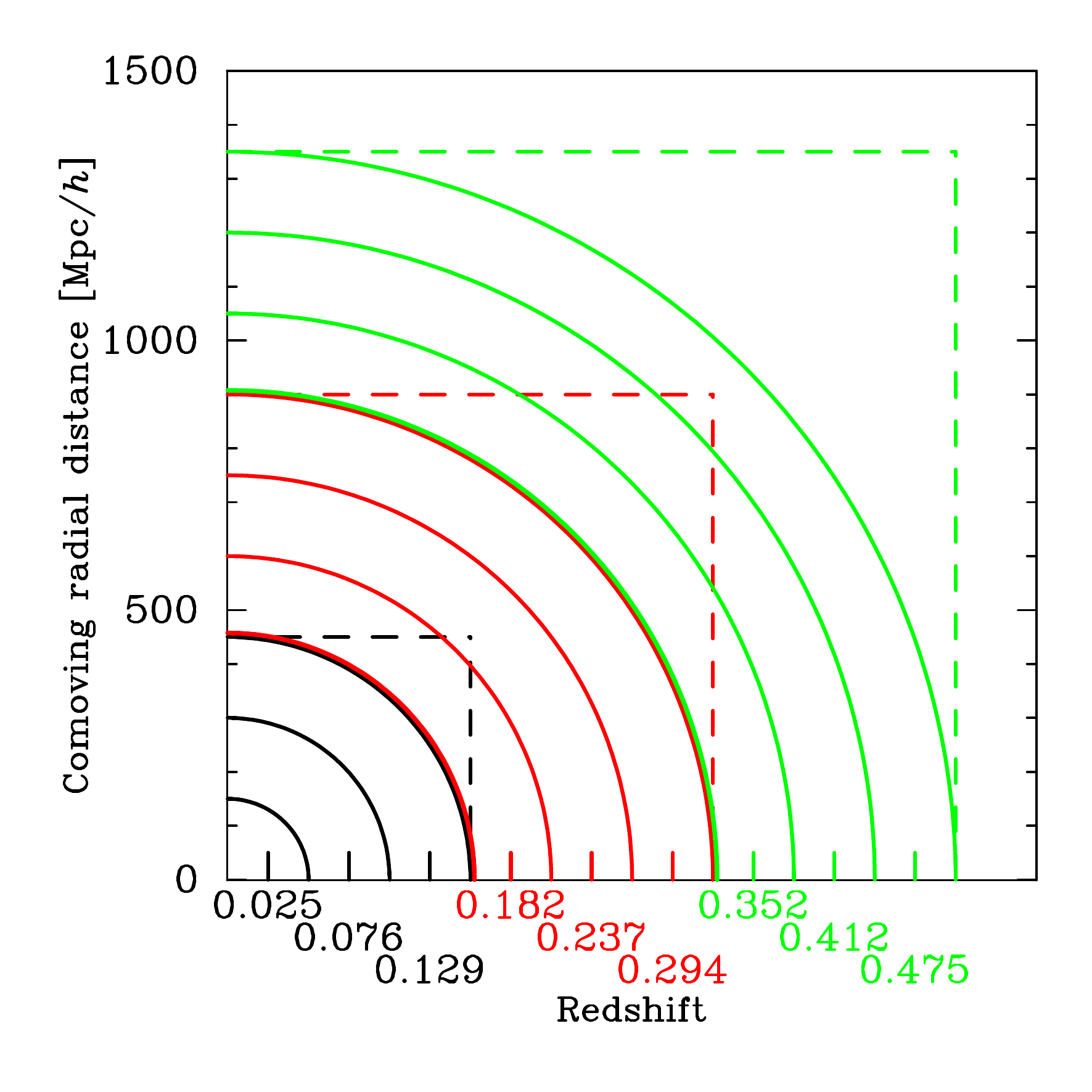}
\caption{
	The configuration of spherical shells of projected matter density.
	The left panel illustrates how we place the outputs of the $N$-body
        simulations. There, the dashed line represents the original box size and
	each of the three shells is taken from the snapshot 
        at different redshift.
	The right panel shows the nested structure of simulation boxes and 
	the configuration of multiple lens shells.
	In the right panel, different colors are used to indicate
        different sets of simulations.
	The corresponding redshift of each shell is also shown 
        at the bottom of the right panel.
	\label{fig:allsky_shell}
	}
\end{figure*}

We briefly summarize our ray-tracing simulations with full-sky coverage.
The detailed description of our ray-tracing simulations is found in 
Appendix~\ref{appendix:raytracing}. 
In our ray-tracing simulation, the light ray path and magnification matrix 
are calculated by the standard multiple lens-plane algorithm.
The multiple lens-plane algorithm on a spherical geometry requires 
contiguous spherical shells of projected matter density.
We thus utilize $N$-body simulations in Section~\ref{subsec:nbody}
to generate a number of thin shells with width of $150\, h^{-1}$Mpc
and produce three shells from a single simulation box 
as shown in the left panel of Figure \ref{fig:allsky_shell}.
In order to extract a target shell region on a lightcone,
we choose an appropriate snapshot at the redshift
that correspond to the comoving distance to the shell from 
a hypothetical observer point (origin).
A set of projected density shells are then configured
using the nested structure of the simulation boxes.
The right panel of Figure \ref{fig:allsky_shell} shows
the configuration of shells used in our ray-tracing 
simulations (only for $z\simlt0.5$).

In this paper, we use {\tt HEALPix} libraries for pixelization 
of a sphere \citep{2005ApJ...622..759G}.
The angular resolution parameter $nside$ is set to be 4096.
The corresponding angular resolution is $\simeq0.86$ arcmin.
For each shell, we calculate the projected mass density map from 
$N$-body particles using Nearest Grid Point method.
Then, we perform the spherical harmonic transformation 
of the density shells 
to calculate the gravitational lensing potential on each shell
via the Poisson equation (see, Eq.~(\ref{eq:phij-Kj})).
The obtained gravitational potential and its derivatives are 
used in the standard multiple lens-plane algorithm.
In our simulation, we follow light ray trajectories  from $z=0$ to $z=1$.
The position of a ray and the magnification matrix $A_{ij}$ at each shell 
are updated
by the recurrence relation as in e.g., \citet{Hilbert2009}.
The initial position of each ray is set to be the center of a pixel 
on {\tt HEALPix} map.
As a ray propagate with deflection, 
the position at each shell deviates from
the position of the {\tt HEALPix} map in general.
We evaluate the potential at each shell
by using an inverse-distant weighted interpolation of the potential field.
The formalism of the multiple lens-plane method on a sphere is found in 
e.g., \citet{2009A&A...497..335T, 2013MNRAS.435..115B}.

\begin{figure*}
\centering \includegraphics[clip, width=0.7\columnwidth, bb= 100 200 858 858]{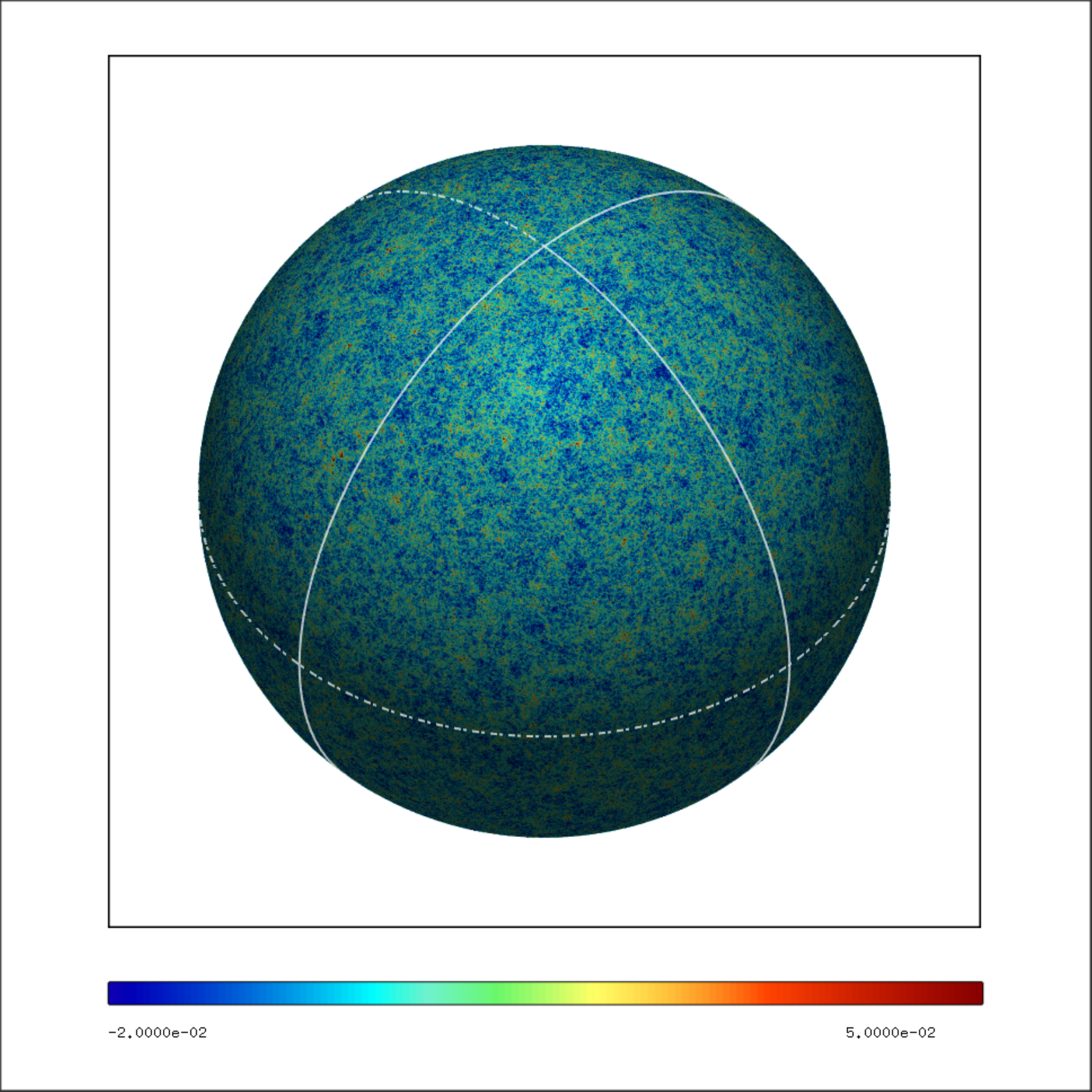}
\caption{
	One of our simulated full-sky convergence maps.
	We perform the Gaussian smoothing with smoothing length of $2.12$ arcmin.
	The shape noise is not included in this figure.
	The red regions represent high convergence, while the blue regions correspond 
	to negative convergence.
	\label{fig:allsky_kappa}
	}
\end{figure*}

We obtain a total of ten all-sky convergence maps
from ten sets of realization of $N$-body simulations.
Note that we use different initial random seeds 
to generate $N$-body simulation data
in order to avoid finding similar structures
along a line of sight through ray-tracing.
Figure \ref{fig:allsky_kappa} shows an example of our 
simulated convergence map on a full-sky.

\subsection{Shape noise and sky mask}
\label{subsec:mask_sim}

\begin{figure*}
\centering \includegraphics[clip, width=0.4\columnwidth, viewport = 20 20 500 600]{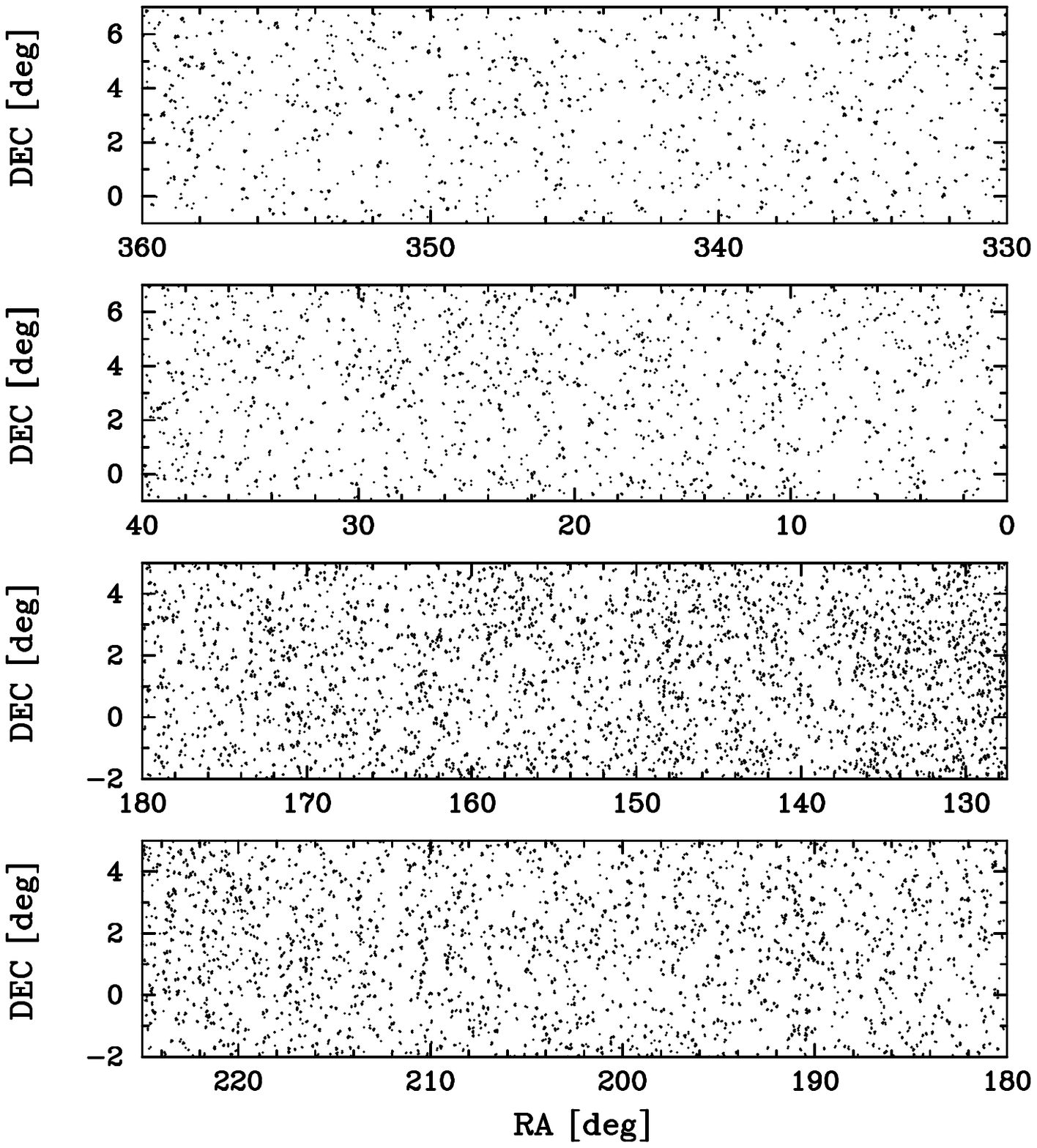}
\centering \includegraphics[clip, width=0.4\columnwidth, viewport = 20 20 530 530]
{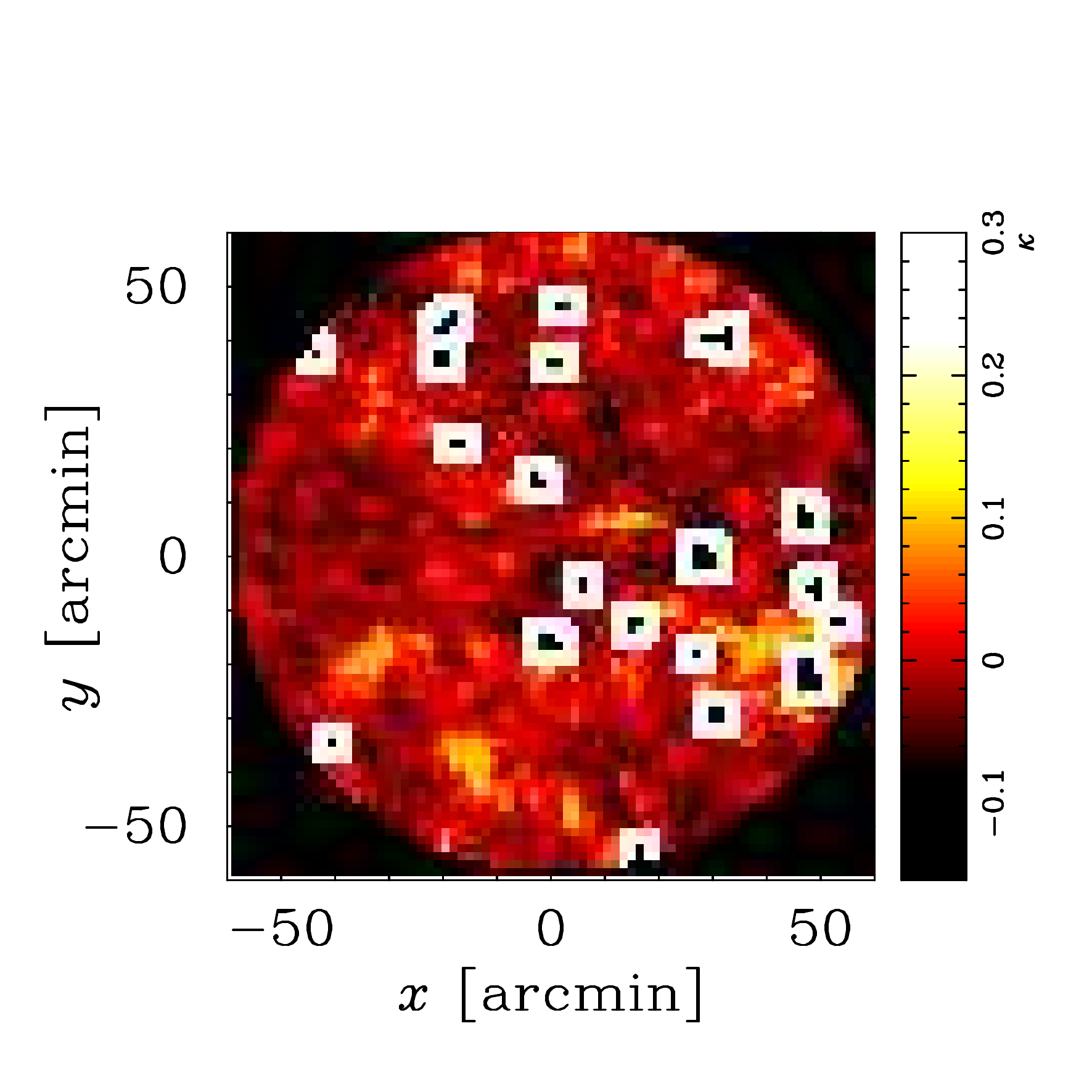}
\caption{
	The configuration of the masked regions.
	In the left panel, black regions show the masks
        on the bright stars with
	the R-band magnitude of $M_{R} < 10$.
	The right panel represents 
	the typical configuration of the masks
        in a circular region with radius of 60 arcmin.
	Black isolated 
         regions show masked regions due to the bright stars and 
	white regions correspond to additional masked region.
	These white regions remove some pixels affected by
	the convolution of survey masks (black regions) 
        with the smoothing filter.
	\label{fig:mask}
	}
\end{figure*}

We generate realistic mock catalogues by adding the effect
of shape noise and sky mask to the obtained convergence maps.
The intrinsic ellipticity of source galaxies is 
the main contaminant of cosmic shear measurement.
We model the shape noise in a convergence map by assuming a two-dimensional 
Gaussian field as follows:
\beqa
\langle \kappa_{\rm N}(\bd{\theta}_{i})
\kappa_{\rm N}(\bd{\theta}_{j})\rangle
= \frac{\sigma_{\gamma}^2}{2n_{\rm gal}\Omega_{\rm pix}} \delta_{ij},
\eeqa
where $\sigma_{\gamma} = 0.4$, $n_{\rm gal}=10\, {\rm arcmin}^{-2}$,
and $\Omega_{\rm pix} = 4\pi/12/4096^2=6.24\times10^{-8}\, {\rm str}$.

It is important to use a realistic sky mask
to study statistics for upcoming lensing surveys.
We here consider masks owing to bright stars.
Among the planned HSC survey regions,
we select two continuous regions with sky 
coverage of $\sim565$ and $\sim680$ squared degrees.
J2000 coordinate of these regions are given by 
$22^{\rm h}00^{\rm m} < {\rm RA} < 2^{\rm h}40^{\rm m}, -1^{\circ} < {\rm DEC} < +7^{\circ}$
and 
$8^{\rm h}30^{\rm m} < {\rm RA} < 15^{\rm h}00^{\rm m}, -2^{\circ} < {\rm DEC} < +5^{\circ}$.
In the following, we consider the bright stars with the R-band 
magnitude of $M_{R} < 17$.
We then select 1,149,871 stars located in these two regions from USNO-A2.0 catalog\footnote{
{\rm http://tdc-www.harvard.edu/catalogs/ua2.html}
}.
In this paper, we assume the following relation between the R-band 
magnitude $M_{R}$
and the effective radius of a halo of bright star $r_{\rm star}$:
\beqa
r_{\rm star} [{\rm arcsec}] = 
\left\{
\begin{array}{ll}
180 & (M_{R} < 9), \\
0.2\times10^{\left[13.75/M_{B}^{0.7}\right]} & ( 9 \le M_{R} \le 17).
\end{array}
\right. 
\label{eq:rstar}
\eeqa
Then, we paste a circular mask with $r_{\rm star}$
around the pixel located at each star using Eq.~(\ref{eq:rstar}).
If $r_{\rm star}$ is less than the 
angular size of our all-sky map ($\sim$ 1 arcmin), 
we simply mask the pixel located in the star.
After the above procedure, we further remove the ``isolated" pixels 
whose surrounding pixels are all labeled as masked pixels.
The final mask configuration generated in this way is 
shown in Figure \ref{fig:mask}.
The left panel displays the masked 
region over the $565+680$ square degrees.
The black dots shows the masked pixels on 
very bright stars $(M_{R} < 10)$
and smaller masked regions are distributed 
in the white region homogeneously 
(but not shown in the figure). 
The mask covers over $\simeq280$ squared degrees in total.
The right panel shows an example of our masked sky simulation 
in a circular region with a radius of $1^{\circ}$.

From a single full-sky simulation,
we make 20 mock HSC convergence maps 
with the masks
by choosing the desired sky coverage 
($565+680=1245$ squared degrees).
We allow to have small overlap regions between the 20 masked maps.
The overlap regions are located near the edge of 
the HSC sky coverage, and thus
we expect this minor compromise does not affect the final results significantly.
Finally, using 10 independent full-sky maps, 
we obtain a total of $20\times10=200$ realizations of mock 
HSC lensing catalogs.

For a smoothed convergence map, we apply the different masking.
As shown in Eq.~(\ref{eq:rstar}), stars with $M_{R} > 12$ 
have smaller mask radii than the pixel size in our simulation.
Therefore, in principle, 
we can extract some information from the pixels where stars 
with $M_{R} > 12$ are located.
Because such faint stars would not affect 
the smoothed convergence map, 
we mask only the bright stars with the R-band magnitude of $M_{R} < 12$. 
There are also ill-defined pixels that are compromised
by the convolution with the smoothing filter.
We remove such ill-defined pixels within 
5 arcmin from the boundary of the mask regions.
Consequently, masked regions on the smoothed convergence ${\cal K}$ maps 
differ from the original masked regions.
Finally, we have a total of $\simeq412$ squared degrees 
as unmasked regions.
The corresponding sky fraction is $\sim$0.01.

\section{RESULT}\label{sec:res}
We present the weak lensing statistics measured from 
the full-sky simulations and also those from 
our 200 mock HSC catalogues.
In the following, we define the threshold of lensing peak 
as ${\cal K}/\sigma_{\rm noise,0}$,
where $\sigma_{\rm noise,0}$ is given by Eq.~(\ref{eq:noise_moment}).
Note that we use this definition also in the case without noise.

\subsection{Ensemble average of statistics}
\label{subsec:res_av}

\subsubsection*{All-sky}

\begin{figure*}
\centering \includegraphics[clip, width=0.4\columnwidth, viewport = 20 20 530 530]{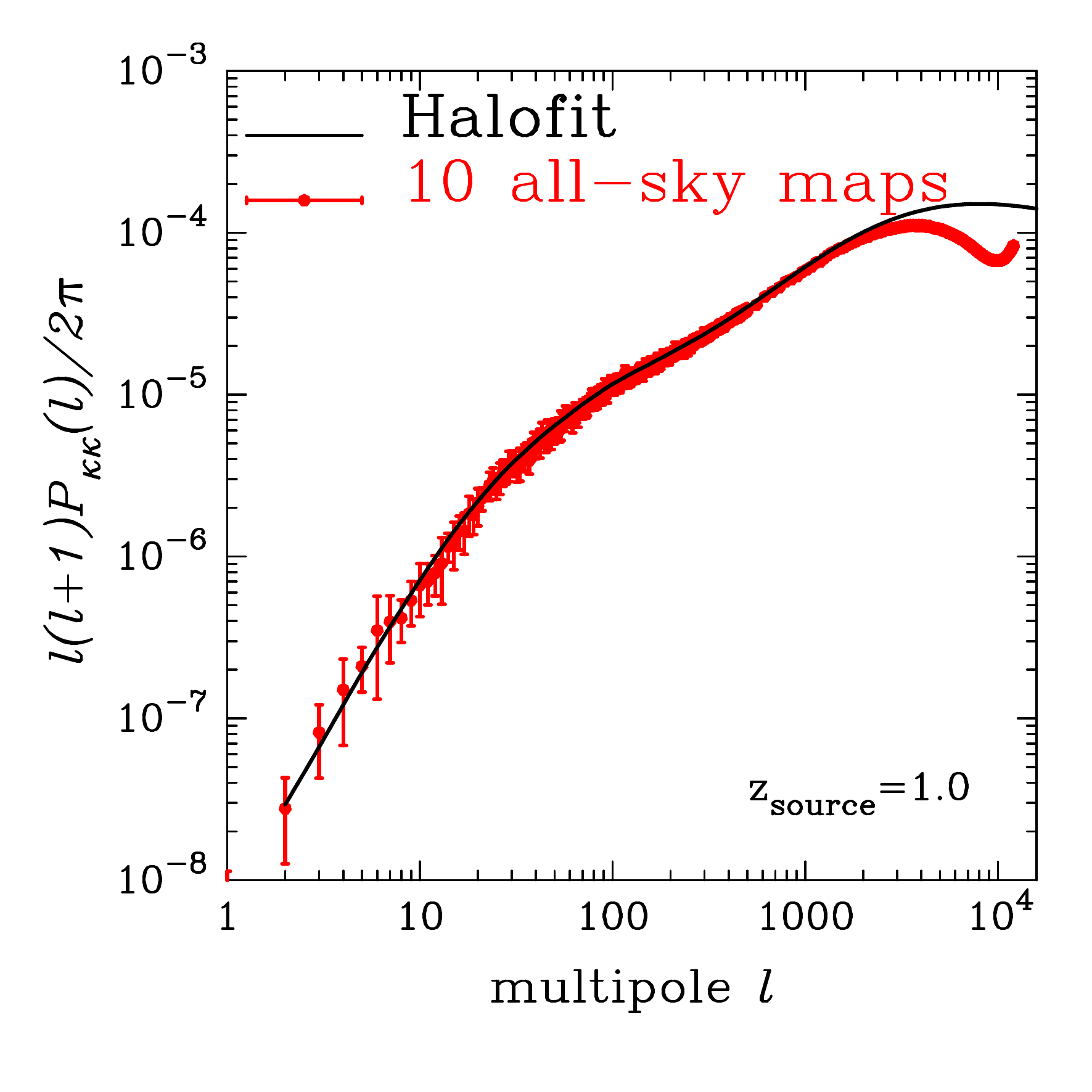}
\centering \includegraphics[clip, width=0.4\columnwidth, viewport = 20 20 530 530]{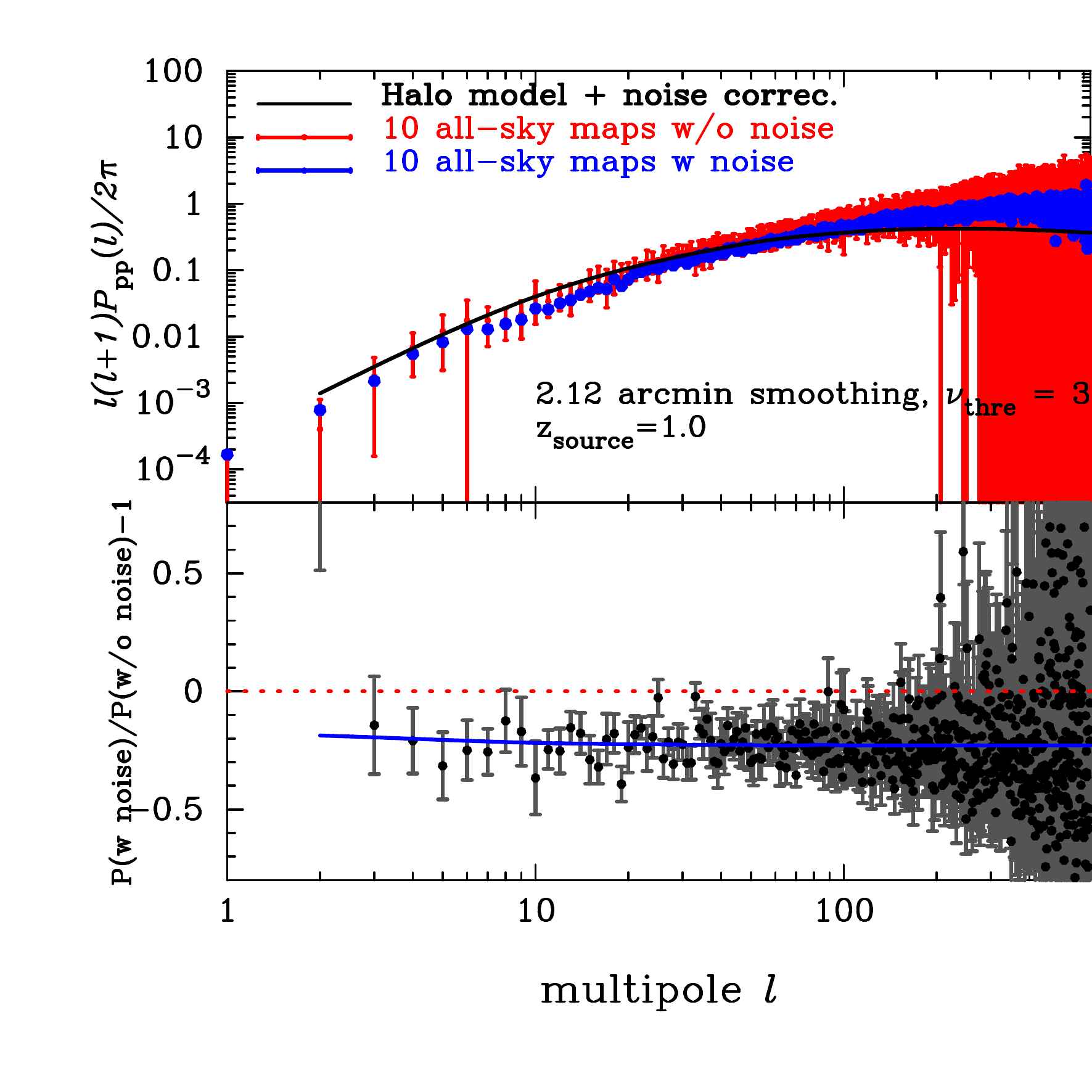}
\centering \includegraphics[clip, width=0.4\columnwidth, viewport = 20 20 530 530]{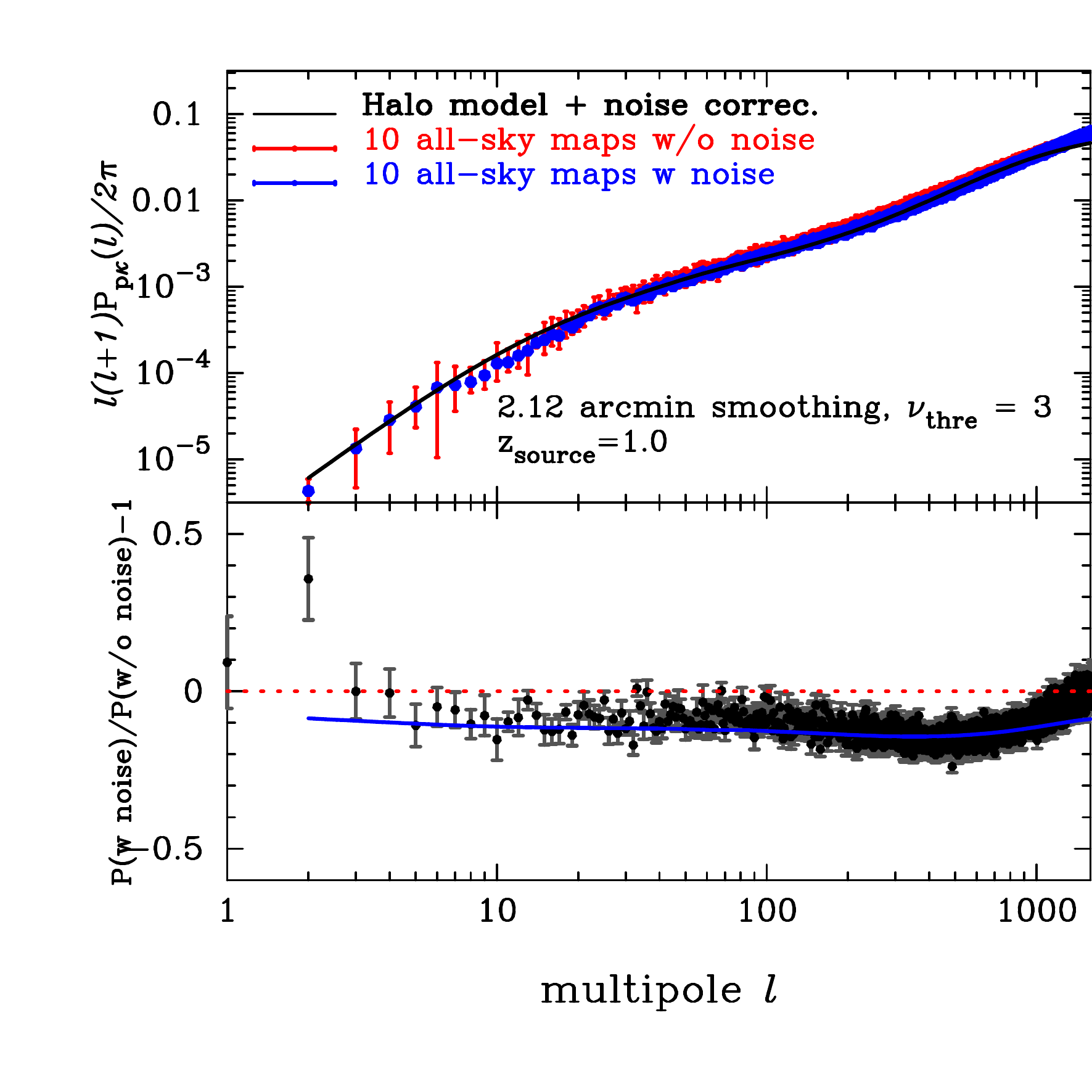}
\centering \includegraphics[clip, width=0.4\columnwidth, viewport = 20 20 530 530]{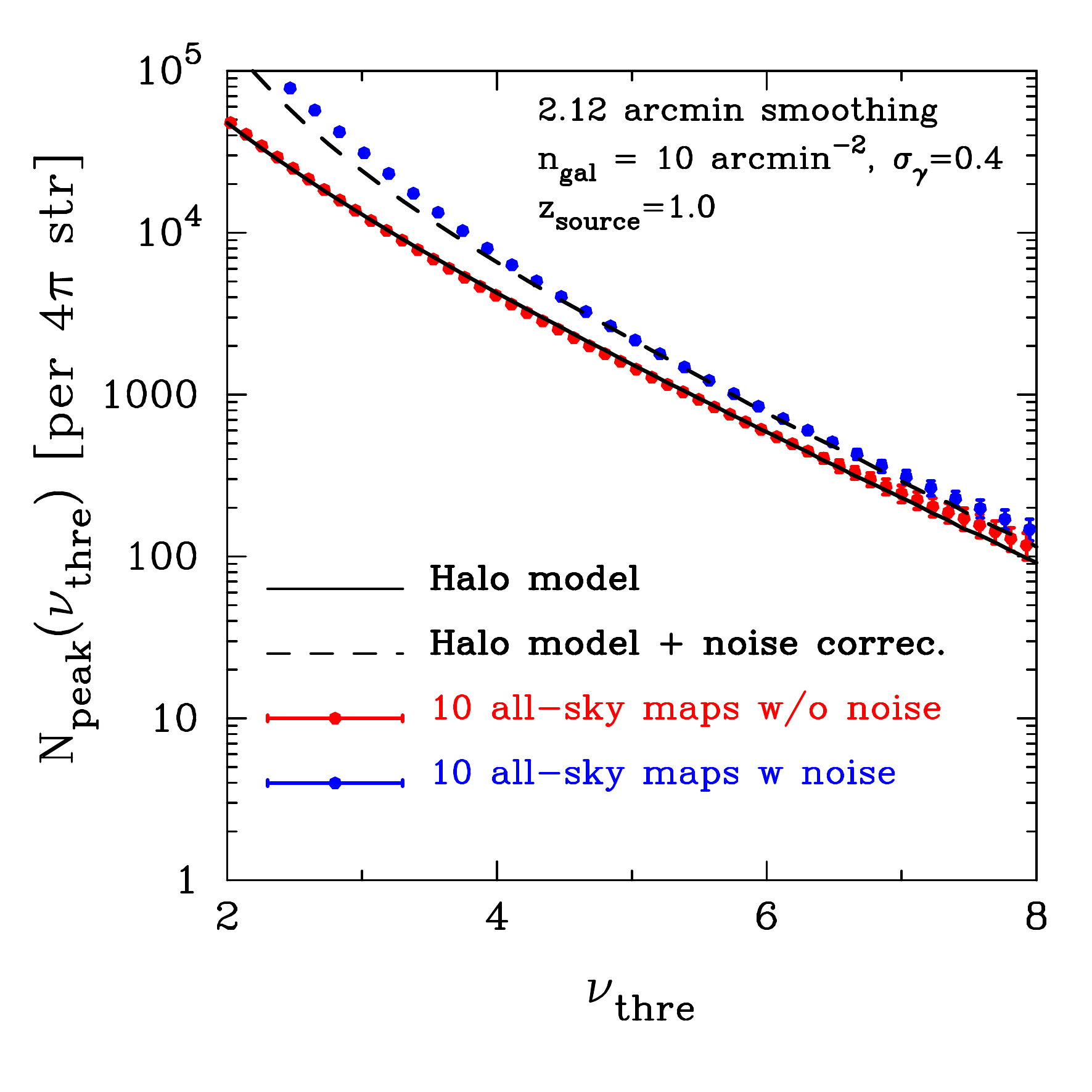}
\caption{
	We compare the statistics measured 
	from the ten all-sky simulations 
        and our model prediction.
	In each panel, the black line is our halo model prediction.
        The red points with error bar 
        represent the measured signal 
	from the 'clean' convergence maps without noise.
	The blue points show the result with noise.
	In the bottom portion in 
	top right and the bottom left panel, 
	we show the ratio of the two.
	The error bars indicate the standard deviation over ten realization.
	Note that the threshold of lensing peaks is defined by ${\cal K}/\sigma_{\rm noise,0}$
	regardless of the presence or absence of noise.
	\label{fig:stat_allsky}
	}
\end{figure*}

We first show the ensemble average statistics over the ten 
full-sky simulations. They can be regarded as the expected values from 
an idealized full-sky observation.
Figure \ref{fig:stat_allsky} summarizes the measurements of
$P_{\kappa\kappa}, P_{{\rm p}{\rm p}}, P_{{\rm p}\kappa}$ and $N_{\rm peak}$.
We selected the lensing peaks with threshold of $\nu_{\rm thre} = 3$ 
(or ${\cal K} \simeq 0.05$) in both the maps with and without shape noise.
To calculate the correlation in harmonic space, 
we correct the pixelisation effect with the pixel window function 
of {\tt HEALPix}.
Overall, our model is in good agreement 
with the measurement from the full-sky simulations.
In the case without shape noise, we calculate
$P_{{\rm p}{\rm p}}$, $P_{{\rm p}\kappa}$ and $N_{\rm peak}$
assuming the one-to-one correspondence between 
lensing peaks and dark matter halos, i.e., 
\beqa
{\rm Prob}({\cal K}_{\rm peak, obs}|\, {\cal K}_{{\rm peak},h}(z, M)) = 
\delta_{D}({\cal K}_{\rm peak, obs}-{\cal K}_{{\rm peak},h}(z, M)).
\eeqa
The results of $P_{{\rm p}{\rm p}}$ and $P_{{\rm p}\kappa}$ 
from the maps with and without noise
show appreciable differences at large angular scales ($\ell \simlt 100$).
This can be explained by the modulation 
of peak height due to the shape noise.
When we select the lensing peaks with a given $\nu_{\rm thre}$, 
we effectively include less massive dark matter haloes as well
in the noisy ${\cal K}$ maps.
Then both $P_{{\rm p}{\rm p}}$ and $P_{{\rm p}\kappa}$ 
are reduced at large angular scales in the noisy maps 
because less massive halos have weaker clustering.
Also, the number count of peaks in the noisy convergence maps
is described well by our model as shown 
in Section~\ref{subsec:WL_stat}.
In the maps with shape noise, we detect more lensing 
peaks for a given threshold,
for example, by a factor of $\sim 3$ for $\nu_{\rm thre}=3$.

\subsubsection*{Masked sky}

\begin{figure*}
\centering \includegraphics[clip, width=0.4\columnwidth, viewport = 20 20 530 530]{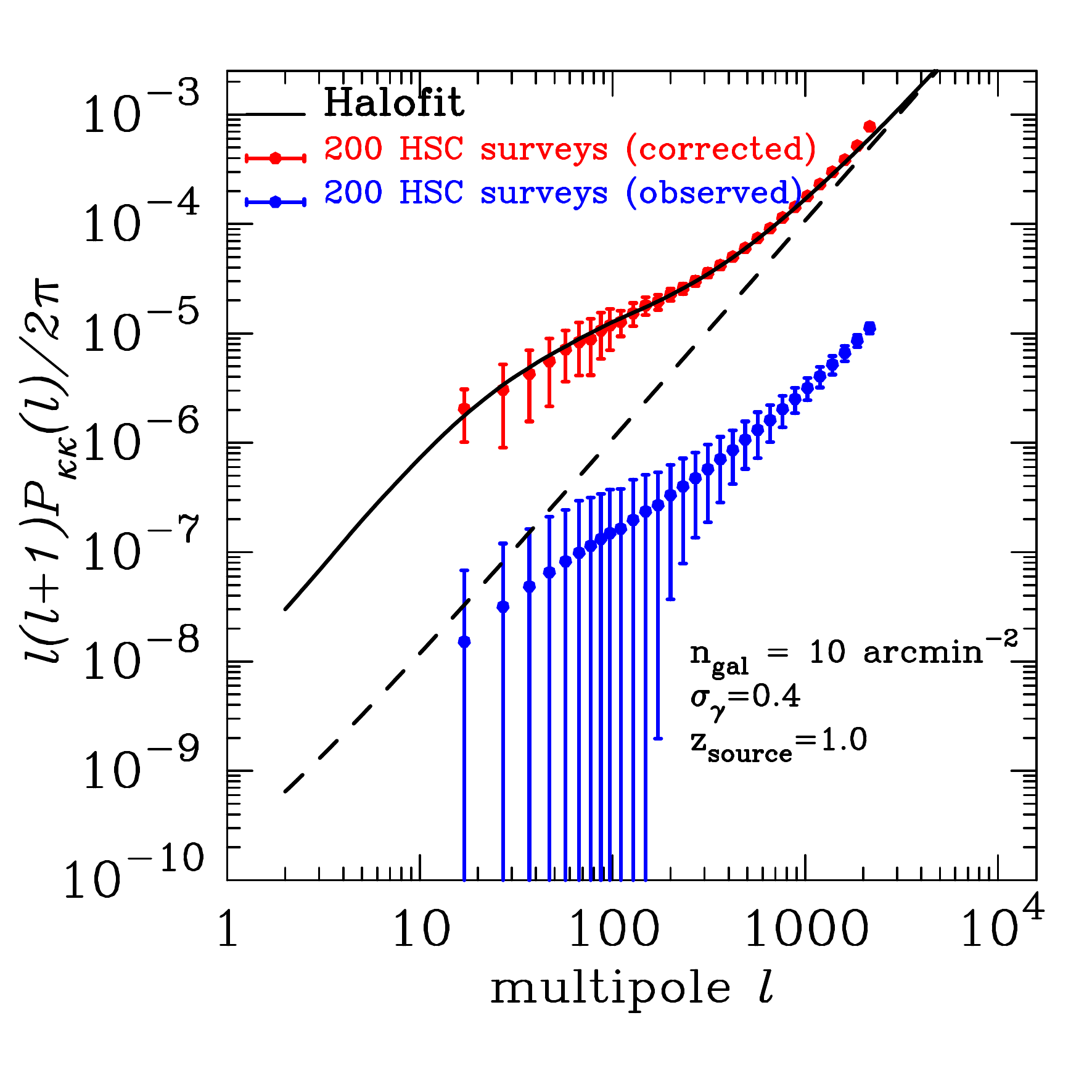}
\centering \includegraphics[clip, width=0.4\columnwidth, viewport = 20 20 530 530]{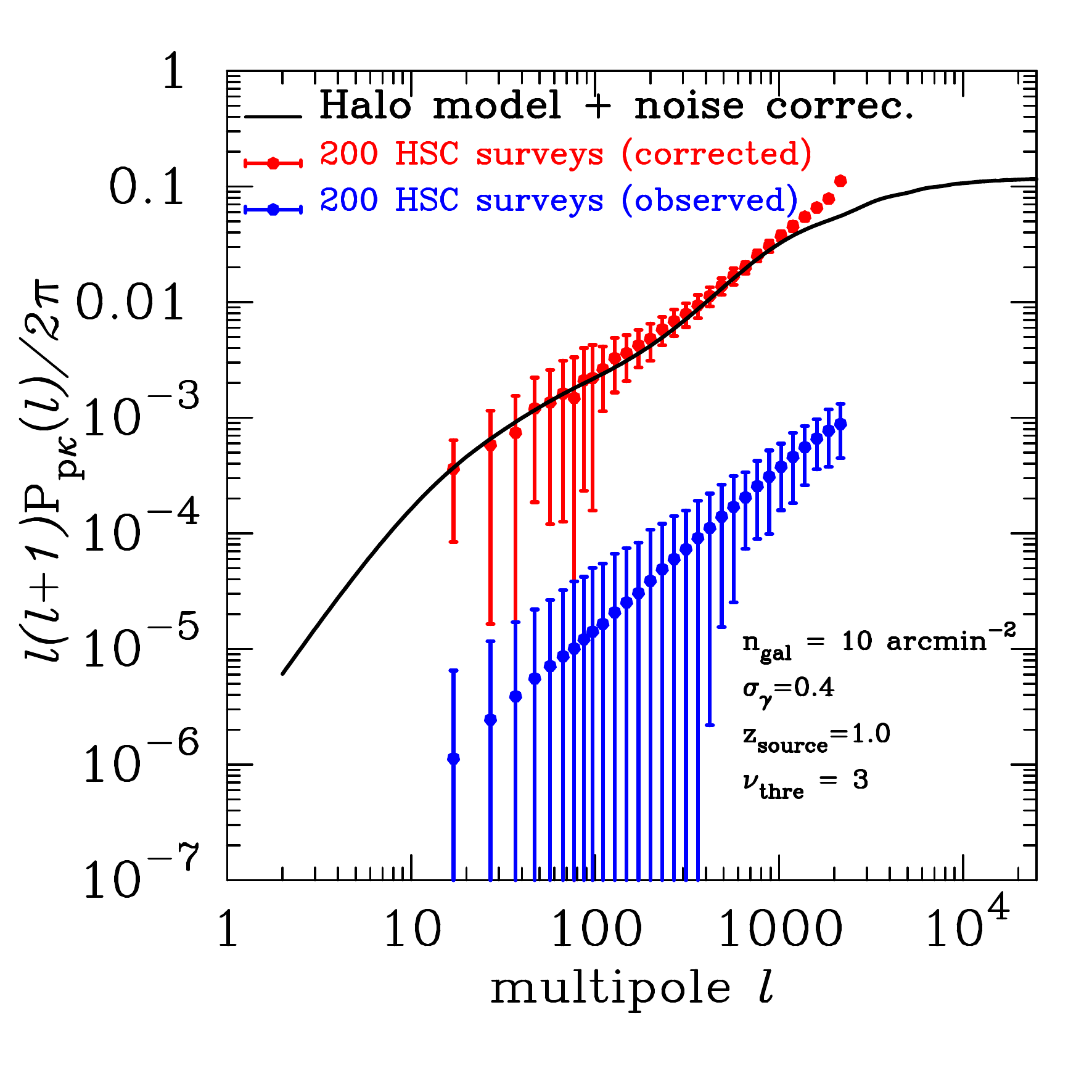}
\caption{
  The auto- and cross-power spectra obtained from 
  200 mock HSC catalogues. The black line is the result of our halo model.
  In the left panel, the dashed line shows the shape noise contribution to convergence power spectrum.
  \label{fig:stat_HSC}
}
\end{figure*}

In the presence of masks, we need the correction 
for the mode-coupling effects in measurements of power spectrum.
We adopt the pseudo-spectrum estimator for this purpose
\citep{2003MNRAS.343..559H, 2004MNRAS.349..603E, 2005MNRAS.360.1262B,2011MNRAS.412...65H}.
The basics of the method is summarized in Appendix~\ref{appendixB}.
In measurements of the binned two spectra $P_{\kappa\kappa}$ and $P_{{\rm p}\kappa}$,
we follow the similar method shown 
in \citet{2005MNRAS.360.1262B, 2011MNRAS.412...65H}.
Let us consider the $b$-th binned power spectrum in the multipole range of
$\ell^{b}_{\rm min} < \ell < \ell^{b+1}_{\rm min}$.
In this case, we can calculate the band-powers $(P_{b})$ as
\beqa
P_{b} = \sum_{b^{\prime}} {\cal M}^{-1}_{bb^{\prime}}
\sum_{\ell}{\cal B}_{b^{\prime}\ell}\tilde{P}(\ell),
\label{eq:pseudo_est}
\eeqa
where $\tilde{P}(\ell)$ is the measured power spectrum on a masked sky 
and ${\cal B}_{b \ell}$ is the binning operator, which is defined by
\beqa
{\cal B}_{b \ell} = 
\left\{
\begin{array}{ll}
\ell (\ell+1)/(2\pi)/(\ell^{b+1}_{\rm min}-\ell^{b}_{\rm min}) 
& (\ell^{b}_{\rm min} < \ell < \ell^{b+1}_{\rm min}), \\
0 & ({\rm otherwise}).
\end{array}
\right. 
\label{eq:bin_op}
\eeqa
Here, we define the binned coupling matrix 
${\cal M}_{bb^{\prime}}$ as
\beqa
{\cal M}_{bb^{\prime}} = \sum_{\ell} {\cal B}_{b \ell}M_{\ell \ell^{\prime}} Q_{\ell^{\prime} b^{\prime}},
\eeqa
where the definition of $M_{\ell \ell^{\prime}}$ is found in Appendix~\ref{appendixB}
and $Q_{b \ell}$ is given by 
\beqa
Q_{b \ell} = 
\left\{
\begin{array}{ll}
2\pi/[\ell(\ell+1)] & (\ell^{b}_{\rm min} < \ell < \ell^{b+1}_{\rm min}), \\
0 & ({\rm otherwise}).
\end{array}
\right. 
\label{eq:bin_op2}
\eeqa
The number of bins is set to be 30.
We perform the binning in linear spacing for the first 10 bins 
as $\ell^{b}_{\rm min} = 2+10b \, (1\le b \le 10)$,
while the remaining 20 bins have logarithmically equal spacing up to
$\ell_{\rm max}=2000$.
The measured and corrected binned power spectra are plotted in
Figure \ref{fig:stat_HSC}.
We use 200 masked sky simulations to calculate the average of 
the binned $P_{\kappa\kappa}$ 
and $P_{{\rm p}\kappa}$.

For cross-correlation, we find $\sim100-200$ lensing peaks 
with $\nu_{\rm thre}=3$ on each simulated HSC map.
In Figure \ref{fig:stat_HSC}, the blue points with error bars 
shows the measured spectrum, and the red points are  
the corrected spectrum 
with the pseudo-spectrum method.
Clearly, we can recover the underlying power spectrum 
with the correction of the mode-coupling due to masks.
The difference in amplitude can be explained approximately
by the effective fraction of sky coverage
$(\sim0.01-0.02)$

\subsection{Covariance}

We use 200 masked sky simulations to calculate covariances 
between the weak lensing statistics of interest.
First, we calculate the convergence power spectrum covariance.
In the flat sky approximation and without masks, 
the covariance of the binned power spectrum is expressed as 
\citep[e.g.,][]{2001ApJ...554...56C}
\beqa
{\rm Cov}[P_{\kappa\kappa}(\ell_{i}), P_{\kappa \kappa}(\ell_{j})]
=
\frac{\delta_{ij}}{2\ell_{i} \Delta \ell f_{\rm sky}}
2P_{\kappa\kappa}^2(\ell_{i})
+\frac{1}{4\pi f_{\rm sky}} 
\int_{\ell_{i}} \, \frac{{\rm d}^2 \bd{\ell}}{A_{s,i}}
\int_{\ell_{j}} \, \frac{{\rm d}^2 \bd{\ell}^{\prime}}{A_{s,j}}
T_{\kappa \kappa \kappa \kappa}(\bd{\ell}, -\bd{\ell}, 
\bd{\ell}^{\prime}, -\bd{\ell}^{\prime}),
\label{eq:flat_sky_cov_PkkPkk}
\eeqa
where $P_{\kappa\kappa}(\ell_{i})$ represents the $i$-th binned power spectrum,
$f_{\rm sky}$ is the fraction of sky covered by the observation, 
$\Delta \ell$ is the bin width in $\ell$ space,
and $A_{s,i}$ is the area of the two-dimensional shell around
the $i$-th bin $\ell_{i}$ in Fourier space.
Here, $T_{\kappa \kappa \kappa \kappa}$ is the tri-spectrum of convergence, 
of which definition and modeling are found in Section~\ref{subsec:cov}.
In practice, we simplify the second term in 
Eq.~(\ref{eq:flat_sky_cov_PkkPkk}) as
\beqa
\int_{\ell_{i}} \, \frac{{\rm d}^2 \bd{\ell}}{A_{s,i}}
\int_{\ell_{j}} \, \frac{{\rm d}^2 \bd{\ell}^{\prime}}{A_{s,j}}
T_{\kappa \kappa \kappa \kappa}(\bd{\ell}, -\bd{\ell}, 
\bd{\ell}^{\prime}, -\bd{\ell}^{\prime})
\rightarrow
T_{\kappa \kappa \kappa \kappa}(\ell_{i}, \ell_{i}, \ell_{j}, \ell_{j}).
\eeqa
In the presence of masked regions,
the estimator of the binned power spectrum 
is given by more complex expressions given in Eq.~(\ref{eq:pseudo_est}).
The mode-coupling due to masked regions 
induces the intricate correlation between different Fourier modes in 
the covariance of $P_{\kappa\kappa}$ 
(the exact expression is found in Appendix~\ref{appendixB}).
Nevertheless,
in the case where the value of masked pixels is set to be either 0 or 1,
we can use the simplified formula of covariances as shown 
in \citet{2004MNRAS.349..603E}.
When we work with harmonic space, the approximated formula 
can be expressed as
\beqa
{\rm Cov}[P_{\kappa\kappa}(\ell_{1}), P_{\kappa \kappa}(\ell_{2})]_{\rm mask}
\simeq
\sum_{\ell^{\prime}_{1} \ell^{\prime}_{2}}
\frac{1}{2\ell^{\prime}_{2}+1} 
(M^{\kappa \kappa})^{-1}_{\ell_{1} \ell^{\prime}_{1}}
(M^{\kappa \kappa})^{-1}_{\ell_{2} \ell^{\prime}_{2}}
M^{\kappa \kappa}_{\ell^{\prime}_{1} \ell^{\prime}_{2}}
\left[
2P_{\kappa \kappa}(\ell^{\prime}_{1}) P_{\kappa \kappa}(\ell^{\prime}_{2})
+
\frac{1}{4\pi}\sqrt{(2\ell^{\prime}_{1}+1)(2\ell^{\prime}_{1}+2)}
T_{\kappa \kappa \kappa \kappa}(\ell^{\prime}_{1}, \ell^{\prime}_{1}, \ell^{\prime}_{2}, \ell^{\prime}_{2})
\right], \label{eq:cov_pseudo_pk}
\eeqa
where $M^{\kappa \kappa}_{\ell \ell^{\prime}}$ 
represents the mode-coupling matrix for convergence.
In Eq.~(\ref{eq:cov_pseudo_pk}), 
we evaluate the term $T_{\kappa \kappa \kappa \kappa}$ 
by the tri-spectrum in harmonic space following 
\citet{2001PhRvD..64h3005H}
because we define $T_{\kappa \kappa \kappa \kappa}$ 
in the flat sky approximation.

\begin{figure*}
\centering \includegraphics[clip, width=0.4\columnwidth, viewport = 20 20 530 530]
{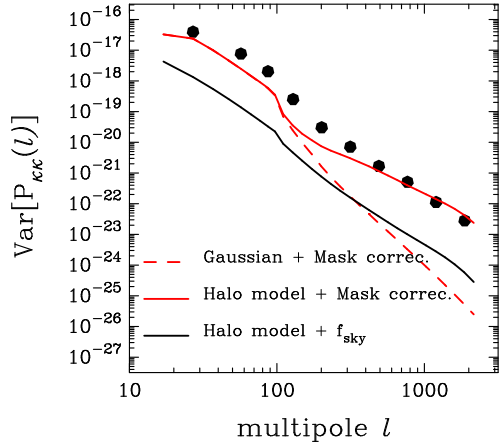}
\caption{
	The variance of the psudo-spectrum estimator of $P_{\kappa \kappa}$
	measured from 200 masked sky simulations.
	The black points show our measurement.
	The three lines represent the different contribution to variance.
	The black line is the simplest halo model with the sky fraction of $f_{\rm sky}=0.023$,
	while the red lines show the halo model covariance with 
	the correction of the effect of masked regions.
	The red dashed line corresponds to the Gaussian model including the effect of masked region.
	For the red solid line, we take into account both non-Gaussianities 
	caused by gravity and masked regions.
	\label{fig:pkkpkk_stat_HSC}
	}
\end{figure*}

We can then examine the validity of 
Eq.~(\ref{eq:flat_sky_cov_PkkPkk})
by using the measured covariance of $P_{\kappa\kappa}$ over 200 masked 
sky simulations.
We use the same binning 
as in Section~\ref{subsec:res_av} 
but reduce the number of bins to 10 by taking average of the binned powers
over nearest $\ell$ bins.
Figure \ref{fig:pkkpkk_stat_HSC} shows the measured
variance of the pseudo-spectrum estimator of $P_{\kappa\kappa}$.
The black point shows the result obtained from the 200 simulations 
and the solid line represents our model of covariance 
in Eq.~(\ref{eq:flat_sky_cov_PkkPkk}) 
with the appropriate value of $f_{\rm sky}$ for our simulations.
Interestingly, although the simple model of 
Eq.~(\ref{eq:flat_sky_cov_PkkPkk}) is expected
to account for non-Gaussianities caused by gravity,
it underestimates the actual covariance by a factor of $\sim 10$.
The corrected covariance components are shown 
in Figure \ref{fig:pkkpkk_stat_HSC}.
The first term and second term in Eq.~(\ref{eq:cov_pseudo_pk})
are plotted as red dashed and red solid line, respectively.
The overall amplitude of the variance 
can be explained by the mask correction,
while the contribution from tri-spectrum dominates at $\ell \simgt 200$.
Clearly, it is problematic to adopt 
the commonly used estimate of covariance 
given by in Eq.~(\ref{eq:flat_sky_cov_PkkPkk}).

\begin{figure*}
\centering \includegraphics[clip, width=0.4\columnwidth, viewport = 20 20 530 530]
{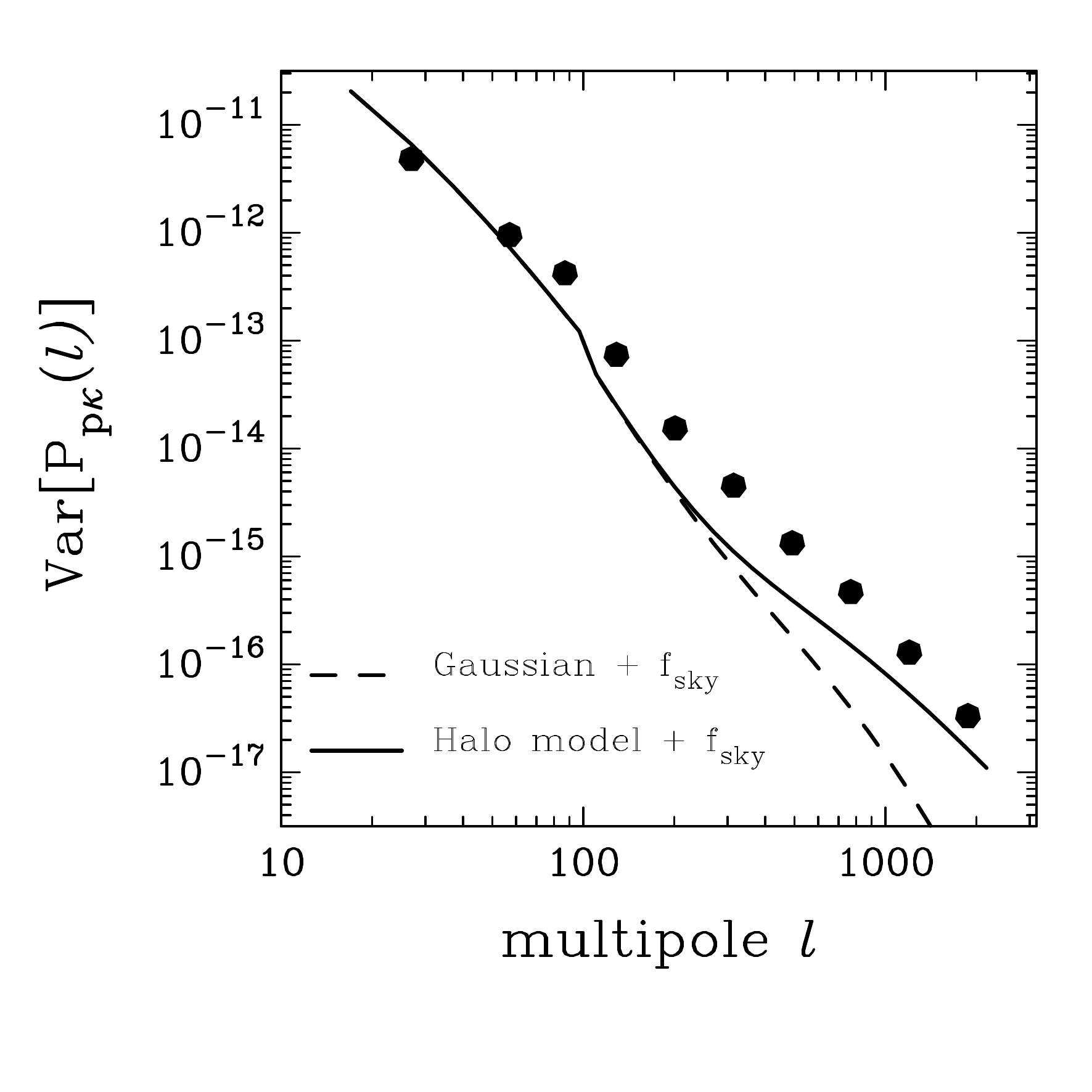}
\centering \includegraphics[clip, width=0.4\columnwidth, viewport = 20 20 530 530]
{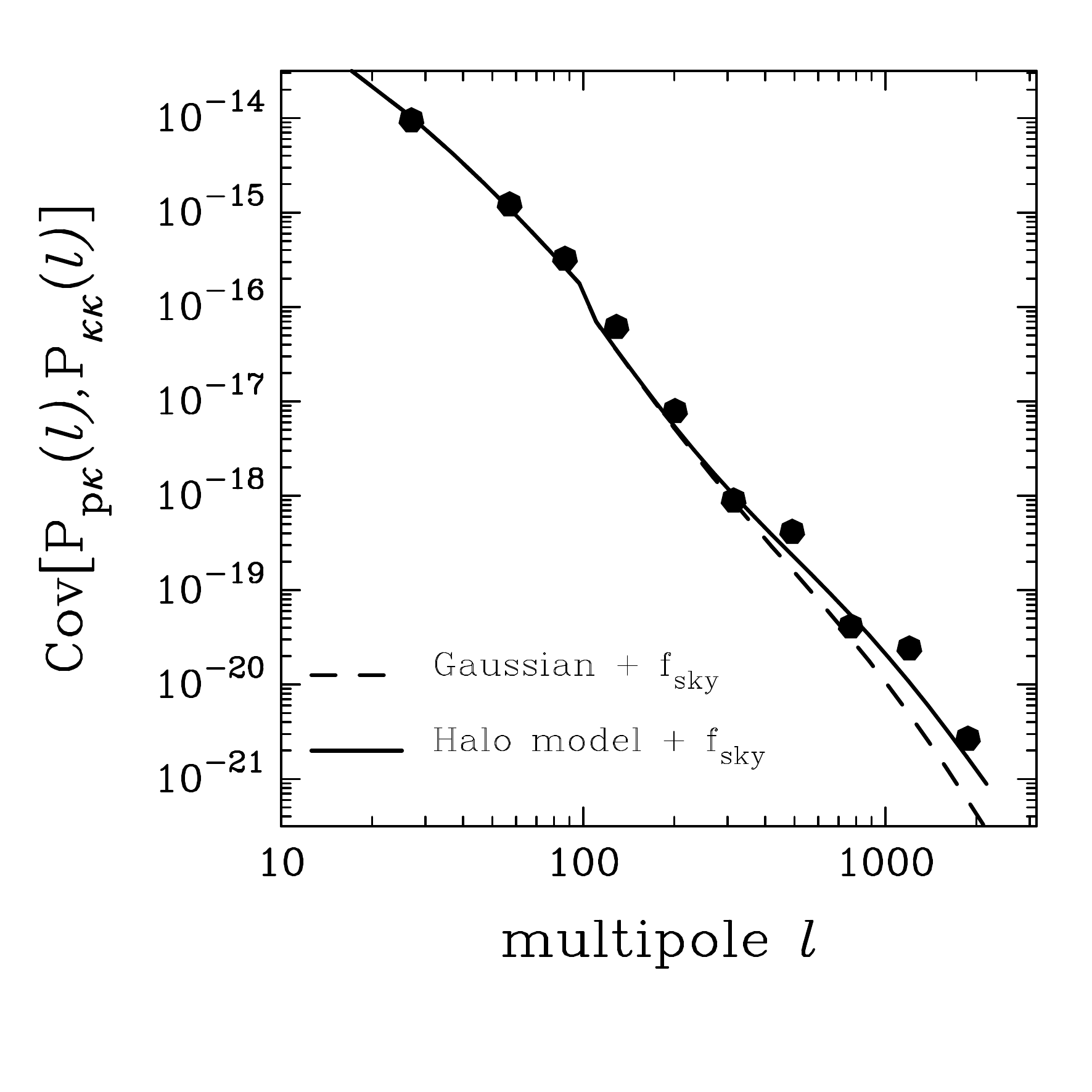}
\centering \includegraphics[clip, width=0.4\columnwidth, viewport = 20 20 530 530]
{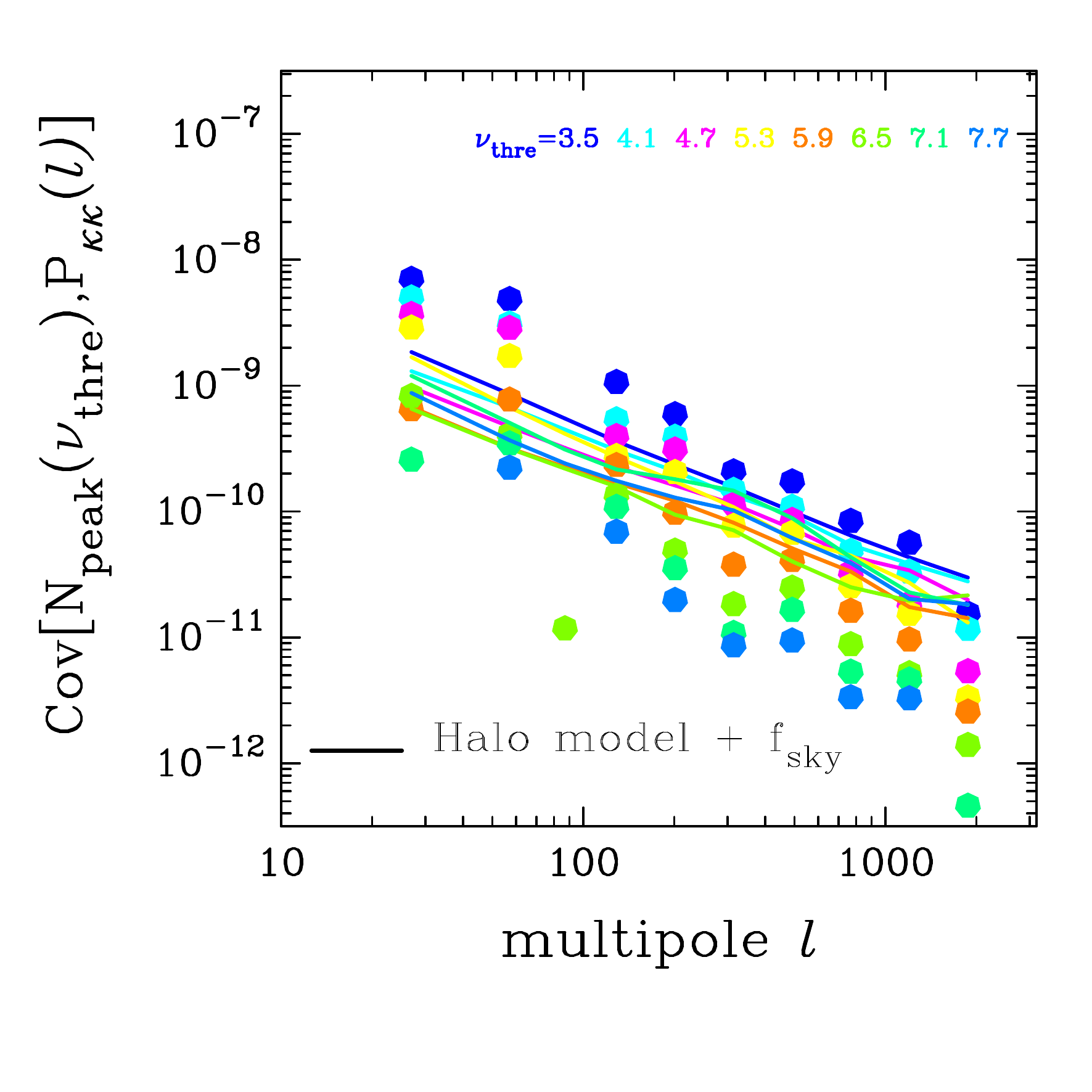}
\centering \includegraphics[clip, width=0.4\columnwidth, viewport = 20 20 530 530]
{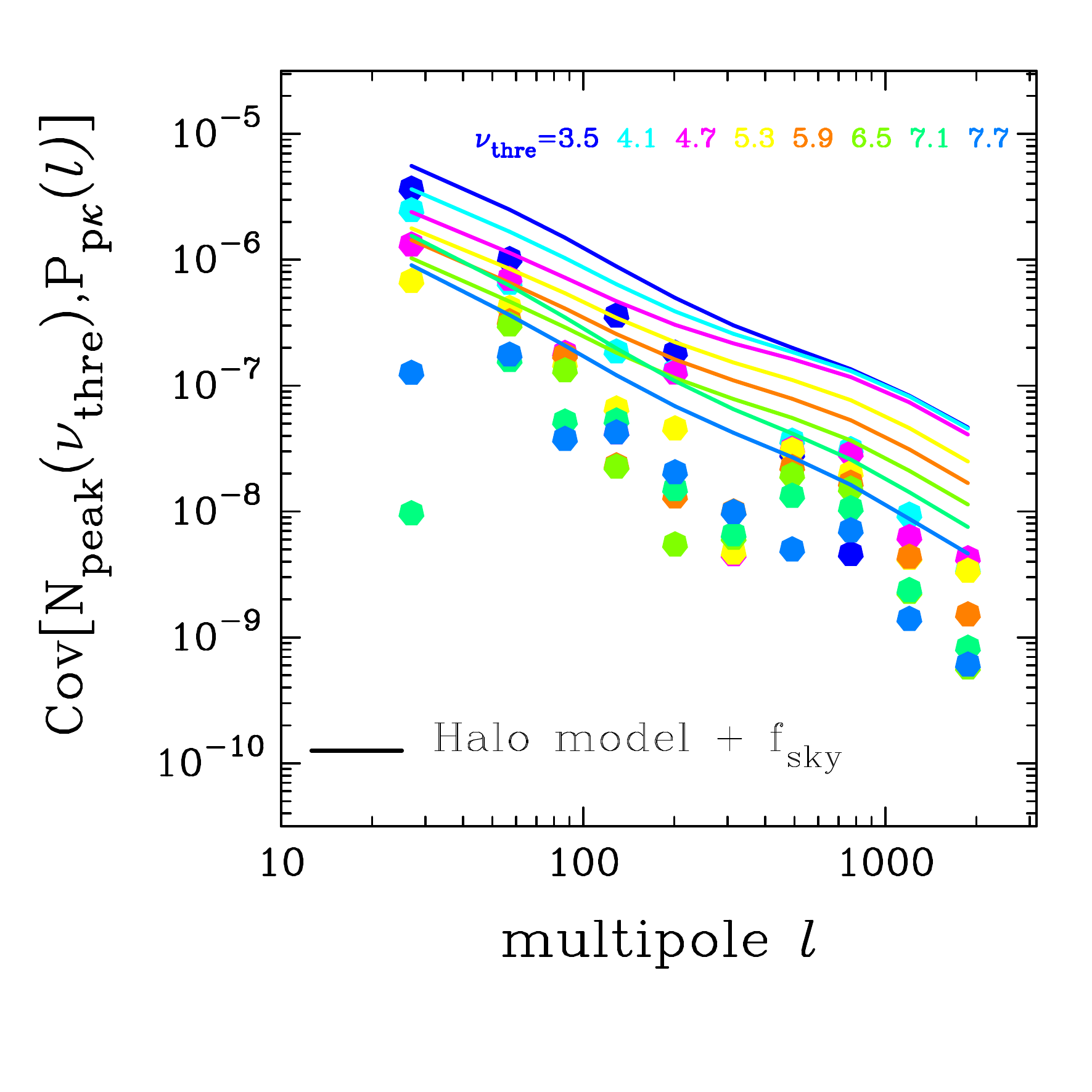}
\caption{
	The cross covariance of weak lensing statistics calculated from the
        200 simulations.
	In each panel, 
        the black or colored points show the measured covariance,
	and the solid line shows the corresponding halo model prediction.
	We define the peaks with $\nu_{\rm thre}$ 
	when measuring $P_{{\rm p} \kappa}$.
	In the bottom two panels, the color indicate the results
        for different
	threshold $\nu_{\rm thre}$ in $N_{\rm peak}$.
	The threshold value is varied in the range $\nu_{\rm thre}=3.5-7.7$.
 	\label{fig:cov_stat_HSC}
	}
\end{figure*}

To perform combined analysis 
of $P_{\kappa\kappa}, P_{{\rm p}\kappa}$ and $N_{\rm peak}$,
we need cross covariance between the statistics.
Unfortunately, it is difficult to model the impact 
of masked region on the cross covariances
in a similar manner to Eq.~(\ref{eq:cov_pseudo_pk}).
Here, we simply compare the measured cross covariances 
with the prediction of our halo model.
In the model, the covariance 
can be derived similarly 
to Eq.~(\ref{eq:flat_sky_cov_PkkPkk}).
For example,  
\beqa
{\rm Cov}[P_{{\rm p}\kappa}(\ell_{i}), P_{{\rm p} \kappa}(\ell_{j})]
&=&
\frac{\delta_{ij}}{2\ell_{i} \Delta \ell f_{\rm sky}}
\left(P_{{\rm p}{\rm p}}(\ell_{i})P_{\kappa\kappa}(\ell_{i})
+P_{{\rm p}\kappa}(\ell_{i})P_{{\rm p}\kappa}(\ell_{i})\right)
+\frac{1}{4\pi f_{\rm sky}} 
\int_{\ell_{i}} \, \frac{{\rm d}^2 \bd{\ell}}{A_{s,i}}
\int_{\ell_{j}} \, \frac{{\rm d}^2 \bd{\ell}^{\prime}}{A_{s,j}}
T_{{\rm p}\kappa {\rm p} \kappa}(\bd{\ell}, -\bd{\ell}, 
\bd{\ell}^{\prime}, -\bd{\ell}^{\prime}),
\label{eq:flat_sky_cov_PpkPpk}
\\
{\rm Cov}[P_{{\rm p}\kappa}(\ell_{i}), P_{\kappa \kappa}(\ell_{j})]
&=&
\frac{\delta_{ij}}{2\ell_{i} \Delta \ell f_{\rm sky}}
2P_{{\rm p}\kappa}(\ell_{i})P_{\kappa \kappa}(\ell_{i})
+\frac{1}{4\pi f_{\rm sky}} 
\int_{\ell_{i}} \, \frac{{\rm d}^2 \bd{\ell}}{A_{s,i}}
\int_{\ell_{j}} \, \frac{{\rm d}^2 \bd{\ell}^{\prime}}{A_{s,j}}
T_{{\rm p}\kappa \kappa \kappa}(\bd{\ell}, -\bd{\ell}, 
\bd{\ell}^{\prime}, -\bd{\ell}^{\prime}),
\label{eq:flat_sky_cov_PpkPkk}
\eeqa
where we have defined
$T_{{\rm p}\kappa {\rm p} \kappa}$,
$T_{{\rm p}\kappa \kappa \kappa}$, and 
the other covariances (i.e. ${\rm Cov}[N_{\rm peak}, N_{\rm peak}]$,
${\rm Cov}[N_{\rm peak}, P_{\kappa \kappa}]$, and 
${\rm Cov}[N_{\rm peak}, P_{{\rm p}\kappa}]$)
in Section~\ref{subsec:cov}.

Figure \ref{fig:cov_stat_HSC} shows 
the measured peak-power cross-covariance.
In constrast to the case of $P_{\kappa \kappa}$, 
the covariance including $P_{{\rm p} \kappa}$ and $N_{\rm peak}$
is less affected by masks:
the simple model that accounts for $f_{\rm sky}$ 
yields a reasonable result with respect to that of the simulations.
We find that the difference is by a factor of $\sim 3$
at most.

\section{CONCLUSION AND DISCUSSION}\label{sec:con}

We have performed all-sky lensing simulations 
to generate a large set of realistic weak lensing mass maps 
with complex masked regions by incorporating the actual 
position of bright stars.
We have used the set of the mock samples
to study the statistical properties of weak lensing convergence 
and convergence peaks in detail.
Full nonlinear covariances between the statistics have been also calculated 
from 200 realization of masked maps. 
We have also developed an analytic halo model that provides
reasonably accurate prediction for the statistics.

When adopting a Gaussian smoothing 
with the full width at half maximum of 5 arcmin,
we can associate weak lensing convergence peaks with dark matter halos 
with mass of $\sim10^{14}\, h^{-1}M_{\odot}$ at $z\sim 0.1-0.2$.
We can also estimate the modulation of peak height due to shape noise
by using a model based on Gaussian peak statistics.
Thus, the abundance of peaks, the angular correlation function,
and the cross-correlation of peaks 
and cosmic shear are all obtained accurately by our halo model 
approach with the shape noise correction.
Furthermore, the halo model can also take the mask effect into account
and indeed produces accurate ensemble average of statistics 
and their covariances.
The impact of masked regions on the covariance can be described by the 
following two effects:
(i) reduction of sky coverage and 
(ii) the mode-coupling effect between different Fourier modes.
We find that the former affects
the overall amplitude of cross-covariance between statistics, 
while the latter is important for the covariance of 
cosmic shear power spectrum $P_{\kappa\kappa}$.
For the masked regions adopted here, ignoring the mode-coupling effect
would induce underestimation of the covariance of $P_{\kappa\kappa}$ 
by a factor of $\sim10$(!).

The number density of source galaxies is an important factor 
in the statistical analysis 
of weak gravitational lensing.
As one may expect, 
a large number density of sources is desired to find clusters with high accuracy
and perform cosmological analysis with selected clusters.
In comparison with the case of $n_{\rm gal} = 10\, {\rm arcmin}^{-2}$,
we have confirmed that the signal-to-noise ratio increase by a factor of $\sim1.5$
in combined analysis 
with $P_{\kappa\kappa}$, $P_{{\rm p}\kappa}$ and $N_{\rm peak}$
even if we ignore the shape noise contaminant (i.e., $n_{\rm gal} \rightarrow \infty$).
This suggests that imaging over a wide area is suitable 
for cosmological analysis with
the lensing statistics even if the number density of 
sources is not significantly increased in such surveys.

We have also examined the validity of our model for two additional cases:
$n_{\rm gal} = 5$ and 30 ${\rm arcmin}^{-2}$.
When the smoothing scale, 
the rms of intrinsic ellipticity, and the source redshift
are all fixed, 
the halo model prediction is in good agreement with the
result of our full-sky simulations 
in the case of $n_{\rm gal}=30\, {\rm arcmin}^{-2}$, 
but the agreement is worse with $n_{\rm gal}=5\, {\rm arcmin}^{-2}$
\footnote{
We expect that the disagreement for small $n_{\rm gal}$ would be 
caused by the offset between the position of a peak and the center of the corresponding halo.
The offset effect would be more important when shape noise increases 
as shown in \citet{2010ApJ...719.1408F}.
}.
Therefore, when we consider the typical value of 
the rms of intrinsic ellipticity and the source redshift, 
our model is expected to be accurate when
$n_{\rm gal} \simgt10\, {\rm arcmin}^{-2}$ with $\sim2$ arcmin 
Gaussian smoothing.

Our model has been successfully applied 
the statistics of a smoothed convergence map
with shape noise.
In principle, one can find an optimal filter function
so that the number of detected clusters is increased
\citep[e.g.][]{2005ApJ...624...59H, 2005A&A...442..851M}.
Our model is based on the assumption that
shape noise in a smoothed lensing map has
Gaussian properties.
The assumption is valid when 
the ellipticities of the source galaxies are uncorrelated and 
when there are a sufficient number of source galaxies 
per pixel (i.e. the central limit theorem).
We expect that, while our model can be 
applied to the general form of filter function, 
it may not be appropriate when there are only 
few source galaxies and/or
when there is no one-to-one correspondence between peaks 
and halos.

Clusters of galaxies are
important targets in future
cosmological surveys.
Selecting galaxy clusters based on cosmic shear measurement 
does not rely on mass estimate nor on calibration with 
additional information such as $X$-ray brightness.
Nevertheless, there remain some systematic effects 
worth mentioning here.

Gravitational lensing causes not only distortion of source images 
but also magnification.
Using magnified galaxies 
in a flux- and size-limited survey would potentially cause 
systematic effect(s) in the weak lensing statistics.
The magnification effects on lensing 
peak statistics have been already 
studied in e.g., \citet{2011ApJ...735..119S}. 
Recent numerical study by \citet{2014PhRvD..89b3515L}
suggests that the magnification effect causes non-negligible 
bias in parameter estimation in the case of LSST.
In order to examine the magnification effect further,
it is essential to run high angular resolution simulations.
This is because the mean number of source galaxies 
on each pixel should be less than
unity to make the one-to-one correspondence between 
a pixel and a (magnified) source galaxy.
In the case of $n_{\rm gal} =10\, {\rm arcmin}^{-2}$,
we should set the pixel size to be 
$1/\sqrt{n_{\rm gal}} \sim 0.3\, {\rm arcmin}$.
We will perform such simulations 
to study the magnification effect in wide-field surveys
in detail.

Another important issues are uncertainties and systematic bias
associated with baryonic effects.
Previous studies
\citep[e.g.,][]{Semboloni2011, 2013MNRAS.434..148S, 2013PhRvD..87d3509Z}
explored the impact of the baryonic component
to two-point statistics of cosmic shear  
and consequently to cosmological parameter estimation.
The baryonic effect is likely important in weak lensing peak 
statistics.
Indeed, \citet{Yang2013} show appreciable baryonic 
effects on peak statistics using a simple model applied 
to dark-matter-only simulations, 
whereas the baryonic effect on higher order convergence statistics 
have been studied with numerical simulations \citep{2015arXiv150102055O}.
Recently, \citet{Mohammed2014} explored halo model approach 
to include the baryonic effect on cosmic shear statistics.

The statistical properties and the intrinsic correlation 
of source galaxies and lensing structures
are still uncertain but could be critical when making a large
lensing mass maps.
Among such correlations, 
source-lens clustering \citep[e.g.,][]{2002MNRAS.330..365H}
and the intrinsic alignment \citep[e.g.,][]{2004PhRvD..70f3526H}
are likely to compromise cosmological parameter estimation.
A promising approach in theoretical studies 
would be associating the source positions
with their host dark matter halos on the light cone. 
This is along the line of our ongoing study using
a large set of cosmological simulations.

Weak gravitational lensing is a promising tool
to probe the dark matter distribution in the universe.
Statistical analysis of a reconstructed mass map can be 
performed to extract precise cosmological information.
The peak statistics considered in the present paper contain 
the information related to massive objects such as clusters of galaxies and
thus have a great potential to probe cosmology and 
constrain the model of 
structure formation simultaneously.
Ongoing/upcoming imaging surveys such as HSC, DES,
and LSST in the near future, will provide the 
largest dark matter map we have never seen before.
We expect our study presented here provides a useful
guide to interprete properly
the reconstructed mass map and to reveal the nature of 
the dark components in the universe.



\section*{acknowledgments}
We would like to thank M. R. Becker for making the source program of 
{\tt CALCLENS} available, and HEALPix team for making {\tt HEALPix}
software publicly available.
This work is supported in part by Grant-in-Aid for
Scientific Research from the JSPS Promotion of Science
(25287050; 26400285). NY acknowledges financial support from JST CREST.
MS is supported by Grant-in-Aid for JSPS Fellows.
Numerical computations presented in this paper were in part carried out
on the general-purpose PC farm at Center for Computational Astrophysics,
CfCA, of National Astronomical Observatory of Japan.

\appendix

\section{HALO-PEAK MATCHING}
\label{appendixH}
\begin{figure*}
\includegraphics[clip, width=0.7\columnwidth, viewport = 20 20 550 550]{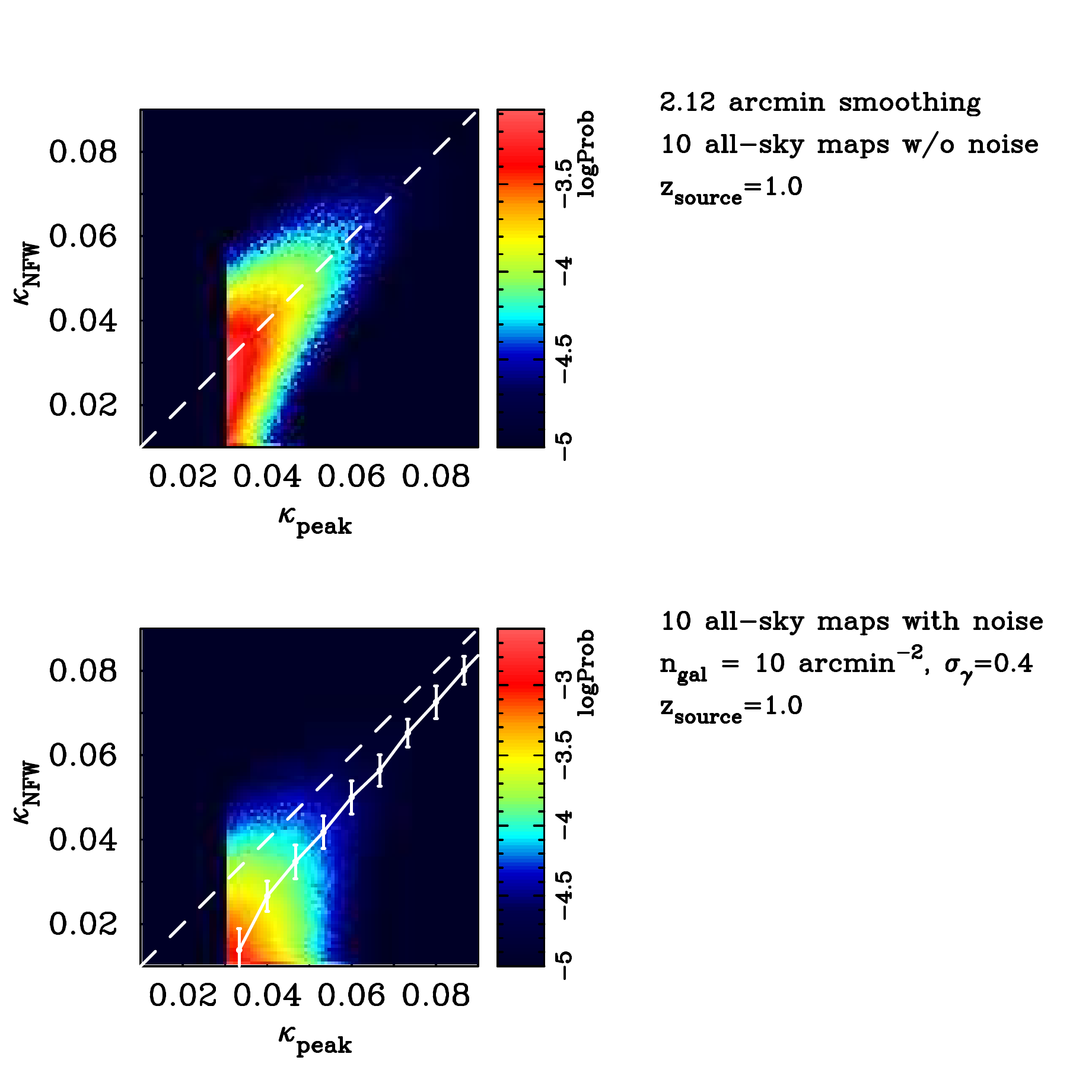}
\caption{
	The correspondence between dark matter halos and lensing peaks.
	The top panel shows the scatter plot of peak height and the expected convergence by
	the matched halos in absence of noise.
	The lower panel corresponds to the case with shape noise.
	In each panel, the horizontal axis represents the peak height and 
	the vertical axis shows the expected convergence of NFW halos.
	\label{fig:halo_peak_comp}
	}
\end{figure*}

In this appendix, we examine the correspondence between dark matter halos 
and the local maximum in lensing convergence map.

In order to generate mock halo catalogs, 
we identify dark matter halos 
in outputs of our $N$-body simulation (see, Section~\ref{subsec:nbody})
using the standard friends-of-friends algorithm
with a linking parameter of $b=0.2$ in units of the mean particle
separation. 
We use dark matter halos with mass greater than $10^{13} \, h^{-1}M_{\odot}$ 
in the following analysis.
This lowest mass corresponds to the mass of $20$ particles 
in the largest simulation. 
Then, the position of dark matter halo in $N$-body simulations 
are arranged in the same way as in the ray-tracing experiments 
in Section~\ref{subsec:RT}.

With our ray-tracing simulations and mock halo catalogs, 
we study the correspondence between halos and the peaks in weak lensing convergence maps.
We first identify the local maxima in the smoothed lensing convergence field 
with source redshift of $z_{\rm source}=1$.
In this appendix, we again adopt the Gaussian smoothing 
with the full width at half maximum of 5 arcmin.
When including the shape noise in convergence maps, 
we set $\sigma_{\gamma}=0.4$ and $n_{\rm gal} = 10 \, {\rm arcmin}^{-2}$.
For selection of peaks, the threshold of peak height is set to be ${\cal K}=0.03$.
This value corresponds to $\sim3\sigma$ in smoothed convergence maps without noise.
For a given position of lensing peak, we search for the matched dark matter halos 
within a radius of 5 arcmin from the peak position.
This search radius is set to be larger than the smoothing scale 
but still smaller than the angular size of massive halos at $z\sim 0.1-0.2$
\citep[also see,][]{Hamana2004}.
When we find several halos in search radius, we regard the matched halo as
the closest halo from the position of peak.
For each matched peak, we estimate the corresponding convergence by using
the universal NFW density profile (see Section~\ref{subsec:WL_select_cluster} in detail).
In the calculation of expected convergence from FoF halos, we simply assume 
that the FoF mass is equal to the virial mass.
In total, we find 632,238 and 1,404,538 pairs of peaks and halos over 10 
noise-less maps and noisy maps, respectively.

Figure \ref{fig:halo_peak_comp} 
shows the scatter plot of peak height in ${\cal K}$ map
and the expected convergence by NFW halos.
The horizontal axis corresponds to peak height, 
while the vertical axis shows the corresponding convergence expected by NFW halos.
Thus, the color map in each panel 
shows the probability of Eq.~(\ref{eq:prob_peak_obs_h}).
We present the line of $y=x$ as the dashed line in each panel.
In lower panel of this figure, we show the effect of the modulation of peak height 
as the solid line with error bars.
The solid line is derived by ${\bar {\cal K}}_{\rm peak, obs}(z, M)$ 
in Eq.~(\ref{eq:mean_peak_height}) and 
the error bars reflect the scatter of ${\bar {\cal K}}_{\rm peak, obs}(z, M)$.
As shown in previous works, we confirm the good correspondence between the matched dark matter halos and lensing peaks in the noise-less maps.
Also, our model as shown in Eq.~(\ref{eq:mean_peak_height}) 
can explain the average relation between peaks and dark matter halos even in the case with noise.

\section{THE CASE OF COMPENSATED GAUSSIAN FILTER}
\label{appendix:compensated}

\begin{figure*}
\centering \includegraphics[clip, width=0.5\columnwidth, viewport = 20 280 320 530]{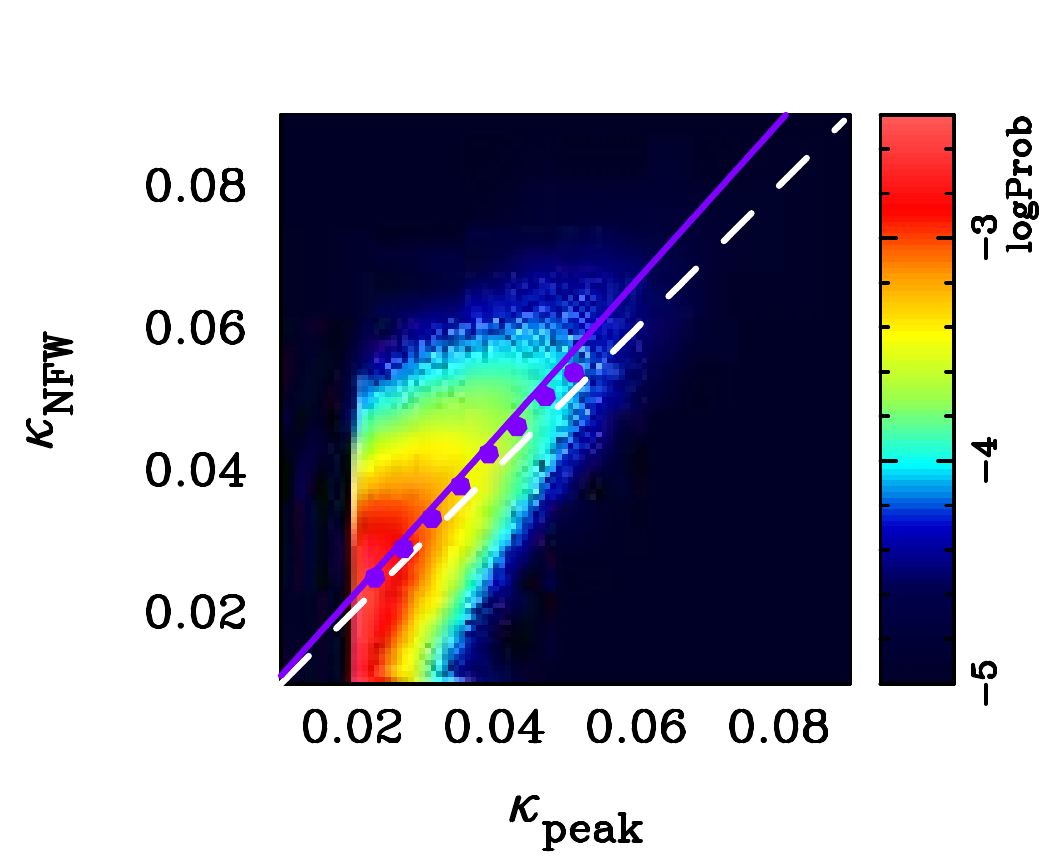}
\centering \includegraphics[clip, width=0.4\columnwidth, viewport = 20 20 530 530]
{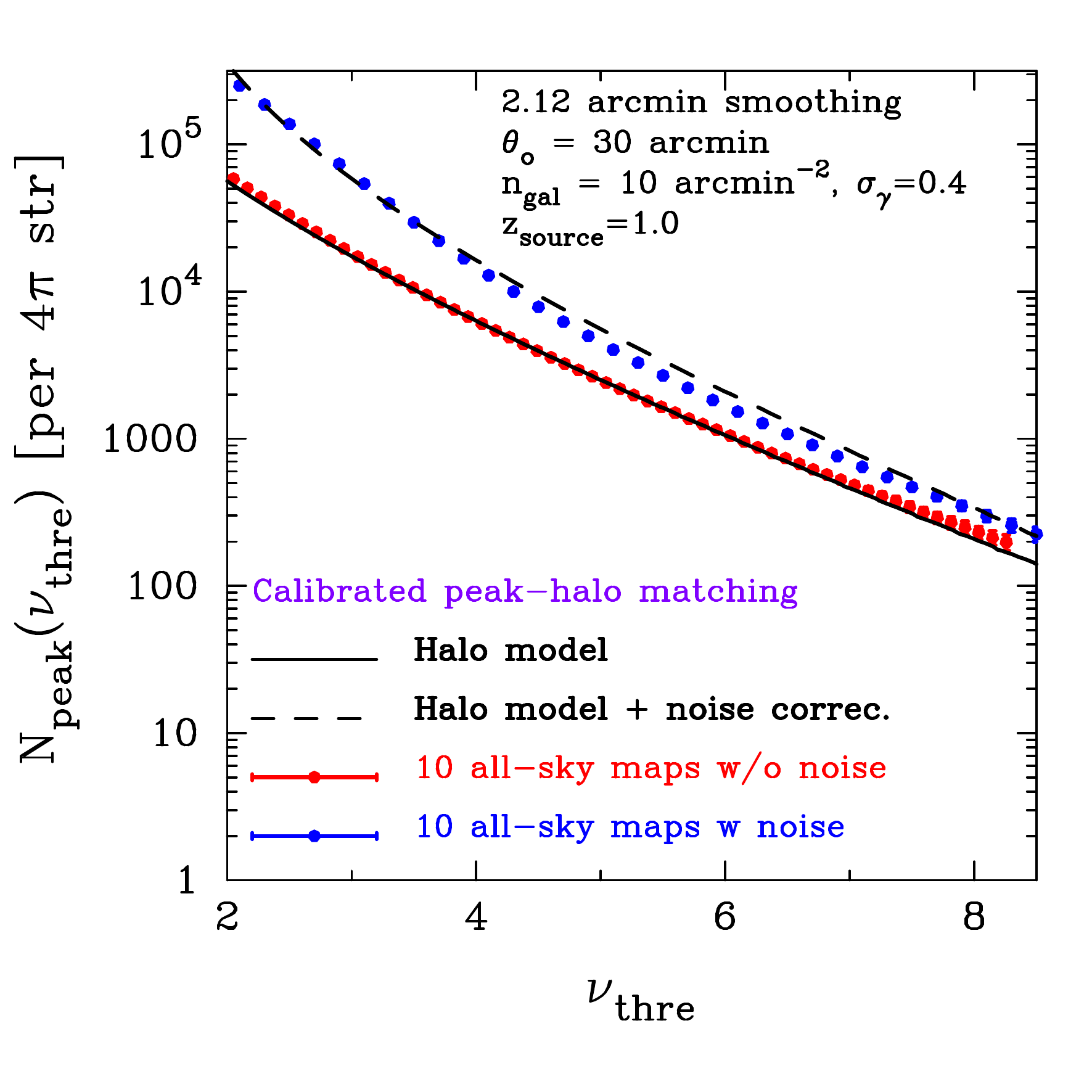}
\caption{
	The property of lensing peaks on smoothed maps by the compensated Gaussian filter.
	The left panel shows the scatter plot of peak height and the expected convergence by
	the matched halos in absence of noise.
	In the left panel, the dashed line shows one-to-one correspondence.
	The purple points represent the mean relation of the expected convergence 
	and measured peak height and the best-fit linear relation is expressed as the purple line.
	The right panel shows the comparison the peak abundance measured 
	from ten full-sky simulations and our model prediction.
	In the right panel, the black line is our halo model prediction.
         The red points with error bar represent the measured signal 
	from the 'clean' convergence maps without noise.
	The blue points show the result with noise.
	We normalize the measured peak height 
	by $\sigma_{\rm noise, 0}$ in both cases with and without noise.
	The error bars indicate the standard deviation over ten realization.
	In the calculation of peak count, we correct the biased relation of 
	$\kappa_{\rm NFW}$ and $\kappa_{\rm peak}$
	as shown in the purple line in the left panel.
	The details of correction are found in the text.
	\label{fig:5arcmin30cut}
	}
\end{figure*}

We examine another practical filter function to construct smoothed 
lensing maps.
In this appendix, we consider $z_{\rm source}=1$,
and we set $\sigma_{\gamma}=0.4$ and $n_{\rm gal} = 10 \, {\rm arcmin}^{-2}$
when including the shape noise in convergence maps. 

We first consider the compensated filter $U_{\rm c}$ based 
on the Gaussian form as
\beqa
U_{\rm c}(\theta) = \frac{1}{\pi \theta_{G}^2}
\exp\left(-\frac{\theta^2}{\theta_{G}^2}\right)
-\frac{1}{\pi \theta_{o}^2}
\left[1-\exp\left(-\frac{\theta_{o}^2}{\theta_{G}^2}\right)\right],
\label{eq:compensated_Gauss}
\eeqa
where $\theta_{o}$ represents the boundary of the filter
and we set $U_{\rm c}$ to be zero for $\theta > \theta_{o}$.
We adopt the smoothing scale of $\theta_{G} = 5/\sqrt{8\ln 2}$ arcmin 
and $\theta_{o}=30$ arcmin.
For the compensated filter function of $U_{\rm c}$,
the noise power spectrum on a smoothed lensing map is expressed as
\citep{VanWaerbeke2000}
\beqa
P_{\cal N}(\ell) = 
\frac{\sigma_{\gamma}^2}{2n_{\rm gal}}|{\tilde U}_{\rm c}(\ell)|^2,
\eeqa
where $\sigma_{\gamma}$ is the rms of the intrinsic ellipticity of sources,
$n_{\rm gal}$ represents the number density of source galaxies, 
and ${\tilde U}_{\rm c}$ is the Fourier transform of $U_{\rm c}$.
As shown in Eq.~(\ref{eq:noise_moment}), 
the noise variance $\sigma_{\rm noise, 0}$ is evaluated by
the integral of $P_{\cal N}(\ell)$ in Fourier space.
In the case of $\theta_{o}=30$ arcmin,
the boundary of the filter changes $\sigma_{\rm noise, 0}$ by about 1\%
compared to the case of the usual Gaussian filter.
Thus, we can safely ignore the difference of $\sigma_{\rm noise, 0}$ 
between the compensated and the usual Gaussian filter for our parameter choice.

In order to investigate the effect of the modification 
of filter function on lensing peak statistics, 
we first study the correspondence between peaks and halos 
by the halo-peak matching analysis
as shown in Appendix~\ref{appendixH}.
When we limit lensing peaks with the height larger than $0.02$, 
we find 1,045,291 matched pairs over ten full-sky maps without shape noise.
The left panel in figure \ref{fig:5arcmin30cut} shows the scatter plot 
of the measured peak height and the expected convergence signal for the spherical NFW halo.
In the calculation of expected signal, 
we simply assume that FoF mass of halos is equal to the virial mass
and use the model of concentration parameter $c_{\rm vir}$ in \citet{Duffy2010}.
We expect that the one-to-one correspondence between peaks 
and halos would still hold. However,
we find biased relation between the mean measured height and expected one
\textit{even in the absence of noise}.
The biased mean relation is shown by the purple points in the left panel 
in figure \ref{fig:5arcmin30cut}. It can also be 
fitted well by linear relation 
of ${\cal K}_{\rm peak} = \alpha {\cal K}_{\rm NFW}+\beta$.
We find the best-fit value of $\alpha$ is 0.9 
while the offset $\beta$ can be approximated to be zero.
A similar relation is also found by \citet{Hamana2012}.
It might be caused as a consequence of various effects such 
as the mismatch of FoF mass and virial mass
defined by spherical over-density.

Even without knowing the origin of the biased relation, 
we can still predict the lensing peak statistics with the compensated filter
by adopting the biased relation in our halo model.
Our approach is simply to replace ${\cal K}_{{\rm peak},h}(z, M)$ 
with $\alpha {\cal K}_{{\rm peak},h}(z, M)+\beta$ 
for the calculation of Eq.~(\ref{eq:prob_peak_obs_h}).
Here, we also assume that ${\cal K}_{{\rm peak},h}$ can be evaluated by 
\beqa
{\cal K}_{{\rm peak}, h}(z, M) = \int {\rm d}^2\theta\, 
U_{\rm c}(\theta;\theta_{G},\theta_{o}) \kappa_{h}(\theta|z,M),
\eeqa
where $\kappa_{h}(\theta|z,M)$ represents the convergence profile of spherical NFW halos
with mass of $M$ and the redshift of $z$.
With the above correction, our model can provide a reasonable fit 
to the measured peak statistics from ten full-sky maps 
as shown in the right panel in figure \ref{fig:5arcmin30cut}.
The right panel shows the measured peak count for the compensated filter 
with and without shape noise. 
In the right panel, the black solid (dashed) line represent our halo model 
in absence (presence) of noise
with the correction for the biased relation of ${\cal K}_{\rm peak}$ and ${\cal K}_{\rm NFW}$.
With the above suitable modifications, 
our model works as long as the one-to-one correspondence 
of peaks and halos holds.
We simply need to calibrate the mean scaling relation of 
${\cal K}_{\rm peak,obs}$ and ${\cal K}_{{\rm peak},h}$ in absence of noise.


\section{FULL SKY RAY-TRACING SIMULATION}
\label{appendix:raytracing}

Here we first summarize basic equations of the multiple-plane
gravitational lensing algorithm, and then describe the 
ray-tracing method through the multiple-plane.
For the former we largely follow \citet{2008ApJ...682....1D}, and for the latter
we adopt one developed by \citet{2009A&A...497..335T}.

Throughout this section, we work on the comoving coordinates.
Thus $\rho$, $\chi$, and $r(\chi)$ denote the comoving matter density,
the radial comoving distance, 
and the comoving angular diameter distance, respectively.

\subsection{Construction of the lensing potentials of multiple-plane}
\label{appendix:raytracing:phi}
The 3-dimensional light-cone matter distribution is composed of the
multiple-layer of shells with a fixed width of $150\, h^{-1}$Mpc taken from
the nested simulation boxes.
The surface matter density field on a sphere for $j$-th shell is defined
by 
\begin{eqnarray}
\label{eq:Delta_Sigma}
\Delta_{\Sigma}^j(\bd{\theta}) = \int_{\rm shell}
{\rm d}\chi~(\rho(\bd{\theta},\chi)-\bar{\rho})r(\chi)^2, 
\end{eqnarray}
where $\bd{\theta}$ denote the angular directions and 
$\bar{\rho}$ corresponds the mean
matter density, and the integration is over the shell width.
We set the lens-planes for each shell at the cone-volume weighted mean
distance, $\chi_j = 0.75
(\chi_{j, \rm max}^4-\chi_{j, \rm min}^4)/(\chi_{j, \rm max}^3-\chi_{j, \rm min}^3)$, 
where $\chi_{j, \rm max}$ and $\chi_{j, \rm min}$ are the farthest and nearest radial
distance to a shell, respectively.
The convergence field for $j$-th shell is given by
\begin{eqnarray}
\label{eq:Kj}
K^j(\bd{\theta}) = {{4 \pi G} \over c^2}
{{\Delta_{\Sigma}^j(\bd{\theta})} \over {a_j r(\chi_j)}}
\end{eqnarray}
where $a_j$ is the scale factor at the lens-plane $\chi_j$.
We use the {\tt HEALPix} \citep{2005ApJ...622..759G} scheme for pixelization
of a sphere, and we make the best use of the {\tt HEALPix} library.
For each shell, we construct the projected mass density map from 
$N$-body particles using Nearest Grid Point method (implemented with
{\tt HEALPix} subroutine {\tt vec2pix\_ring}). Using the volume ($V_{\rm
  sim}$) and total number of $N$-body particles ($N_{\rm part}$) of the
$N$-body simulations, the total number of pixels ($npix$), the number of $N$-body
particles within $i$-th pixel ($n_{{\rm pix},i}$) and its mean value
($\bar{n}_{\rm pix}$), the convergence field is given by
\begin{eqnarray}
\label{eq:Kj-sim}
K^j(\bd{\theta}_i) = {{3 \Omega_m} \over {2 a_j r(\chi_j)}}
\left( {{H_0} \over c }\right)^2
{{V_{\rm sim}} \over {N_{\rm part}}} {{npix} \over {4 \pi}} 
(n_{{\rm pix},i} - \bar{n}_{\rm pix}).
\end{eqnarray}
Having the convergence field being ready, we expand it in spherical
harmonics to have its coefficients, $K_{lm}^j$, using the {\tt HEALPix}
subroutine {\tt map2alm}. 
Then, the spherical harmonics coefficients for the lensing
potential $\phi^j$ can be obtained via
\begin{eqnarray}
\label{eq:phij-Kj}
\phi_{lm}^j = {2\over {l(l+1)}} K_{lm}^j \mbox{~~ for $l\neq 0$,}
\end{eqnarray}
and $\phi_{lm}^j=0$ for $l= 0$. 
This gives us the lensing potential field on a sphere, and its 1st and
2nd derivatives relate to the gravitational lensing deflection field
($\alpha_i^j=-\nabla_{\hat{n}_i} \phi^j $) and the optical tidal matrix  
($U_{ik}^j=-\nabla_{\hat{n}_i} \nabla_{\hat{n}_j}\phi^j $), respectively.
Note that $\nabla_{\hat{n}_i}$ ($i=1,2$) denotes the angular derivative.
In an actual computation, we utilize the {\tt HEALPix} subroutine 
{\tt alm2map\_der}.

\subsection{Light ray propagation}
\label{appendix:raytracing:raytracing}

Let us first describe the method to trace the ray trajectory using the
multiple-plane algorithm, for which we basically follow one developed by  
\citet{2009A&A...497..335T}.
A virtual observer is located at the center of the nested simulation
boxes. Rays are traced backward from the observer point with the initial
ray directions being set on {\tt HEALPix} pixel centers. 
Thus the ray positions on the 1st (closest to the observer) is exactly
at the {\tt HEALPix} pixel centers.
At each lens-plane, the ray directions are deflected according to
$\alpha_i^j$, and the ray positions on the next lens-plane
are computed using the method described in Appendix A of
\citet{2009A&A...497..335T}. Note that ray positions on the $j$-th 
($j>1$) lens-plane are not exactly at the pixel center due to the lensing
deflections, 
however the lensing fields ($\alpha_i^j$, $U_{ik}^j$) are
only computed at the pixel centers.
In order to evaluate the lensing fields at an arbitrary position
($\bd{\theta}$), we adopt the inverse distance weighted interpolation from
nearest four pixel values;
\begin{eqnarray}
\label{eq:interpol}
\alpha_i^j(\bd{\theta})=
{{\sum_{k=1}^4 w_k \alpha_i^j(\bd{\theta}_k\rightarrow \bd{\theta})}
\over
{{\sum_{k=1}^4 w_k}}},
\end{eqnarray}
where $w_k = 1/|\bd{\theta}-\bd{\theta}_k|$, and
$\alpha_i^j(\bd{\theta}_k \rightarrow \bd{\theta})$ is the deflection angle at
the pixel center $\bd{\theta}_k$ but after parallel transporting to the
ray position $\bd{\theta}$ (in order to take into account the change in the
local ($\mbox{\boldmath $e$}_\theta , \mbox{\boldmath $e$}_\phi$) basis).
In the actual computation, the parallel transport of the vector and
tensor (for $U_{ik}^j$ described below) is implemented by the
rotation of them by an angle between two coordinate bases at
$\bd{\theta}_k$ and $\bd{\theta}$ \citep[see Appendix C
  of][]{2013MNRAS.435..115B}.

Having evaluated the ray positions at a lens-plane, we are able to
compute the lensing magnification matrix ($A_{ik}^j$) using the recurrence
relation \citep{Hilbert2009,2013MNRAS.435..115B};
\begin{eqnarray}
\label{eq:recurr}
A_{ik}^{j+1} &=& 
\left( 
1 - {{r(\chi_j)} \over {r(\chi_{j+1})}}
{{r(\chi_{j+1}-\chi_{j-1})} \over {r(\chi_j - \chi_{j-1})}}
\right) A_{ik}^{j-1}
+
{{r(\chi_j)} \over {r(\chi_{j+1})}}
{{r(\chi_{j+1}-\chi_{j-1})} \over {r(\chi_j - \chi_{j-1})}}
A_{ik}^{j}
-
{{r(\chi_{j+1}-\chi_{j})} \over {r(\chi_{j-1})}}
U_{im}^j A_{mk}^{j},\nonumber \\
A_{ik}^{1} &=& \delta_{ik}, \\
A_{ik}^{0} &=& \delta_{ik},\nonumber
\end{eqnarray}
for $j\ge 1$, and note that in our notation, the lens-plane closest to
the observer is $j=1$. 
The optical tidal matrix, $U_{ik}^j$, in the above relation is evaluated
at the ray position on each lens-plane in the same interpolation scheme
as eq. (\ref{eq:interpol}), and then again parallel transporting to the
unperturbed ray position (i.e., the initial ray direction), because the
observed magnification matrix should be evaluated in the local basis of
the image position.

We choose the source-plane at an arbitrary redshift $z_s$ (and thus the
correspondence radial distance to the source-plane $\chi_s$), and
evaluate
the source position on the source-plane and the magnification matrix by
using the above methods but replacing, e.g., $\chi_{j+1} \rightarrow
\chi_s$. 

\subsection{Image positions of haloes}
\label{appendix:raytracing:halo}

The 3-dimensional light-cone distribution of dark matter haloes is generated in the
same manner as for the matter distribution.
The spatial position of a halo is converted into the angular position
$\bd{\theta}_S^{\rm halo}$, where the subscript
``${}_S$'' means the {\it source position}.
We search for the corresponding {\it image position} $\bd{\theta}_I^{\rm
  halo}$ in the following manner.
First, we search for the nearest ray to the halo source position on the
lens-plane of the shell where the halo is located. 
The displacement vector between the angular positions of the halo and
the nearest ray is computed, $\bd{\Delta \theta_S} =
\bd{\theta}_S^{\rm halo} - \bd{\theta}_S^{\rm ray}$.
This vector is parallel transported to the image position of the nearest
ray $\bd{\theta}_I^{\rm ray}$, and we denotes it by $\bd{\Delta \theta_I}$
Then the image position of the halo is given by
$\bd{\theta}_I^{\rm halo} = \bd{\theta}_I^{\rm ray} + \bd{\Delta
  \theta_I}$. The last step is valid if the difference in the
lensing deflection angles between ray-trajectory to the halo and the
nearest ray is very small.
The statistical properties of differences in the lensing deflection
angles between nearby two rays (the, so-called, the lensing excursion
angle) were studies in \citet{Hamana2001,2005MNRAS.356..829H}.
They found that the root-mean-square (rms) value of the lensing excursion
angles of rays for $z_s=1$ with the separation of 1 arcmin is
$\sim 1$ arcsec. 
This value can be considered as the typical error in
$\bd{\theta}_I^{\rm halo}$.
Considering the fact that the pixel scale of the current ray-tracing
simulation is $\sim 1$ arcmin, we may conclude that the above
approximation is reasonably valid.
However it should be noticed that for rays gone through a strong
lensing region, the excursion angle can be much larger than the rms
value, and thus $\bd{\theta}_I^{\rm halo}$ may not be very accurate.
There is room for improvement on this issue that we leave for future
work.

\section{STATISTICAL PROPERTY OF PSEUDO-SPECTRUM ESTIMATORS}
\label{appendixB}
In this appendix, we summarize the statistical property of pseudo-spectrum estimators.
The pseudo-spectrum method is a powerful framework to construct the power spectrum 
of an underlying random field on limited sky 
\citep[e.g.,][]{2003MNRAS.343..559H, 2004MNRAS.349..603E, 2005MNRAS.360.1262B}.

Let us consider the two random fields in each direction in the sky:
convergence field $\kappa(\Omega)$ and number density field of lensing peaks $p(\Omega)$.
These two fields would commonly be expanded in spherical harmonic as follows:
\beqa
\kappa(\Omega) = \sum_{\ell m}\kappa_{\ell m}{\cal Y}_{\ell m}(\Omega),
\eeqa
where 
${\cal Y}_{\ell m}(\Omega)$ represents the spherical harmonics and we can define $p_{\ell m}$
for the random field $p(\Omega)$ similarly.
The inverse transform is then given by
\beqa
\kappa_{\ell m} = \int {\rm d}\Omega \, \kappa(\Omega) {\cal Y}_{\ell m}(\Omega),
\eeqa
and the similar relation can be adopted for $p_{\ell m}$.

The effect of finite sky coverage for each field is characterized as
\beqa
\tilde{\kappa}(\Omega) &=& W^{\kappa}(\Omega)\kappa(\Omega), \\
\tilde{p}(\Omega) &=& W^{p}(\Omega)p(\Omega),
\eeqa
where $W^{\kappa}$ and $W^{p}$ are the window function of sky masking 
for $\kappa$ and $p$, respectively\footnote{
In practice, $W^{p}$ are not equal to $W^{\kappa}$.
This is because peaks of convergence field are defined by that of \textit{smoothed} convergence map.
When the area with mask $W^{\kappa}$ 
is smoothed, there would exist ill-defined pixels due to the convolution between 
$W^{\kappa}$ and a filter function for smoothing.
Therefore, we need to remove the ill-defined pixels to find peaks.
This procedure makes the effective sky coverage of $W^{p}$ 
smaller than that of $W^{\kappa}$.
}.
Thus, the harmonic modes in presence of masked region is expressed as
\beqa
\tilde{X}_{\ell m} = 
\sum_{\ell^{\prime} m^{\prime}} 
W^{X}_{\ell m \ell^{\prime}m^{\prime}} X_{\ell^{\prime} m^{\prime}},
\eeqa
where $X=\kappa$, $p$ and 
$W^{X}_{\ell m \ell^{\prime}m^{\prime}}$ is defined by
\beqa
W^{X}_{\ell m \ell^{\prime}m^{\prime}}
= \int {\rm d}\Omega \, {\cal Y}_{\ell^{\prime}m^{\prime}}(\Omega)
W^{X}(\Omega){\cal Y}^{*}_{\ell m}(\Omega).\label{eq:conv_harmonic}
\eeqa
The estimators of power spectra on limited sky is defined by
\beqa
\tilde{\bd{P}}(\ell) = \frac{1}{2\ell + 1}\sum_{m}
\langle \tilde{\bd{X}}_{\ell m}\tilde{\bd{X}}_{\ell m}^{\dagger}\rangle,
\eeqa
where $\tilde{\bd{X}}_{\ell m} = (\tilde{\kappa}_{\ell m}, \tilde{p}_{\ell m})$
and $\langle \tilde{\bd{X}}_{\ell m}\tilde{\bd{X}}_{\ell m}^{\dagger} \rangle$ represents
the following set of power spectra:
\beqa
\langle \tilde{\bd{X}}_{\ell m}\tilde{\bd{X}}_{\ell m}^{\dagger} \rangle 
= \left(
\begin{array}{cc}
\langle \tilde{\kappa}_{\ell m} \tilde{\kappa}^{*}_{\ell m} \rangle &  \langle \tilde{\kappa}_{\ell m} \tilde{p}^{*}_{\ell m} \rangle \\
\langle \tilde{p}_{\ell m} \tilde{\kappa}^{*}_{\ell m} \rangle & \langle \tilde{p}_{\ell m} \tilde{p}^{*}_{\ell m} \rangle \\
\end{array}
\right).
\eeqa
For the underlying field $\bd{X}=(\kappa, p)$, we can define the power spectra $\bd{P}(\ell)$ as
\beqa
\bd{P}(\ell) = \langle \bd{X}_{\ell m}\bd{X}_{\ell^{\prime} m^{\prime}}^{\dagger} \rangle \delta_{\ell \ell^{\prime}}\delta_{m m^{\prime}}. \label{eq:def_power}
\eeqa
Using Eqs.~(\ref{eq:conv_harmonic}) and (\ref{eq:def_power}), 
we can find the relation between 
$\tilde{\bd{P}}(\ell)$ and ${\bd{P}}(\ell)$ as follows:
\beqa
\tilde{\bd{P}}(\ell) &=& \frac{1}{2\ell + 1}
\sum_{m}\sum_{\ell^{\prime} m^{\prime}}
\bd{W}_{\ell m \ell^{\prime}m^{\prime}} \bd{P}(\ell^{\prime})
(\bd{W}_{\ell m \ell^{\prime}m^{\prime}})^{\dagger}, \\
&=&\sum_{\ell^{\prime}} \bd{M}_{\ell \ell^{\prime}} \bd{P}(\ell^{\prime}),
\label{eq:relation_tildaP2trueP}
\eeqa
where $\bd{W}_{\ell m \ell^{\prime}m^{\prime}}$ is defined by
\beqa
\bd{W}_{\ell m \ell^{\prime}m^{\prime}}
= \left(
\begin{array}{cc}
W^{\kappa}_{\ell m \ell^{\prime}m^{\prime}} & 0  \\
0 & W^{p}_{\ell m \ell^{\prime}m^{\prime}} \\
\end{array}
\right).
\eeqa
The matrix $\bd{M}_{\ell \ell^{\prime}}$ represents the mode coupling effect due to 
masked region on power spectra, which is given by 
(in terms of $\bd{P}(\ell) = (P_{\kappa\kappa}(\ell), P_{{\rm p}\kappa}(\ell), P_{{\rm pp}}(\ell))^{T}$),
\beqa
\bd{M}_{\ell \ell^{\prime}}
= \left(
\begin{array}{ccc}
M^{\kappa \kappa}_{\ell \ell^{\prime}} & 0  & 0 \\
0 & M^{p \kappa}_{\ell \ell^{\prime}} & 0 \\
0 & 0  & M^{p p}_{\ell \ell^{\prime}} \\
\end{array}
\right),
\eeqa
where 
\beqa
M^{XY}_{\ell \ell^{\prime}} 
&=& \sum_{m m^{\prime}} 
W^{X}_{\ell m \ell^{\prime}m^{\prime}}
\left[W^{Y}_{\ell m \ell^{\prime}m^{\prime}}\right]^{*}, \\
&=& \frac{2\ell^{\prime}+1}{4\pi}\sum_{L}{\cal W}^{XY}_{L}
\left(
\begin{array}{ccc}
\ell & \ell^{\prime} & L \\
0 & 0  & 0 \\
\end{array}
\right)^2,
\eeqa
and ${\cal W}^{XY}_{\ell}$ is given by
\beqa
{\cal W}^{XY}_{\ell} 
&=& 
\sum_{m} w^{X}_{\ell m}\left(w^{Y}_{\ell m}\right)^{*}, \\
w^{X}_{\ell m} 
&=& 
\int {\rm d}\Omega\, W^{X}(\Omega){\cal Y}_{\ell m}(\Omega).
\eeqa
Hence, we can construct the estimator of $\bd{P}(\ell)$ from $\tilde{\bd{P}}(\ell)$ as
\beqa
\hat{\bd{P}}(\ell) = \sum_{\ell^{\prime}} \bd{M}^{-1}_{\ell \ell^{\prime}} 
\tilde{\bd{P}}(\ell^{\prime}),
\eeqa
where $\hat{\bd{P}}(\ell)$ is so-called psuedo-spectrum estimators.

Next, we consider the covariance of the pseudo-spectrum estimators.
The covariance of $\hat{\bd{P}}(\ell)$ is defined by
\beqa
{\rm Cov}[\hat{P}_{XY}(\ell), \hat{P}_{MN}(\ell^{\prime})]
= \langle \hat{P}_{XY}(\ell)\hat{P}_{MN}(\ell^{\prime})\rangle-
\langle \hat{P}_{XY}(\ell)\rangle \langle \hat{P}_{MN}(\ell)\rangle,
\eeqa
where $X, Y, M, N$ is set to be $\kappa$ or $p$. 
When the underlying field follows non-Gaussian and there exist no masked regions,
the covariance can be expressed as
\beqa
{\rm Cov}[P_{XY}(\ell), P_{MN}(\ell^{\prime})]_{\rm all-sky}
= 
\frac{\delta_{\ell \ell^{\prime}}}{2\ell+1}
\left[P_{XM}(\ell)P_{YN}(\ell^{\prime})+P_{XN}(\ell)P_{YM}(\ell^{\prime})\right]
+
\frac{1}{2\ell+1}\frac{1}{2\ell^{\prime}+1}
\sum_{m m^{\prime}}\langle X_{\ell m}Y^{*}_{\ell m} 
M_{\ell^{\prime} m^{\prime}}N^{*}_{\ell^{\prime} m^{\prime}}\rangle_{c},
\label{eq:cov_allsky}
\eeqa
where the first term of the right-hand side in Eq.~(\ref{eq:cov_allsky}) represents
the Gaussian contribution to the covariance matrix and the second term corresponds 
to the contribution of four-point correlation function 
due to non-Gaussianity in the underlying field.
On the other hand, in presence of masked region,
the covariance of the pseudo-spectrum estimators is expressed as
(see also, e.g., \citet{2005MNRAS.360.1262B})
\beqa
{\rm Cov}[\hat{P}_{XY}(\ell_1), \hat{P}_{MN}(\ell_2)] 
&=& 
{\rm Cov}[\hat{P}_{XY}(\ell_1), \hat{P}_{MN}(\ell_2)]_{\rm NG} 
+
\sum_{\ell^{\prime}_{1} \ell^{\prime}_{2}}
(M^{XY})^{-1}_{\ell_{1} \ell^{\prime}_{1}}
(M^{MN})^{-1}_{\ell_{2} \ell^{\prime}_{2}}
\nonumber \\
&&
\, \, \, \, \, \, \, \, \, \, \, \,
\times
\sum_{\ell^{\prime\prime}_{1} \ell^{\prime\prime}_{2}}
\left[
P_{AD}(\ell^{\prime\prime}_{1})P_{BC}(\ell^{\prime\prime}_{2})
{\cal X}^{[XA, ND, MC, YB]}_{\ell^{\prime}_{1} \ell^{\prime}_{2} \ell^{\prime\prime}_{1} \ell^{\prime\prime}_{2}}
+
P_{AC}(\ell^{\prime\prime}_{1})P_{BD}(\ell^{\prime\prime}_{2})
{\cal X}^{[XA, MC, ND, YB]}_{\ell^{\prime}_{1} \ell^{\prime}_{2} \ell^{\prime\prime}_{1}
\ell^{\prime\prime}_{2}}
\right],
\label{eq:cov_maskedsky}
\eeqa
where 
all the possible combinations of $A, B, C$ and $D$ are taken into account 
in Eq.~(\ref{eq:cov_maskedsky}) and 
${\cal X}^{[XA, ND, MC, YB]}_{\ell \ell^{\prime} \ell_{1} \ell_{2}}$ is given by 
\beqa
{\cal X}^{[XA, ND, MC, YB]}_{\ell \ell^{\prime} \ell_{1} \ell_{2}}
= \frac{1}{(2\ell+1)(2\ell^{\prime}+1)}
\sum
W^{XA}_{\ell m \ell_{1} m_{1}}
\left(W^{ND}_{\ell^{\prime} m^{\prime} \ell_{1} m_{1}}\right)^{*}
W^{MC}_{\ell^{\prime} m^{\prime} \ell_{1} m_{1}}
\left(W^{YB}_{\ell m \ell_{2} m_{2}}\right)^{*}.
\label{eq:coupling_WWWW}
\eeqa
The summation in Eq.~(\ref{eq:coupling_WWWW}) is taken 
over all $m$, $m^{\prime}$, $m_{1}$, $m_{2}$.
Here, $W^{XY}_{\ell m \ell^{\prime} m^{\prime}}$ 
denotes 
$W^{p \kappa} = W^{\kappa p} = 0$,
$W^{\kappa \kappa} = W^{\kappa}$, and $W^{p p} = W^{p}$ 
with Eq.~(\ref{eq:conv_harmonic}). 
The non-Gaussian term in Eq.~(\ref{eq:cov_maskedsky}) is defined by
\beqa
{\rm Cov}[\hat{P}_{XY}(\ell_1), \hat{P}_{MN}(\ell_2)]_{\rm NG}
&=& \frac{1}{(2\ell_{1}+1)(2\ell_{2}+1)}
\sum_{\ell^{\prime}_{1} \ell^{\prime}_{2}}
(M^{XY})^{-1}_{\ell_{1} \ell^{\prime}_{1}}
(M^{MN})^{-1}_{\ell_{2} \ell^{\prime}_{2}}
\nonumber \\
&\times&
\sum 
W^{XA}_{\ell_{1} m_{1} \ell^{\prime}_{1} m^{\prime}_{1}}
\left(W^{YB}_{\ell_{1} m_{1} \ell^{\prime \prime}_{1} m^{\prime \prime}_{1}}\right)^{*}
W^{MC}_{\ell_{2} m_{2} \ell^{\prime}_{2} m^{\prime}_{2}}
\left(W^{ND}_{\ell_{2} m_{2} \ell^{\prime \prime}_{2} m^{\prime \prime}_{2}}\right)^{*}
\langle 
A_{\ell^{\prime}_1 m^{\prime}_1}
B^{*}_{\ell^{\prime \prime}_1 m^{\prime \prime}_1}
C_{\ell^{\prime}_2 m^{\prime}_2}
D^{*}_{\ell^{\prime \prime}_2 m^{\prime \prime}_2}
\rangle_{c},
\label{eq:cov_maskedsky_NG}
\eeqa
where the second summation in Eq.~(\ref{eq:cov_maskedsky_NG})
is over all $m_{i}$, $m^{\prime}_{i}$, $m^{\prime \prime}_{i}$,
$\ell^{\prime}_{i}$, $\ell^{\prime \prime}_{i} \, (i = 1, 2)$
and the values of $A, B, C$ and $D$.
Eqs.~(\ref{eq:cov_maskedsky}) 
and (\ref{eq:cov_maskedsky_NG}) clearly show that 
the complicated masked regions on sky would induce the \textit{additional} 
mode-coupling of the covariance matrix of the pseudo-spectrum estimators.

\clearpage


\bibliographystyle{mn2e}
\bibliography{bibtex_library}

\newpage

\end{document}